%% file: main.tex
\pdfoutput=1 
\documentclass[useAMS,usenatbib]{mn2e}

\usepackage[letterpaper,margin=0.635in]{geometry}
\usepackage[ansinew]{inputenc}
\usepackage{amsfonts}
\usepackage{amsmath}
\usepackage{enumerate}
\usepackage{enumitem}
\usepackage{graphicx,float}
\usepackage{epsfig}
\usepackage{color}
\usepackage{float}
\usepackage[normalem]{ulem}
\usepackage{comment}
\usepackage{times}
\usepackage{epstopdf}

\usepackage{amssymb}
\usepackage{pdflscape}

\def\Mpc{\, h^{-1} \, {\rm Mpc}}

\newcommand{\magauto}{{\tt MAG\_AUTO}\ }
\newcommand{\magdetmodel}{{\tt MAG\_DETMODEL}\ }

\newcommand{\beq}{\begin{equation}}
\newcommand{\eeq}{\end{equation}}
\newcommand{\beqa}{\begin{eqnarray}}
\newcommand{\eeqa}{\end{eqnarray}}

\title[Galaxy clustering, photo-$z$ and systematics in DES-SV]{Galaxy clustering,
  photometric redshifts and diagnosis of systematics \\in the DES Science Verification data}
\input{authorlist}

\begin{document}
\pagerange{\pageref{firstpage}--\pageref{lastpage}} \pubyear{0000}

\maketitle

\label{firstpage}

\begin{abstract}
We study the clustering of galaxies
detected at $i<22.5$ in the Science Verification observations of the
Dark Energy Survey (DES). Two-point correlation functions are measured
using $2.3\times 10^6$ galaxies over a contiguous 116 deg$^2$ region in five bins of photometric redshift width $\Delta z = 0.2$ in the range $0.2 < z < 1.2.$
The impact of photometric redshift errors are assessed
by comparing results using a template-based photo-$z$ algorithm (BPZ)
to a machine-learning algorithm (TPZ).  A companion paper (Leistedt et
al 2015) presents maps of several observational variables (e.g. seeing, sky
brightness) which could
modulate the galaxy density.  Here we characterize and mitigate systematic
 errors on the measured clustering which arise from these observational variables, in
addition to others such as Galactic dust and stellar contamination.
After correcting for systematic effects we measure galaxy bias over a broad range of linear scales relative to mass clustering predicted from the Planck $\Lambda$CDM model, finding agreement with CFHTLS measurements with $\chi^2$ of 4.0 (8.7) with 5 degrees of freedom for the TPZ (BPZ) redshifts.  
We test a ``linear bias" model, in which the galaxy clustering is a
fixed multiple of the predicted non-linear dark-matter clustering.   
The precision of the data allow us to determine that 
the linear bias model describes the observed galaxy clustering to $2.5\%$
accuracy down to scales at least $4$ to $10$ times 
smaller than those on which linear theory is expected to be
sufficient. 
\newline

\end{abstract}


\input{intro}

\input{Sec1}

\input{Sec2}
\input{Sec3}

\input{Sec4}
\input{Sec5}

\input{conclusions}

\input{acknowledgments}

\bibliographystyle{mn2e}
\bibliography{main}
 
\renewcommand{\thesubsection}{\Alph{subsection}}

\section*{Appendix}
\input{appendix.tex}

\bsp
\input{affiliations}
\label{lastpage}

\end{document}

%% file: authorlist.tex
\author[M.~Crocce et al.]{
\parbox{\textwidth}{
\Large M.~Crocce$^{1}$\thanks{crocce@ice.cat}, J.~Carretero$^{1,2}$, A.~H.~Bauer$^{1}$, A.~J.~Ross$^{3}$, I.~Sevilla-Noarbe$^{4,5}$, T.~Giannantonio$^{6,42}$, F.~Sobreira$^{7,8}$, 
J.~Sanchez$^{4}$, E.~Gaztanaga$^{1}$, M.~Carrasco~Kind$^{5,10}$, C.~S{\'a}nchez$^{2}$,
C.~Bonnett$^{2}$, A.~Benoit-L{\'e}vy$^{9}$, R.~J.~Brunner$^{5,10}$, A.~Carnero~Rosell$^{8,11}$, R.~Cawthon$^{12}$, P.~Fosalba$^{1}$, W.~Hartley$^{13}$, E.~J.~Kim$^{5}$, B.~Leistedt$^{9}$, R.~Miquel$^{2}$, H.~V.~Peiris$^{9}$, W.J.~Percival$^{15}$, R.~Rosenfeld$^{8,16}$, E.~S.~Rykoff$^{17,18}$, E.~S{\'a}nchez$^{4}$, T.~Abbott$^{19}$, F.~B.~Abdalla$^{9,43}$, S.~Allam$^{7}$, M.~Banerji$^{6,20}$, G.~M.~Bernstein$^{21}$, E.~Bertin$^{22,23}$, D.~Brooks$^{9}$, E.~Buckley-Geer$^{7}$, D.~L.~Burke$^{17,18}$, D.~Capozzi$^{15}$, F.~J.~Castander$^{1}$, C.~E.~Cunha$^{17}$, C.~B.~D'Andrea$^{15}$, L.~N.~da Costa$^{8,11}$, S.~Desai$^{24,25}$, H.~T.~Diehl$^{7}$, T.~F.~Eifler$^{21,26}$, A.~E.~Evrard$^{27,28}$, A.~Fausti Neto$^{8}$, E.~Fernandez$^{2}$, D.~A.~Finley$^{7}$, B.~Flaugher$^{7}$, J.~Frieman$^{7,12}$, D.~W.~Gerdes$^{28}$, D.~Gruen$^{29,30}$, R.~A.~Gruendl$^{5,10}$, G.~Gutierrez$^{7}$, K.~Honscheid$^{3,31}$, D.~J.~James$^{19}$, K.~Kuehn$^{32}$, N.~Kuropatkin$^{7}$, O.~Lahav$^{9}$, T.~S.~Li$^{33}$, M.~Lima$^{8,34}$, M.~A.~G.~Maia$^{8,11}$, M.~March$^{21}$, J.~L.~Marshall$^{33}$, P.~Martini$^{3,35}$, P.~Melchior$^{3,31}$, C.~J.~Miller$^{27,28}$, E.~Neilsen$^{7}$, R.~C.~Nichol$^{15}$, B.~Nord$^{7}$, R.~Ogando$^{8,11}$, A.~A.~Plazas$^{26}$, A.~K.~Romer$^{36}$, M.~Sako$^{21}$, B.~Santiago$^{8,37}$, M.~Schubnell$^{28}$, R.~C.~Smith$^{19}$, M.~Soares-Santos$^{7}$, E.~Suchyta$^{3,31}$, M.~E.~C.~Swanson$^{10}$, G.~Tarle$^{28}$, J.~Thaler$^{38}$, D.~Thomas$^{15}$, V.~Vikram$^{39}$, A.~R.~Walker$^{19}$, R.~H.~Wechsler$^{17,18,40}$, J.~Weller$^{24,29,30}$, J.~Zuntz$^{41}$ 
\begin{center} (The DES Collaboration) \end{center}}
\\
Affiliations are listed at the end of the paper}

%% file: intro.tex
\section{Introduction}
\label{sec:intro}

Vast galaxy surveys trace the large-scale structure (LSS) of the Universe at late times and therefore complement and improve the wealth of information already provided by cosmic microwave background (CMB) and supernovae experiments. In particular our understanding and characterization of the cosmic accelerated expansion. Imaging from multi-band photometry, e.g., SDSS \citep{2000AJ....120.1579Y}, PanSTARRS \citep{2000PASP..112..768K}, KiDS \citep{2013Msngr.154...44J}, HSC \citep{2012SPIE.8446E..0ZM} 
and the planned LSST \citep{2003NuPhS.124...21T}, provides the angular positions and detailed color information of the galaxies. From this color information, photometric redshifts (photo-$z$) can be measured for each galaxy providing distance estimates that have typically  low resolution. While obtaining detailed radial information requires a spectroscopic redshift survey, e.g., 2dF \citep{2001MNRAS.328.1039C}, VVDS \citep{2005A&A...439..845L}, WiggleZ \citep{2010MNRAS.401.1429D} and BOSS \citep{2013AJ....145...10D}, obtaining spectra is more time consuming. Hence the data volume available to a photometric redshift sample will naturally far exceed that of a spectroscopic one. Many LSS studies 
have already obtained constraints on cosmological parameters from photo-$z$ samples, such as the sum of neutrino masses, the matter/energy content of the Universe, and the nature of dark energy \citep{Pad07,Thomas10,2011MNRAS.417.2577C,2012ApJ...761...14H,2012MNRAS.419.1689C,2013MNRAS.435.3017D}.

The Dark Energy Survey (DES, \cite{2005IJMPA..20.3121F}) will image 5000 deg$^2$ of the South Galactic Cap to a depth of $i < 24$, recording 300 million galaxies in 5 broadband filters ($grizY$), thereby providing high quality photometric redshifts up to redshift $z=1.4$ \citep{2014MNRAS.445.1482S}. The DES camera (DECam, \citealt{2015arXiv150402900F,2012PhPro..37.1332D}) was installed and commissioned in the second semester of 2012. A Science Verification (SV) period of observations followed, lasting from November 2012 to February 2013. The DES observations officially started in late August 2013.

In this paper, we analyze the clustering of galaxies observed during the SV period. The SV region has been observed to match the nominal 24th mag depth in the $i$- band expected for the full DES survey, and we can therefore use the data to characterize the properties of the galaxy samples that DES can reliably observe. We perform detailed systematic tests on the galaxy samples and develop methods to  mitigate systematic effects on the measured clustering. We asses the impact on clustering measurements from uncertainties in photometric redshift estimation by obtaining our samples using two different photo-$z$ algorithms, one based on a template fitting method and another on a machine learning one. Each of these methods has advantages and disadvantages, which we discuss. Our resulting clustering measurements allow us to characterize the evolution of bias as a function of redshift, to assess the validity of linear bias models and to provide baseline clustering measurements that other DES studies using SV data can use to compare against (e.g., \citealt{GiannantonioCMBLens,balrog}). We were able to complete this work without the benefit of a large number of simulated DES SV galaxy samples, but future DES LSS studies (covering larger volumes) will rely on such simulations. 

The outline of the paper is as follows: we introduce the clustering estimators and the theory used throughout the paper in Section~\ref{sec:angular_clustering}, describe our data set in Sections~\ref{sec:data} and \ref{sec:benchsample}, and present the maps of potential systematics in Section~\ref{sec:sysmaps}. In Section~\ref{sec:systematics} we discuss the extent and mitigation of possible systematics, summarize and discuss our results in Section~\ref{sec:results}, before concluding in Section~\ref{sec:conclusions}.

%% file: Sec1.tex
\section{Angular Clustering: Theory and Estimators}
\label{sec:angular_clustering}

In this section we review the modeling of the angular correlation function and its covariance matrix, in photo-$z$ bins. We also describe the algorithms used to estimate these quantities from the data.

\subsection{Two-point Angular Correlations: Modeling}
\label{sec:theory_modeling}

Galaxy clustering can be modeled starting from the dark matter overdensity field at the angular position $ \mathbf{\hat n}$ and at redshift $z$: $\delta( \mathbf{\hat n}, z)$. If we assume a linear bias model, the projected overdensity of our galaxy sample is given by,
\begin{equation}
\delta_g(\mathbf{\hat n}) = \int_0^{\infty} b(z) \, \frac {dn}{dz} (z) \, \delta(\mathbf{\hat n}, z) \, dz \, .
\end{equation}
where $dn/dz$ is the probability of detecting a galaxy at a given redshift (i.e. the normalised redshift distribution)
and $b(z)$ is the bias. The angular two-point correlation function between redshift bins $A$ and $B$ is then defined as,
\begin{eqnarray}
w_{AB}(\theta) &\equiv& \langle \delta_g(\mathbf{\hat n}) \delta_g(\mathbf{\hat n}+\hat{\bf \theta}) \rangle =  \nonumber \\
& = & \int dz_1 \, \phi_A(z_1) \int dz_2 \, \phi_B(z_2)
\, \xi(r_{12}(\theta),\bar{z})
\label{eq:wtheta1}
\end{eqnarray}
where $r^2_{12}=r(z_1)^2+r(z_2)^2- 2 r(z_1) r(z_2) \cos (\theta)$ with $r(z)$ being the comoving distance to redshift $z$ (assuming a flat Universe), and
\begin{equation}
\phi(z)=\frac{D(z)}{D(\bar{z})} b(z) \frac{dn}{dz}(z),
\nonumber
\end{equation}
where $D(z)$ is the linear growth factor. Equation (~\ref{eq:wtheta1}) should in principle be written in terms of a spatial correlation function $\xi$ that evolves with redshift. Instead we evaluate the spatial correlation at some mean redshift $\bar{z}$ and encode the growth evolution relative to this redshift into the selection function $\phi$ (this is exact at the linear level) which we make proportional to $D(z)/D(\bar{z})$. In turn, rather than assuming a parametric function for $b(z)$, we will measure a single bias value at each tomographic photo-$z$ bin. Lastly, the way we estimate the redshift distribution of the sample, $dn/dz$, is described in Sec.~\ref{sec:benchsample} and shown in Fig.~\ref{fig:dndz}.

We calculate the linear and nonlinear power spectra using CAMB (Lewis et al. 2000). For nonlinear dark matter clustering we use the recently re-calibrated {\tt Halofit} prescription \citep{Takahashi2012} built into CAMB. We then Fourier transform these into configuration space to obtain $\xi$ and evaluate Eq.~(\ref{eq:wtheta1}).
Even though the effect of redshift space distortions is negligible for the angular scales considered in this paper,  we do include them in our predictions (e.g. see formulae in \citealt{2011MNRAS.414..329C}).

Throughout the paper we assume a fiducial flat $\Lambda$CDM+$\nu$ (one massive neutrino) cosmological model based on \textit{Planck} 2013 + \textit{WMAP} polarisation + ACT/SPT + BAO, with parameters \citep{Planck13cosmo}
: $\omega_b = 0.0222$, $\omega_c = 0.119$, $\omega_{\nu} = 0.00064$, $h = 0.678$,   $\tau = 0.0952$, $n_s = 0.961$ and $A_s = 2.21 \cdot 10^{-9}$ at a pivot scale $\bar k = 0.05$ Mpc$^{-1}$ (yielding $\sigma_8=0.829$ at $z=0$) where $h \equiv H_0 / 100 \,{\rm km}\,{\rm s}^{-1} \,{\rm Mpc}^{-1}$ and $\omega_i \equiv \Omega_i h^2$ for each species $i$.

\subsection{Two-point Angular Correlations: Estimators}
\label{sec:2ptCF}

In this paper we use two different estimators for angular correlation functions.  One is a direct pair-count algorithm particularly suited to investigating small-scale clustering, and another based on pixelized maps using {\tt {\tt HEALPix}} which is particularly useful to investigate the cross-correlation of our galaxy sample with maps of potential systematics (such as observing conditions) in an efficient way at the expense of decreased angular resolution.

The pair-counting algorithm uses the publicly available tree code ATHENA \footnote{\tt www.cosmostat.org/software/athena} to measure the angular correlation function with the standard \cite{LS} estimator,
\begin{equation}
w(\theta) = \frac{DD-2DR+RR}{RR}
\label{eq:wls}
\end{equation}
where $DD$, $DR$ and $RR$ represent the number of pairs of objects, taken from the galaxy catalogue ($D$) or from a randomly generated sample covering the angular footprint ($R$), which are separated by a scale $\theta$ within a bin size $\Delta\theta$.
The random points are distributed uniformly over the footprint. As we discuss in Sec.~\ref{sec:completeness} our angular footprint considers only regions of the survey that are deeper than the flux limit in the sample definition, therefore the distribution of random points is uniform in regions where the sample is complete.

The pixel-based estimator for the correlation between maps 1 and 2 is given by,
\begin{equation}
w_{1,2}(\theta) = \sum_{i=1}^{i=N_{\mathrm{pix}}}\sum_{j=i}^{j=N_{\mathrm{pix}}}\frac{(N_{i,1}-\bar{N_1})(N_{j,2}-\bar{N_2})}{\bar{N_1}\bar{N_2}}\Theta_{i,j},
\label{eq:wpix}
\end{equation}
where $N_{\mathrm{pix}}$ is the number of pixels used, $N_i$  is the value in pixel $i$ of the quantity of interest (either the number of galaxies or, e.g., the seeing), $\bar{N}$ represents the mean over all pixels, and $\Theta_{i,j}$ is 1 if pixels $i,j$ are separated by a scale $\theta$ within a bin size $\Delta\theta$ and is zero otherwise. The sum only runs through those pixels that fall within the footprint. 

\subsection{Covariance Matrix}
\label{sec:covariance}

In order to estimate the covariance matrix for our $w(\theta)$ measurements we combine two different approaches: a modeling of the off-diagonal elements and the jackknife (JK) method for the variance.

The jackknife technique, which has been widely applied to angular galaxy clustering measurements at small scales (e.g., \citealt{Scranton02,Myers06,Ross09,2012A&A...542A...5C}) works by splitting the footprint into $N_{\mathrm{JK}}$ small equal-area sections and recomputing $w(\theta)$ with each section removed. The covariance matrix between $w(\theta)$ measurements at two angular separations between two redshift bins, $A$ and $B$, is estimated by,
\begin{eqnarray}
\indent{\rm Cov}^{JK, AB}_{\theta_i,\theta_j} &=& \frac{N_{\mathrm{JK}}-1}{N_{\mathrm{JK}}}\sum_{k=1}^{N_{\mathrm{JK}}}\Delta w^A[\theta_i]\,\Delta w^B[\theta_j], \label{eq:JKcovariance} \\
\Delta w^{A(B)} [\theta] &=& \left(w^{A\,(B)}_k[\theta]-\bar{w}^{A\,(B)}[\theta]\right),
\label{eq:none}
\end{eqnarray}
where $w_k$ is the correlation function measured with the $k$-th JK region removed and $\bar{w}$ is 
the mean of the $N_{\mathrm{JK}}$ jackknife $w(\theta)$'s.  
In our case we use $40$ JK regions, in such a way that each patch is $\sim 2\,{\rm deg}$ across, matching our largest scales of interest. 

Theoretical estimates of the covariance for angular clustering have also been explored and tested (e.g. \cite{2003moco.book.....D,2007MNRAS.381.1347C,2011MNRAS.414..329C,2011MNRAS.415.2193R}). They rely on the assumption that errors scale as the square root of the survey area $\mathcal{C}_{\theta_i,\theta_j} \sim f_{sky}^{-1} \mathcal{C}^{\rm full\,sky}_{\theta_i,\theta_j}$. The covariance for a full-sky survey is then easily derived considering that angular power spectrum measurements $C_\ell$ are uncorrelated, ${\rm Cov}_{\ell \ell^{\prime}}={\rm  Var}(C_{\ell})\delta_{\ell\ell^{\prime}}$
with ${\rm Var}(C_{\ell}) = 2 \, C_{\ell}^2 / (2\ell+1)$. One can then use a Legendre transform to express this error in real space for $\mathcal{C}^{\rm full\,sky}_{\theta_i,\theta_j}$ and arrive to, 
\begin{equation}
{\rm Cov}^{TH}_{\theta_i,\theta_j} =
\frac{2}{f_{sky}}\sum_{\ell\ge 0}
\frac{2\ell+1}{(4\pi)^2}P_{\ell}(\cos\theta_i) P_{\ell}(\cos
\theta_j) (C_\ell+1/{\bar n})^2
\label{eq:CovAuto}
\end{equation}
for auto-correlations in one bin, and 
\begin{eqnarray}
{\rm Cov}^{TH, AB}_{\theta_i,\theta_j} &=&
\frac{1}{f_{sky}}\sum_{\ell\ge 0}
\frac{2\ell+1}{(4\pi)^2}P_{\ell}(\cos\theta_i) P_{\ell}(\cos
\theta_j) \left[\left(C^{AB}_\ell\right)^2\right. \nonumber \\
&+&\left.\left(C^{AA}_{\ell}+1/{\bar n_a} \right) \left(C^{BB}_{\ell}+1/{\bar n_b} \right)\right]
\label{eq:CovFull}
\end{eqnarray}
for cross-correlations between bins A and B. In Eqs.~(\ref{eq:CovAuto},\ref{eq:CovFull}) we have included the standard shot-noise contribution
arising in the variance of the $C_{\ell}$ estimates 
(${\bar n}$ is the number of objects per steradian).
The angular power spectrum between bins A and B is given by
\begin{equation}
C^{AB}_\ell = \frac{2}{\pi} \int dk k^2 P(k,\bar{z}) W^A_\ell(k) W^B_\ell(k)
\label{eq:cl}
\end{equation}
with the kernel defined for each bin by
\begin{equation}
W_\ell(k) = \int dz \, b(z) \, \frac{dn}{dz}(z) \, \frac{D(z)}{D(\bar{z})} \,  j_\ell(k r(z))
\label{eq:Wl}
\end{equation}
As before, we evaluate the power spectrum in Eq.~(\ref{eq:cl}) at some mean redshift and then compute the growth within $W_\ell$ with respect to this reference point. In Eq.~(\ref{eq:Wl}) each bin is characterized by its $dn/dz$ and $b$, while $j_\ell$ are the spherical Bessel functions. 

Measurements of the correlation function at different angular scales are considerably correlated. The JK estimation is limited by the number of independent samples one can build out of the given footprint and hence yields noisy off-diagonal correlations. Obtaining good estimates of the off-diagonal elements is important, as the correlation $w(\theta)$ between adjacent $\theta$ bins are typically greater than 0.9. For the diagonal elements of the covariance matrix, we expect the JK estimation to be adequate and favor it over the theoretical estimation as it is extracted from the data itself and hence can trace effects such as non-linearities, the effect of the mask, and residual systematic variation. The theory estimate on the other hand is adequate for off-diagonal elements because it avoids the intrinsic noise from the limited number of realizations. 
Therefore we combine both approaches by defining our covariance matrix as, 
\begin{equation}
{\rm Cov}^{MIX}_{\theta_i,\theta_j} = {\rm Corr}^{TH}_{\theta_i,\theta_j} \sigma_{JK}(\theta_i) \sigma_{JK}(\theta_j)
\label{eq:mixcov}
\end{equation} 
where $\sigma^2_{JK}(\theta_i) \equiv \mathcal{C}^{JK}_{\theta_i,\theta_i}$ from Eq.~(\ref{eq:JKcovariance}) and
${\rm Corr}^{TH}_{\theta_i,\theta_j}$ is the correlation matrix (or reduced covariance matrix) defined by
\begin{equation}
{\rm Corr}^{TH}_{\theta_i,\theta_j} = \frac{{\rm Cov}^{TH}_{\theta_i,\theta_j}}{ \sigma_{TH}(\theta_i) \sigma_{TH}(\theta_j)}
\end{equation}
from Eqs.~(\ref{eq:CovAuto},\ref{eq:CovFull}). We refer to this as the ``mixed'' approach.

The methodology we use to model our covariance matrix is further justified by the results presented in a companion paper \citep{GiannantonioCMBLens}, where we present a discussion of various estimators for the covariance matrix of auto-correlations. As naively expected, it is found that the JK method performs better than theory errors for diagonal correlations (i.e. variance) when both are compared to a covariance matrix derived from an N-body simulation.
But under-performs compared to theory ones for the off-diagonal elements. In Appendix \ref{sec:appendix_b} we show that on large scales both the JK and theory approaches yield similar results for the variance, and for derived best-fit biases.

%% file: Sec2.tex
\section{The DES Science Verification (SVA-1) photometric sample} 
\label{sec:data}

\subsection{The Science Verification dataset}

The Science Verification (SV) observations, taken over 78 nights in 2012 and 2013, provided science-quality data for more than 250 deg$^2$~at close to the nominal depth of the survey. The SV data footprint was chosen to contain areas already covered by several deep spectroscopic galaxy surveys, including VVDS \citep{2005A&A...439..845L}, zCOSMOS \citep{2007ApJS..172...70L}, and ACES \citep{2012MNRAS.425.2116C}, which together provide a calibration sample for the DES photometric redshifts \citep{2014MNRAS.445.1482S}. In addition to these, two large contiguous regions (termed SPT-E and SPT-W, due to their partial coverage of the South Pole Telescope \citep{SPTref} fields) were observed.

\subsection{Image processing: the DESDM pipeline}

The raw images taken each night by DECam are sent for processing to the National Center for Supercomputer Applications (NCSA\footnote{http://www.ncsa.illinois.edu/}, 
Urbana-Champaign, United States) using the DES Data Management (DESDM) pipeline described in \cite{2011arXiv1109.6741S} and \cite{2012SPIE.8451E..0DM}. We briefly describe the process below. 

These single-epoch images are 'detrended' by applying bias and flat-fields corrections. Then, cross-talk, pupil, illumination, fringing and non-linearity corrections are carried out. Bad pixels are flagged and traced in a map while satellite trails, cosmic rays and saturation effects from bright stars are detected and flagged. The astrometric calibration is performed using \texttt{SCAMP} \citep{ScampSoft}, fitting the best local-to-celestial reference system transformation using the single-epoch image positions of bright sources. One then relates them to reference celestial positions, correcting for distortions from the large field-of-view. Nightly absolute calibration is performed using reference stars which 
are observed every night, to obtain a photometric solution for that period including the  zero point, a color, and airmass terms for each CCD.

The resulting reduced images are then re-mapped into a uniform pixel grid in which the coaddition of said images is performed. The purpose of this process is to obtain increased depth.

Lastly, the catalog creation stage uses the \texttt{SExtractor} \citep{SextractorSoft} software , which detects objects on the coadded images for each band and stores them as unique single object entries at a database at NCSA. During this procedure, the coadded astrometric and photometric information is measured and stored as well. The astrometry is accurate to 0.2$^{\prime\prime}$. A global photometric relative calibration step is performed for this dataset using the exposure overlap across different fields, which provides repeated observations of the same stars so a minimization procedure can be performed \citep{Glazebrook94}. Observations from PreCAM \citep{Kuehn13} provide a grid of standards to which this relative calibration can be anchored to in order to minimize large scale variations\footnote{This calibration is improved using the Stellar Locus correction described below.}. Finally, observations of spectrophotometric standards ties this photometry to the absolute AB system \citep{Tucker07}. This final data set represents the ``coadd'' catalog.

\subsection{The Gold catalog}

\begin{figure}
\begin{center}
\includegraphics[width=0.5\textwidth]{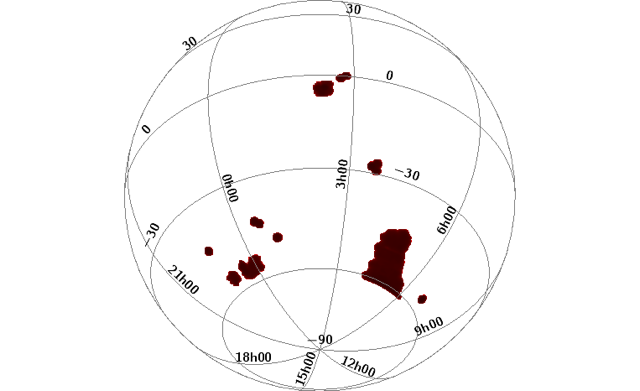} 
\caption{Footprint of the SVA1 Gold catalog. In this analysis, we only use data from the largest compact footprint occupying the lower-right, known as `SPT-E'.}
\label{fig:footprint}
\end{center}
\end{figure}

The coadd catalog produced by DESDM of the SV dataset was analyzed and tested resulting in the generation of the \textit{SVA1 Gold} catalog (Rykoff et al., in prep.)
The footprint of the catalog in equatorial coordinates is shown in Fig.\ref{fig:footprint}.

The basic additions of this value added catalog include:
\begin{itemize}
\item Incorporating satellite trail and other artifact information to mask out specific areas;
\item Removing areas where the colors are severely affected by stray light in the images and areas with a small exposure count (at the borders of the footprint);
\item Removing the area below declination of $-61^{\circ}$ to avoid the increased stellar contamination in our galaxy catalog due to the Large Magellanic Cloud (LMC), as well as the area dominated by the nearby star R Doradus.
\item Applying an additional Stellar Locus correction (\cite{SLRref}) to tighten the calibration even further, ensuring the agreement of stellar colors with respect to those of reference stars. After this correction, magnitudes are calibrated to 2 per cent accuracy, with a 1 per cent accuracy in colors (Rykoff et al. in prep.).
\end{itemize}

After this selection, about 244 deg$^2$ remain in the catalog.

\subsubsection{Survey mask}
\label{sec:limiting_depth_map}

Concurrently with the generation of the catalog, another pipeline builds 
a nominal \texttt{MANGLE}\footnote{http://space.mit.edu/\textasciitilde molly/mangle/} mask \citep{MollyMangle} that takes into account the DECam CCD pointings and properties of the sky during each night. Artifacts such as airplane or satellite trails, cosmic rays, etc., and areas near bright stars are masked out. An estimation of the depth for the remaining regions  
 is calculated as the magnitude at which an object is measured with signal-to-noise of $10$ in a $2$ arcsecs aperture.  
This aperture-based depth must be converted into a total-magnitude depth corresponding to \texttt{SExtractor}'s $\magauto$ measurements used in the galaxy selection.  This conversion, described in Rykoff et al. (in prep.), takes into account properties of the data such as the seeing, which affects the relationship between a fixed aperture and a total magnitude measurement. The resulting 10$\sigma$ depth map is then translated to an averaged weighted pixelized map of resolution given by $N_{\mathrm{side}}=4096$ (an angular scale of 0.74 arcmin$^2$) using the \texttt{HEALPix} software \citep{HealpixSoft}. These maps will be a key component of our final masks (see below). See also the appendix of \cite{balrog} for a description of how the mask is constructed.

\section{The LSS bench-mark sample}
\label{sec:benchsample}

From the \textit{SVA1 Gold} catalog described above 
we select a sample of galaxies that enables robust clustering measurements by applying further magnitude and color cuts and restricting our analysis to the largest contiguous area overlapping the SPT-E field\footnote{Note that we are being very conservative with potential contamination from the LMC by removing one extra stripe at $-61 < dec\,{\rm [deg]} < -60$.},
\begin{align}
60  & < ra\, [{\rm deg}] < 95  \nonumber \\
-60 & < dec \, [{\rm deg}] < -40 
\end{align}
We then focused on an flux limited sample defined by
\begin{equation}
18 < i < 22.5,
\end{equation}
where $i$ refers to the \texttt{SExtractor}'s $\magauto$ quantity. In Sec.~\ref{sec:completeness} we discuss that this sample is complete in regions of the survey deeper than $i = 22.5$ magnitudes. Therefore we will only consider such regions as our baseline footprint, covering an area of $116.2\,{\rm deg}^2$.
Within this footprint we have $2,333,294$ objects in the bench-mark LSS sample with number density $n_g=5.6\,{\rm arcmin}^{-2}$.

We also perform the following additional color cuts
\begin{align}
-1 < g - r < 3  \nonumber \\
-1 < r - i < 2 \nonumber\\
\,-1 < i - z < 2  ,
\end{align}
in order to remove outliers in color space. While $\magauto$ measures galaxy properties independently for each band, $\magdetmodel$ applies a consistent morphological model across all bands, yielding more accurate galaxy colors.  We therefore use $\magdetmodel$ when making the above color cuts on the sample.

\subsection{Star-galaxy separation}
\label{sec:stargalaxyseparation}

\begin{figure}
\centering
\includegraphics[width=0.44\textwidth]{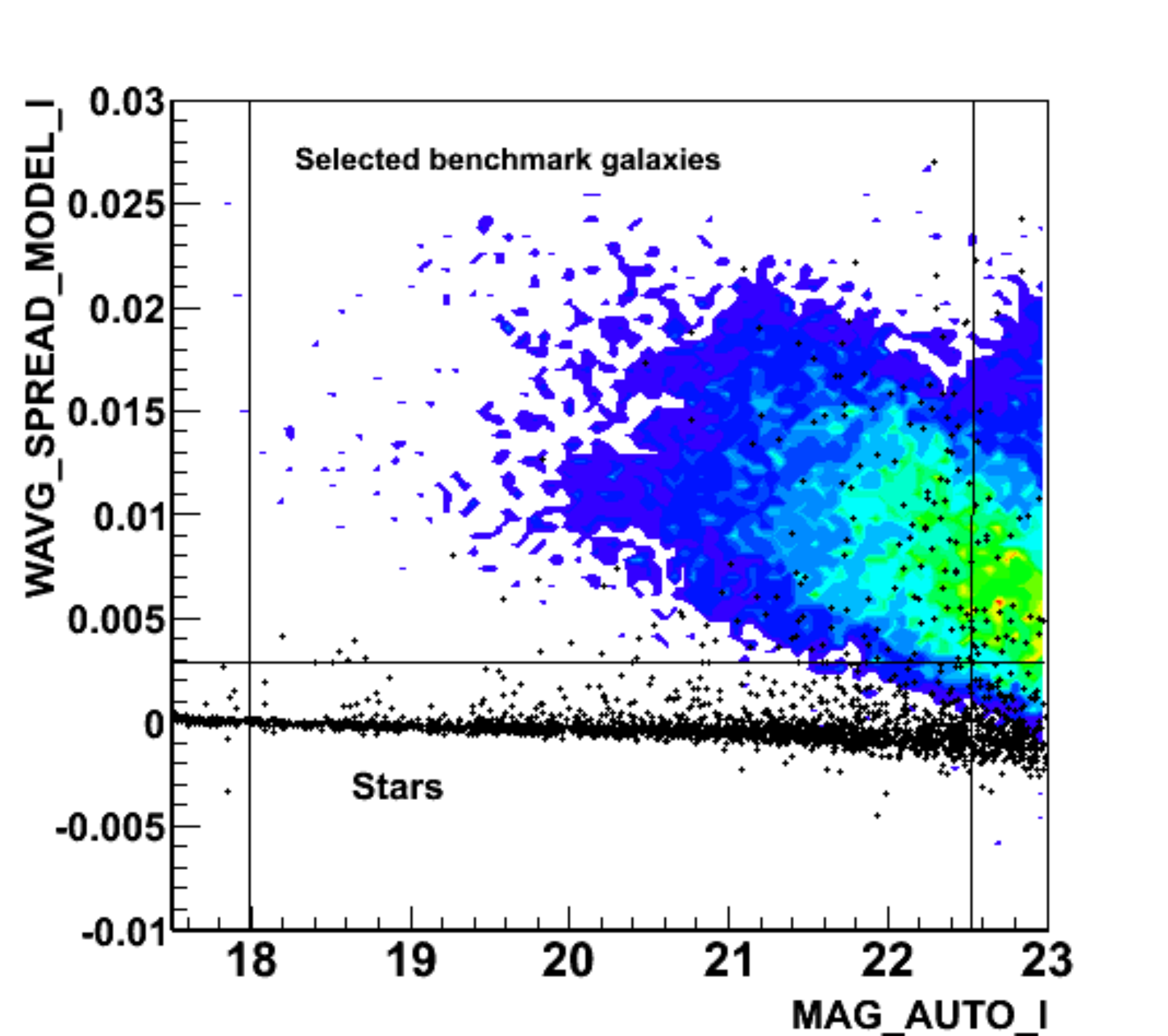}
\caption{Star-galaxy separator distribution of true stars and galaxies vs. magnitude from a combined spectroscopic \citep{2014MNRAS.445.1482S} and space imaging sample \citep{2007ApJS..172..219L}. Color contour corresponds to the true galaxy locus whereas true stars are depicted as dots. Magnitude cuts for the sample used in this work are shown as vertical lines, together with the selection of stars and galaxies using the $\texttt{WAVG\_SPREAD\_MODEL}$ classifier in horizontal lines.} 
\label{fig:stargal_distribution}
\end{figure}

The selection of galaxies (as opposed to stars) from the Gold catalog is achieved using a star-galaxy classifier that relies on the \texttt{SExtractor} \texttt{SPREAD\_MODEL} parameter \citep{2012ApJ...757...83D,2013arXiv1306.5236S} which measures the light concentration of each object using a linear discriminant analysis at the pixel level. This value is calculated for all single epoch images contributing to the final coadded object and the classifier is built using a weighted average expression of the individual \texttt{SPREAD\_MODEL} values. The final value is then termed \texttt{WAVG\_SPREAD\_MODEL}.
The star-galaxy separation is done selecting objects with, 
\begin{equation}
\texttt{WAVG\_SPREAD\_MODEL}> 0.003
\end{equation}

In order to assess the efficacy of this cut, we use DES observations over the COSMOS field, in which we can make use of the measurements from the ACS instrument of the Hubble Space Telescope in that area as a source of 'truth' information for morphological separation of stars and galaxies \citep{2007ApJS..172..219L}, and also the DES observations that overlap with the spectroscopic samples that were used in \citet{2014MNRAS.445.1482S} (which provide truth information based on the spectra). The distribution of $\texttt{WAVG\_SPREAD\_MODEL}$ as a function of $i$-band magnitude for these objects is shown in Fig.~\ref{fig:stargal_distribution}.
This data allow us to determine a contamination level of $1.5\%$ from stars, with a cut efficiency in the galaxy sample of $96\%$ (for DES objects with $i < 22.5$). The impact of stellar contamination on clustering measurements is discussed in Section \ref{sec:stellcont}.

\subsection{Photometric redshift estimation and redshift binning} 
\label{sec:photoz}

A priori, different photo-$z$ codes will lead to different estimates of galaxy redshifts, which then will propagate into theory predictions and eventually into the fitted parameters. There exist two main approaches for photometric redshift estimation: template fitting methods and machine learning ones. Each of these approaches have their advantages and disadvantages, in particular they depend differently on how the spectroscopic set is used, which can be used for training a machine learning approach or to derive fitting priors on the template-based one. Their differences also depend on  the distribution of the magnitude errors, the survey depth, the observing conditions and the galaxy population among others. 

A direct comparison of clustering observables and derived quantities such as bias between these two approaches is a good way of assessing the impact of photometric redshift estimation on clustering measurements. To our knowledge this test has not been done in clustering analyses in the literature. In this paper, we employ two different algorithms to estimate photometric redshift: BPZ (Bayesian Photometric Redshifts), a well known template-fitting  based method \citep{2000ApJ...536..571B,2006AJ....132..926C}  and TPZ \citep{2013MNRAS.432.1483C,2014MNRAS.442.3380C}, a high-performing machine learning algorithm for DES data. 

\begin{table}
\begin{center}
\begin{tabular}{ccc} 
 \hline
  Photo-$z$ bin & $N_{\rm BPZ}$ & $N_{\rm TPZ}$ \\
 \hline
 $0.2 < z < 0.4$ & $684,416$ & $441,791$  \\ 
 $0.4 < z < 0.6$ & $759,015$ & $721,696$  \\ 
 $0.6 < z < 0.8$ & $494,469$ & $586,510$ \\
 $0.8 < z < 1.0$ & $270,077$ & $361,937$  \\ 
 $1.0 < z < 1.2$ & $55,954$ & $93,958$    \\ 
 \hline
\end{tabular}
\end{center}
\caption{The number of galaxies in each of our photo-$z$ bins, when using the template based method BPZ ($N_{\rm BPZ}$) and when using the machine learning method TPZ ($N_{\rm TPZ}$).}
\label{tab:ngal}
\end{table}

\subsubsection{Template Fitting method}

BPZ\footnote{http://www.stsci.edu/$\sim$dcoe/BPZ/} compares the broad-band galaxy spectral energy distribution to a set of galaxy templates until a best fit is obtained, which determines both the galaxy spectral type and its photometric redshift. The details and capabilities of BPZ on early DES data are presented in \citet{2014MNRAS.445.1482S}, where it shows the best performance among template-based codes. The primary set of templates used contains the \citet{1980ApJS...43..393C} templates, two starburst templates from \citet{1996ApJ...467...38K} and two younger starburst simple stellar population templates from \citet{2003MNRAS.344.1000B}, added to BPZ in \citet{2006AJ....132..926C}. We calibrate the Bayesian prior by fitting the empirical function $\Pi(z,t|m_0)$ proposed in \cite{2000ApJ...536..571B}, using a spectroscopic sample matched to DES galaxies and weighted to mimic the photometric properties of the DES-SV sample used in this work.
 
\subsubsection{Machine Learning method}

TPZ\footnote{http://lcdm.astro.illinois.edu/research/TPZ.html} \citep{2013MNRAS.432.1483C,2014MNRAS.442.3380C} is a machine learning algorithm that uses prediction trees and random forest techniques to produce robust photometric redshift probability density functions (PDFs) together with ancillary information for a given galaxy sample. The prediction tree is built by asking a sequence of questions that recursively split the input data taken from the spectroscopic sample, frequently into two branches, until a terminal leaf is created that meets a stopping criterion (e.g., a minimum leaf size or a variance threshold). By perturbing the data using their magnitude errors and by taking bootstrapping samples, many (600 in this case) uncorrelated trees can be created whose results are aggregated to construct each individual photo-$z$ PDF.

\citet{2014MNRAS.445.1482S} contains detailed information about the application of TPZ and BPZ to DES Science Verification data including the derivation of photometric redshifts used in this paper. It also
presents a comparison of these two algorithms to numerous other existing photo-$z$ methods. They both proved to be among the best performing codes in the tests presented in that work, motivating their use in further science analyses using DES-SV data. The 68th percentile widths (corresponding to the scatter in the photo-$z$ solution) were found to be 0.078 for TPZ and 0.097 for BPZ, with 3$\sigma$ outlier fractions being 2 per cent for both algorithms, for detailed metrics see \citet{2014MNRAS.445.1482S}. 


\subsubsection{Redshift binning}

Because we wish to study the evolution of galaxy clustering with redshift, we split our sample in five photo-$z$ bins of width $\Delta z = 0.2$, from $z_{phot}=0.2$ up to $z_{phot}=1.2$. For this analysis we use two photometric redshift estimation algorithms which provide photo-$z$ probability density functions (PDF). One is a standard template-fitting based code, BPZ, which we take as our default photo-$z$ algorithm. For robustness and cross-validation we also use a machine learning based algorithm, TPZ. Both are discussed in more detail in Sec.~\ref{sec:photoz}. The number of galaxies in each photo-$z$ bin is given in Table \ref{tab:ngal}.

\begin{figure}
\begin{center}
\includegraphics[width=0.4\textwidth]{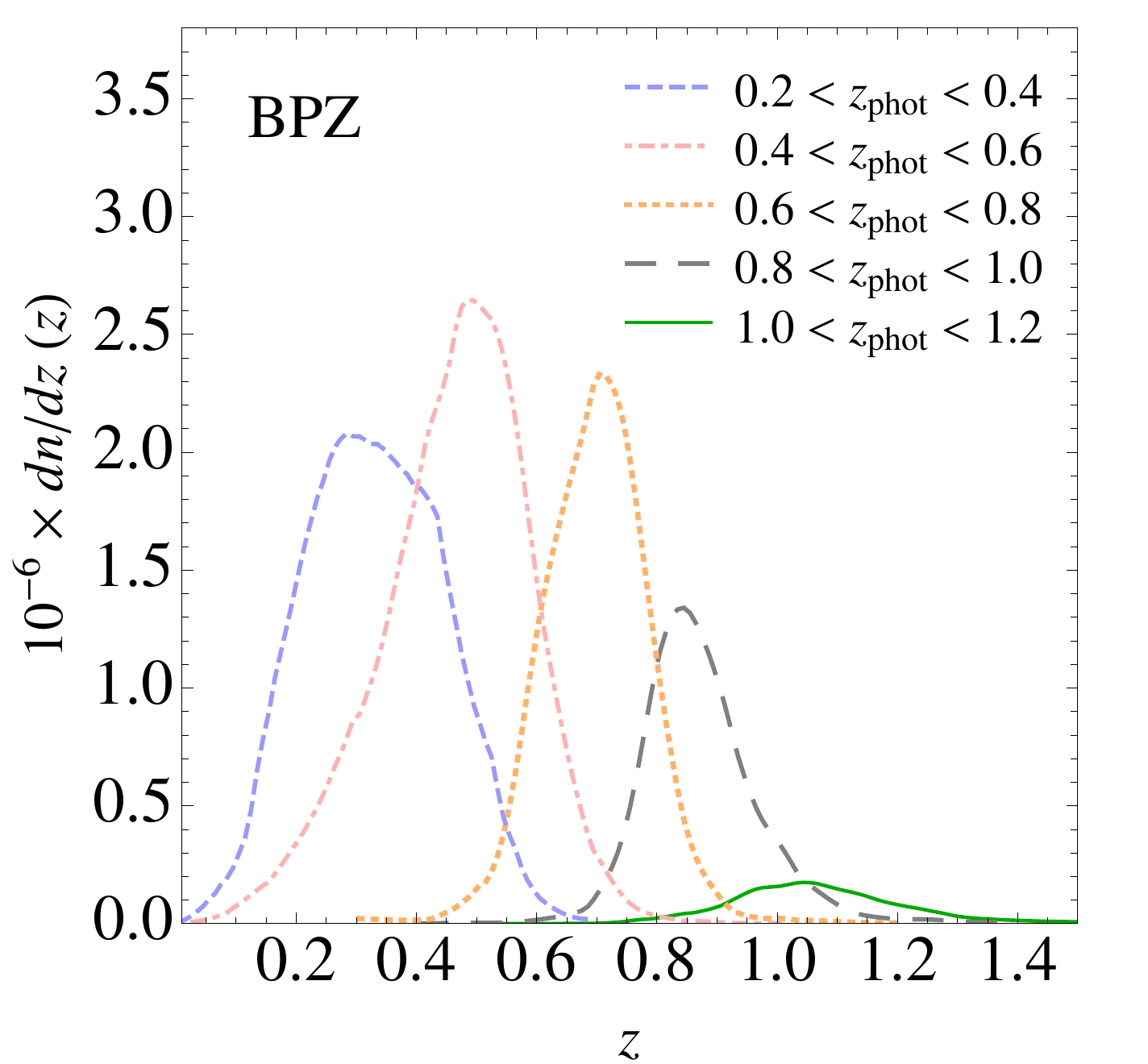} 
\includegraphics[width=0.4\textwidth]{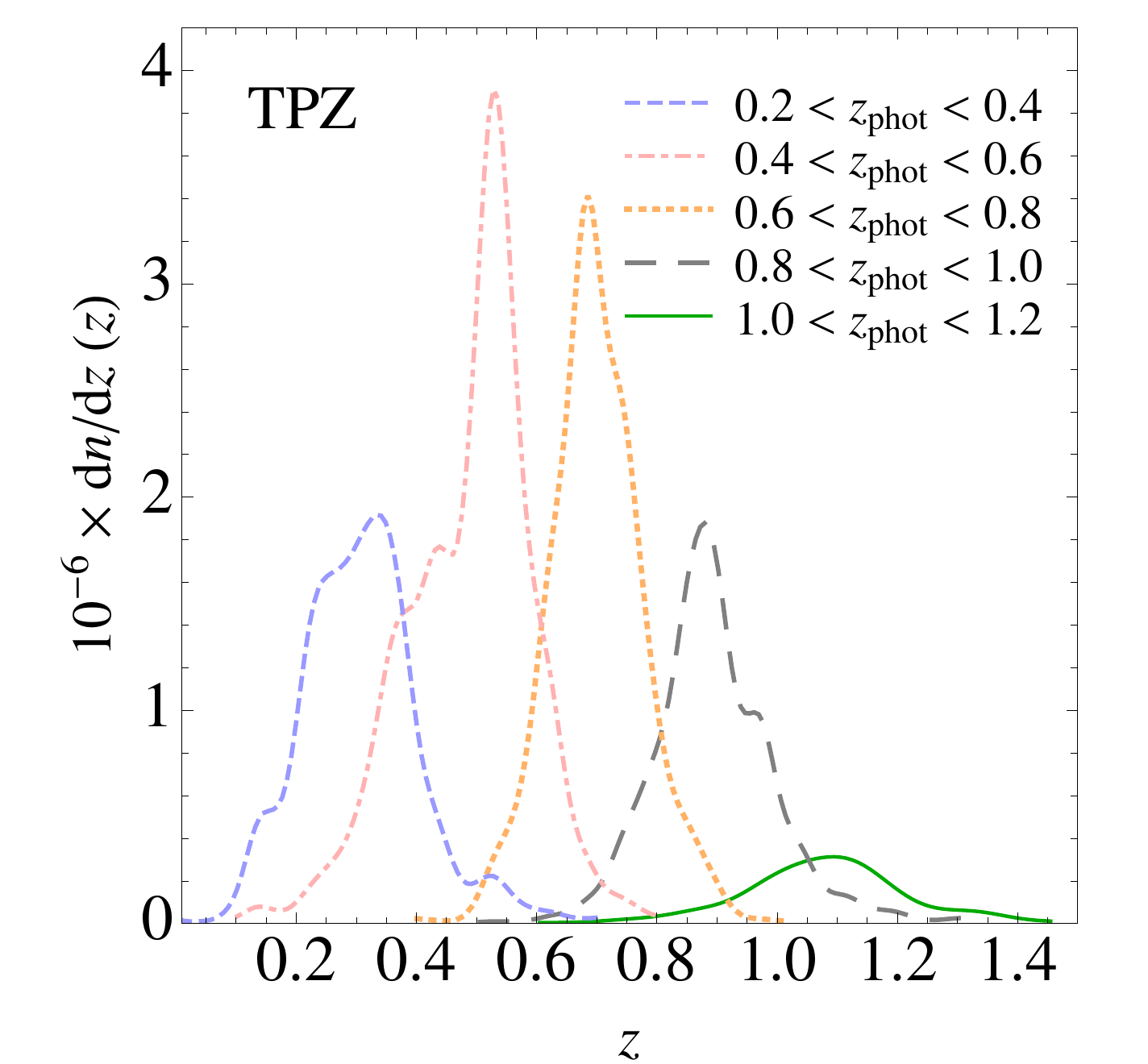} 
\caption{Redshift distributions obtained from stacking the photo-$z$ PDF of objects selected in five top-hat photo-$z$ bins using either the template based estimation (BPZ, top panel) or the machine learning one (TPZ, bottom panel).}
\label{fig:dndz}  
\end{center}
\end{figure}

In Fig.~\ref{fig:dndz} we show the estimated ``true'' redshift distribution corresponding to each of these five bins for BPZ and TPZ; we will use these in Eq. \ref{eq:wtheta1} to make model predictions for the clustering analysis. These were computed by stacking the individual photo-$z$ PDFs of each galaxy in the binned sample. The accuracy of this approach in DES-SV data is discussed in \citet{2014MNRAS.445.1482S} and its performance in simulated data in a companion paper, \cite{LeistedtMAPS}.

As can be seen by inspecting Table \ref{tab:ngal}, there are some large differences in the numbers of galaxies within the same $z_{phot}$ bin but using different photo-$z$ algorithms. These differences can be understood when also considering the reconstructed $n(z)$ of each sample. For instance, in the $0.2 < z_{phot} < 0.4$ bin, the distribution is considerably wider for BPZ. Thus, the bin has more objects in it, as objects that are truly at higher redshift (and occupy a larger volume) are mis-estimated to occupy the low redshift bin. Therefore, despite their differences, we expect each set of galaxies to effectively probe the mean clustering properties of the galaxies at the effective redshift of the sample, when properly analyzed within the context of the $dn/dz$.

\subsection{Survey depth and baseline angular mask}
\label{sec:completeness}

\begin{figure*}
\centering
\includegraphics[width=0.65\textwidth]{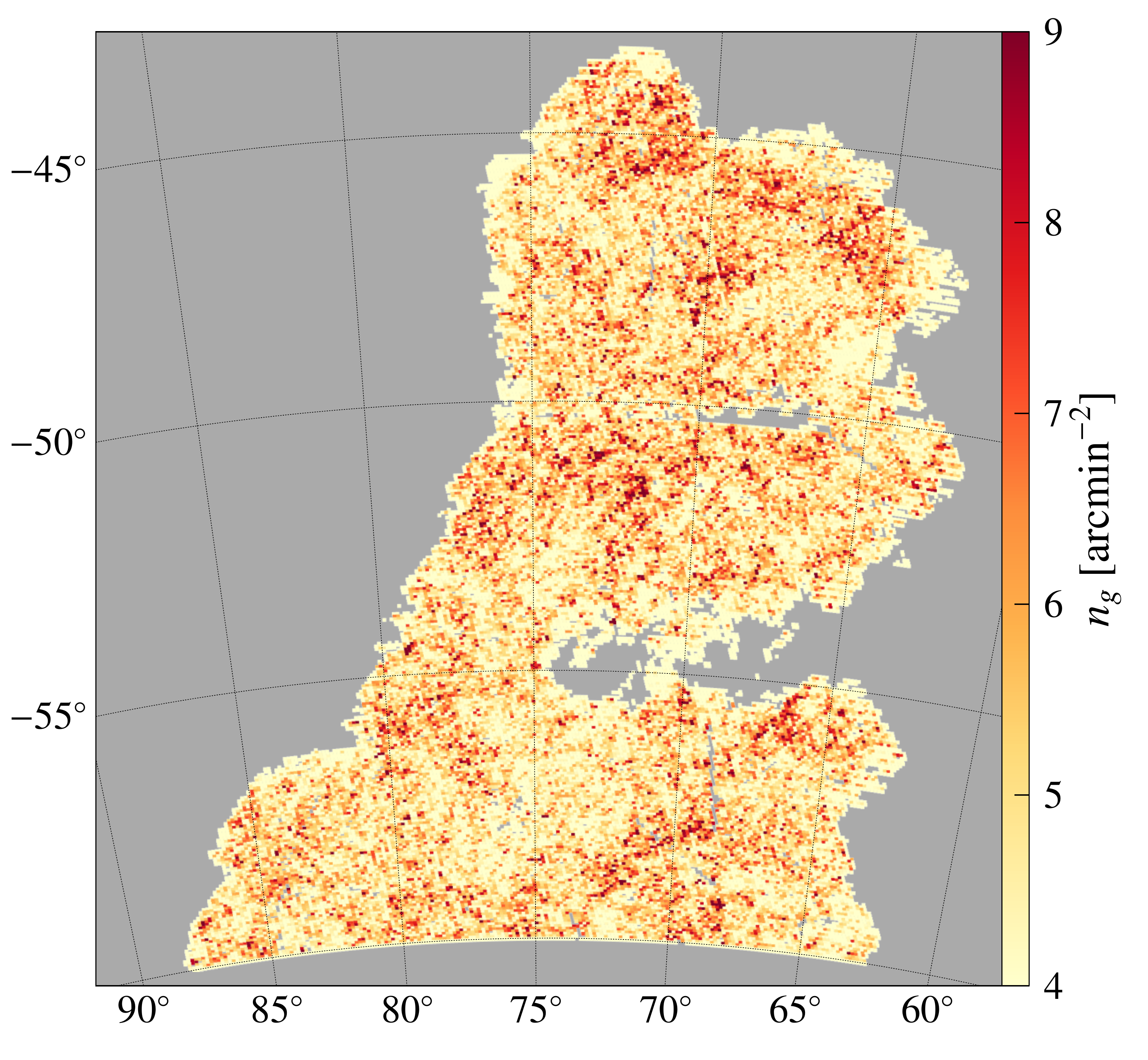}
\caption{The distribution of LSS bench-mark sample galaxies over the angular footprint defined by regions with survey limiting magnitude in the $i$-band $>22.5$. The sample is selected to be flux limited to $i \le 22.5$ and has a mean density of 5.6 arcmin$^{-2}$. All the regions considered provide at least $S/N \> 10$ measurements for objects at $i$-band $ = 22.5$. This choice balances concerns between using the maximum depth and area possible, as described in the text. The x-axis (y-axis) corresponds to right ascension (declination) measured in degrees.}
\label{fig:benchmark_footprint}
\end{figure*}

Within any given redshift range, minimization of the expected uncertainty on clustering measurements is a trade-off between area and number of objects (see, e.g., \citealt{2011MNRAS.414..329C}). 
A fainter flux-limited sample maximizes the number density of observed galaxies, but one should discard areas where this limit is not met in order to secure the completeness of the sample. The available SV area decreases by approximately 20 deg$^{2}$ for limiting magnitudes in the range $20 < \magauto <22.5$ but exhibits a sharp decrease of deeper regions. 
For example, from the approximately 150 deg$^{2}$ SV area overlapping SPT-E region, only about 80 deg$^{2}$ have a limiting magnitude of $i =23$.

Thus, we set the flux-limit of $i = 22.5$ for the definition of the sample and consider only the regions with $i$-band limiting magnitude $\magauto>22.5$ in our clustering analysis
\footnote{We have checked that the mean number density of objects in the sample does not change any longer if we restrict to deeper areas.}.
The resulting angular footprint, after combining with the survey mask of Sec.~\ref{sec:limiting_depth_map}, is shown in Fig.~\ref{fig:benchmark_footprint}. It occupies a contiguous area of 116.2 deg$^{2}$. 
 We consider this footprint as our baseline mask which will be used for all subsequent clustering measurements in this work. But we note that further masking may be due in some photo-$z$ bins to mitigate systematic effects, this will be the subject of Sec.~\ref{sec:systematics}.

We also note that our area and sample are very close to the ones used in the CHFTLS analysis of \cite{2012A&A...542A...5C}; with an area of 133 deg$^{2}$ to a limiting depth of $i  = 22.5$. However our footprint is contiguous, which makes the impact of the integral constraint on large scales negligible as opposed to measurements in the four separated fields of CFHTLS. The sample we analyze is the largest contiguous area demonstrated to have a reliable $i$-band depth of 22.5 for extra-galactic sources, despite being less than $3\%$ of the size of the final DES footprint.

%% file: Sec3.tex
\section{Potential Sources of Systematics and Map Projections}
\label{sec:sysmaps}

Quantities that may modulate the detection efficiency of galaxies and produce spurious galaxy correlations
have been recorded and mapped so that any systematic effects can be empirically studied. The maps used in this paper are presented below.

Galaxy catalogs can be affected by the time-dependent fluctuations in the observing conditions of the survey. These fluctuations affect the depth in non-trivial ways because they occur in the single-epoch images, which are coadded and post-processed before extracting the galaxy properties. In other words, the transfer function between the raw images and final catalogs is a complicated function of the input single-epoch images and the coadding and source-extraction pipelines, also coupled to the galaxy density field and astrophysical foregrounds. 

A significant effort is dedicated in DES to understanding the transfer function from intrinsic to observed quantities (see, e.g., the BCC-UFig framework, \citealt{Chang2014bccufig} and also the {\tt Balrog} framework, \citealt{balrog}), which will be critical to precisely identify sources of systematics and eliminate them in the cosmological analyses. Such techniques can in principle be used to account for the observational effects that cause fluctuations in the observed galaxy density. These methods are still maturing (though see \citealt{balrog} for a working example of applying such techniques to LSS measurements), however, and thus for the DES-SV data we instead test the observed galaxy density against a large number of maps of potential sources of systematics (e.g., the mean seeing). These maps are then used to run null-tests and correct for systematic shifts in the clustering measurements, as detailed in subsequent sections. This requires one to project the properties of the single-epoch images onto the sky, accounting for the geometrical overlap and also weighting due to the coadding procedure. This projection is fully detailed in \cite{LeistedtMAPS}, where the full set of maps, their potential applications, and a pedagogical example using catalogs extracted from the BCC-UFig simulated images are presented. 

\begin{figure*}
\begin{center}
\includegraphics[width=0.24\textwidth]{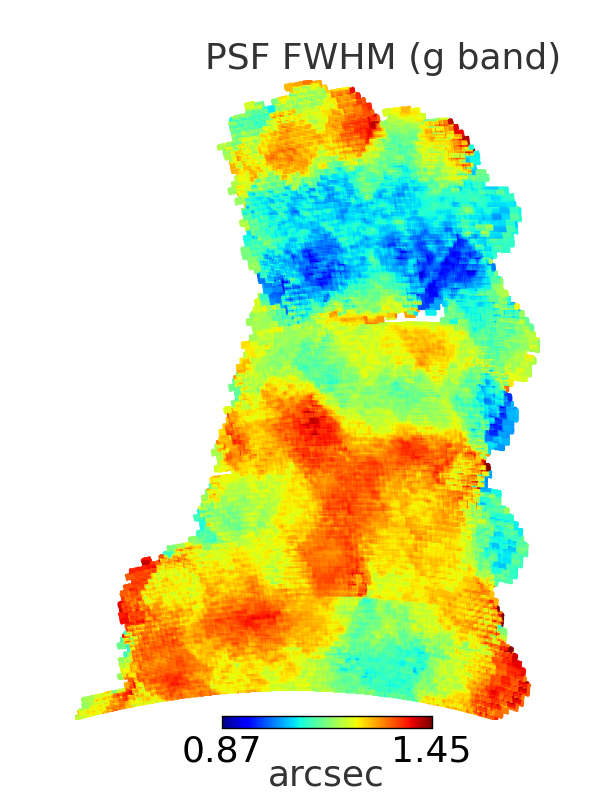} 
\includegraphics[width=0.24\textwidth]{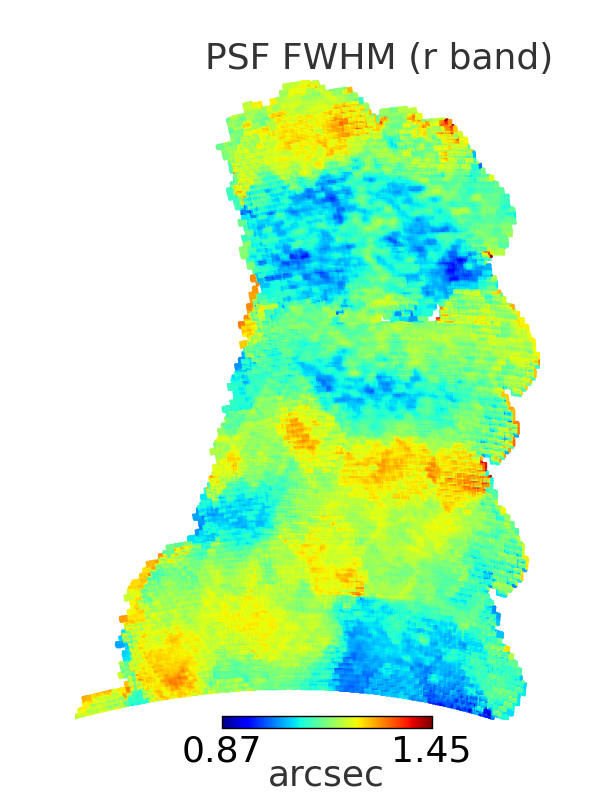} 
\includegraphics[width=0.24\textwidth]{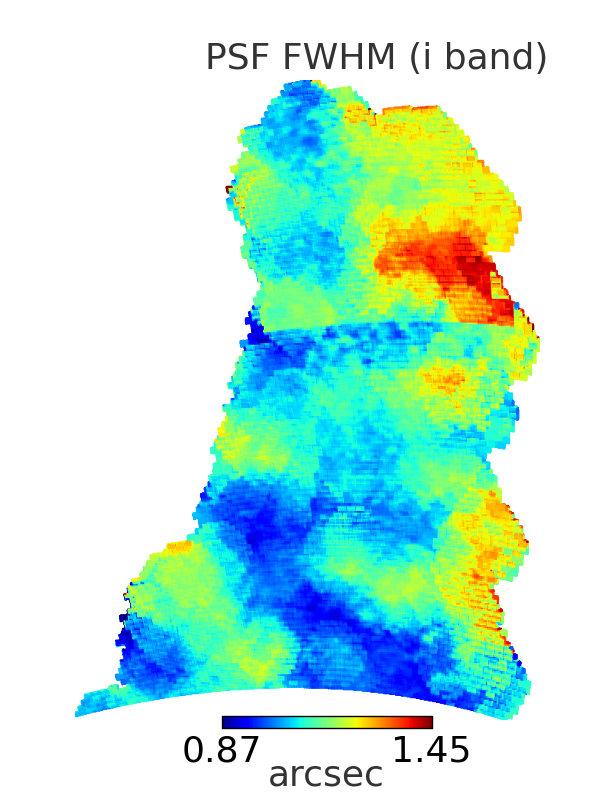} 
\includegraphics[width=0.24\textwidth]{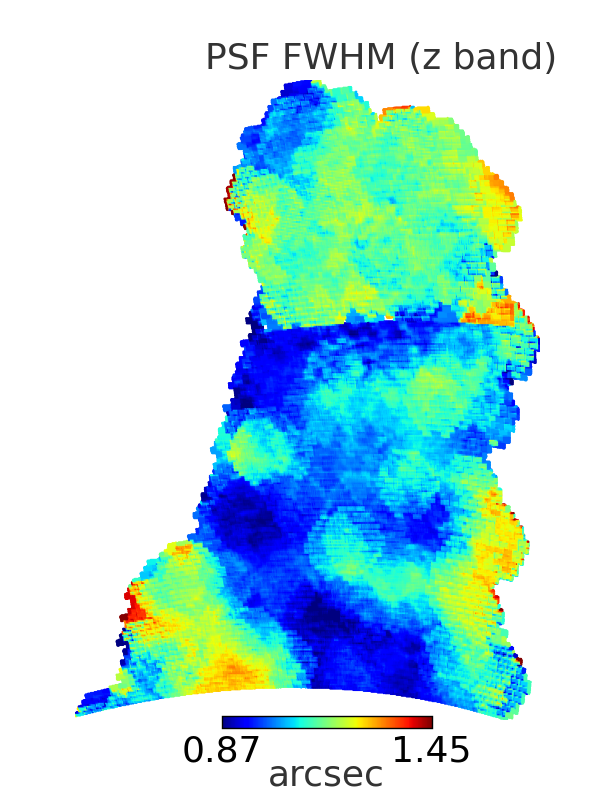} 
\includegraphics[width=0.24\textwidth]{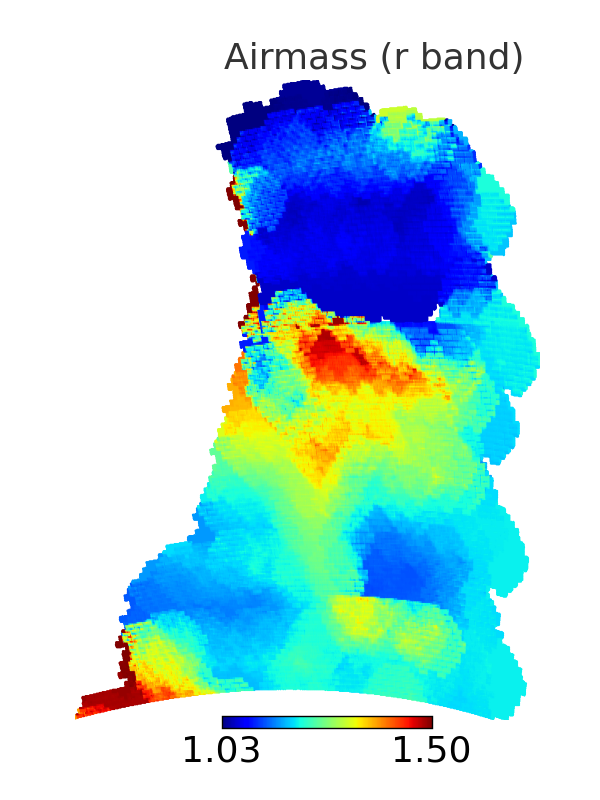} 
\includegraphics[width=0.24\textwidth]{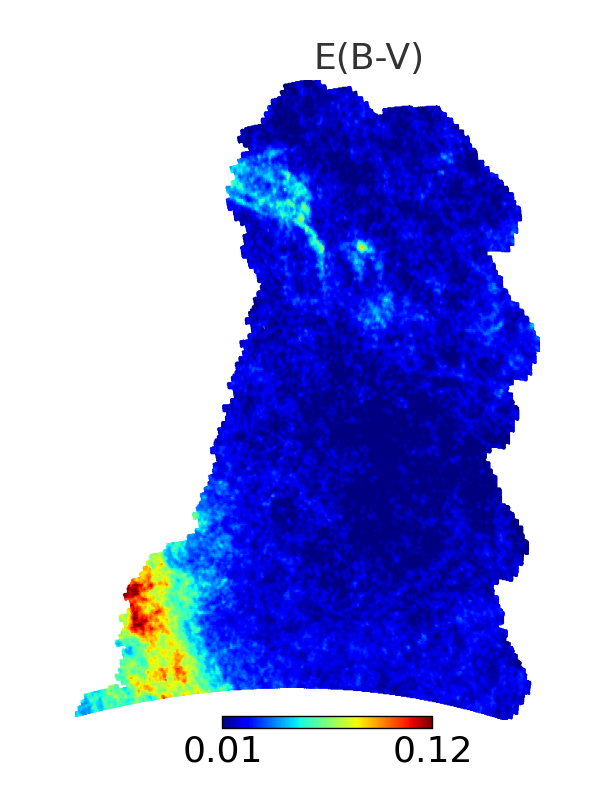} 
\includegraphics[width=0.24\textwidth]{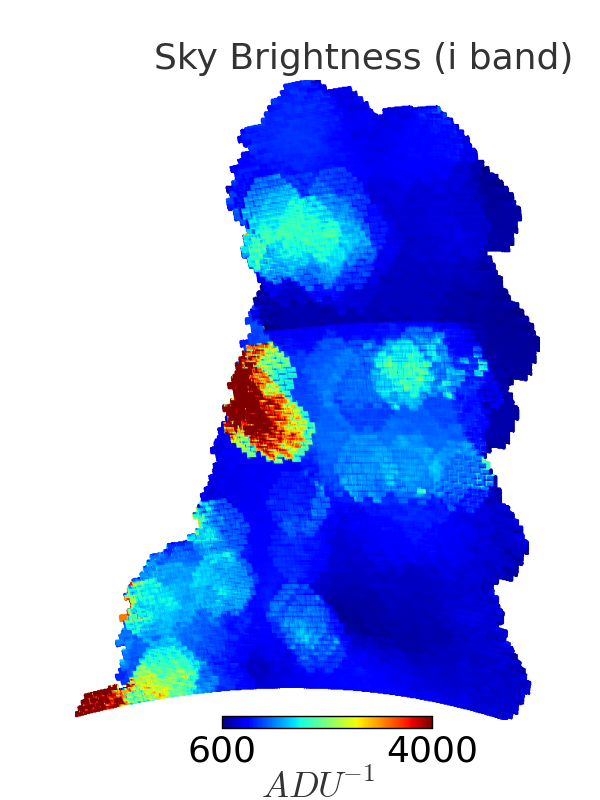} 
\includegraphics[width=0.24\textwidth]{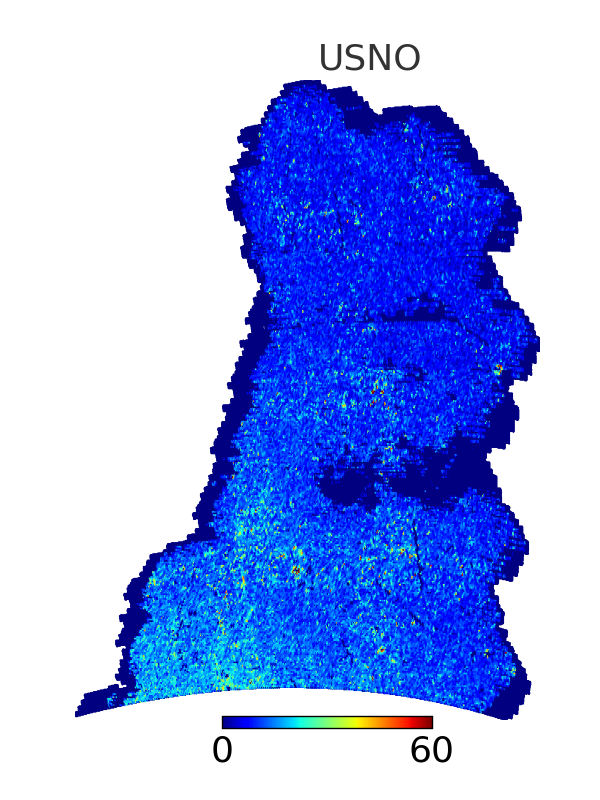} 
\caption{Maps of potential sources of systematic errors that can lead to spatial variations in the number of observed galaxies (through depth fluctuations or sample contamination) or degrade the data quality itself (impacting the determination of magnitudes, colors or photo-z). The values at each pixel are computed as single statistical estimators of the single-epoch values for the images contributing to that map pixel.} 
\label{fig:syst_maps}
\end{center}
\end{figure*}

For this paper, we consider single-epoch properties that are known to affect the depth and therefore might affect the clustering measurements. Each of these observational properties is mapped on the sky into high-resolution (Nside = 4096) {\tt HEALPix} format maps, and we reduce the full set of values in each pixel into a weighted mean using inverse-variance weights (as in the coadd and {\tt mangle} pipelines). The four quantities used in this work, computed in the {\it grizY} bands, are (see \citealt{LeistedtMAPS} for a full description):

\begin{itemize}
{\item {\tt FWHM}: the mean seeing (in pixel units), measured as the full width at half maximum of the flux profile.} 
{\item {\tt airmass}: the mean airmass. } 
{\item {\tt skybrite}: the mean sky brightness. } 
{\item {\tt skysigma}:  the mean sky background noise derived as flux variance per amplifier in each CCD chip\footnote{In principle, the sky brightness and the mean sky background should be strongly correlated. We test against both as a validation that this process has worked properly.}.}
\end{itemize}

We will consider only the mean quantities; in the future it might be important to test against maps of the variance of these properties as well. In addition to the above maps, we consider the following:
\begin{itemize}

{\item {\tt Galactic dust extinction}: As described above, the data are calibrated using the technique of stellar locus regression, which in addition to correcting for instrumental and atmospheric effects also removes reddening due to Galactic extinction. To test for residual effects of Galactic dust on the photometry, we include the \cite{1998ApJ...500..525S} dust map as a potential systematic, pixelized with {\tt HEALPix} resolution Nside=4096. }
\newline

{\item {\tt USNO}: Contamination of our galaxy sample by stars will affect the measured clustering; stellar density increases towards the galactic plane, following a gradient that causes significant clustering signal on large scales \citep{2011MNRAS.417.2577C, Ross11,2012ApJ...761...14H}. Furthermore, because the footprint used in this work neighbors the Large Magellanic Cloud, stellar contamination may introduce more complicated spurious clustering. To investigate this effect, we include a map of stellar density across the field, as measured by the USNO-B1 catalog \citep{2003AJ....125..984M}. We take USNO stars with B magnitude brighter than 20; deeper than this limit the depth of the USNO catalog varies across the field. Although this catalog is brighter than our sample, we expect the stars in our sample to trace the same galactic density distribution as those in the USNO catalog.  Due to low statistics we use a coarser map {\tt HEALPix} resolution of Nside=256.\footnote{We also tested results against a sample of fainter DES detected stars and found no significant differences.}  } 
\newline

{\item {\tt depth$\_$mask}: As described in section \ref{sec:completeness}, we cut our sample at a magnitude of $\magauto <22.5$ and use a footprint where the data provides $>10\sigma$ measurements at that depth.  We include the depth map as a potential systematic in order to probe possible inaccuracy of the map and incompleteness at the faint end of the sample. } 
\newline

{\item {\tt chi2$\_$psf$\_$fit}: The coadded image on which we base our photometry can have complicated structure, as the number of input images varies across the field and each input image has a unique PSF. It is therefore difficult for the photometric reduction to characterize the image and its PSF across an entire coadd unit.  Poor characterization of the coadd image quality at the location of a star will cause the star to be poorly fit by the PSF model. In locations of a poor PSF fit to stars, galaxy photometry will also be adversely affected by the inaccuracy of the estimate of the object's size. We therefore use a map of the average $\chi^2$ of objects' fits to the PSF model to test against this potential source of systematic uncertainty. We only consider very bright objects, with $16 < i < 18$, to ensure that the large majority of them are stars.} 
\newline

{\item {\tt chi2$\_$detmodel$\_$fit}: {\tt MAG\_DETMODEL} magnitudes are measured by fitting a S\'{e}rsic profile, convolved with the PSF model, to the object image. Poor {\tt MAG\_DETMODEL} fits over a region of data would indicate photometric systematic errors in the same manner as poor PSF fits, as described above. We therefore also test our data against a map of the average $\chi^2$ of the {\tt detmodel} magnitude fits;  we expect this map to be somewhat degenerate with the {\tt chi2$\_$psf$\_$fit} map. }

\end{itemize}
We show some of the most relevant maps in Fig.~\ref{fig:syst_maps}. 

%% file: Sec4.tex
\begin{figure*}
\centering
\includegraphics[width=0.4\textwidth]{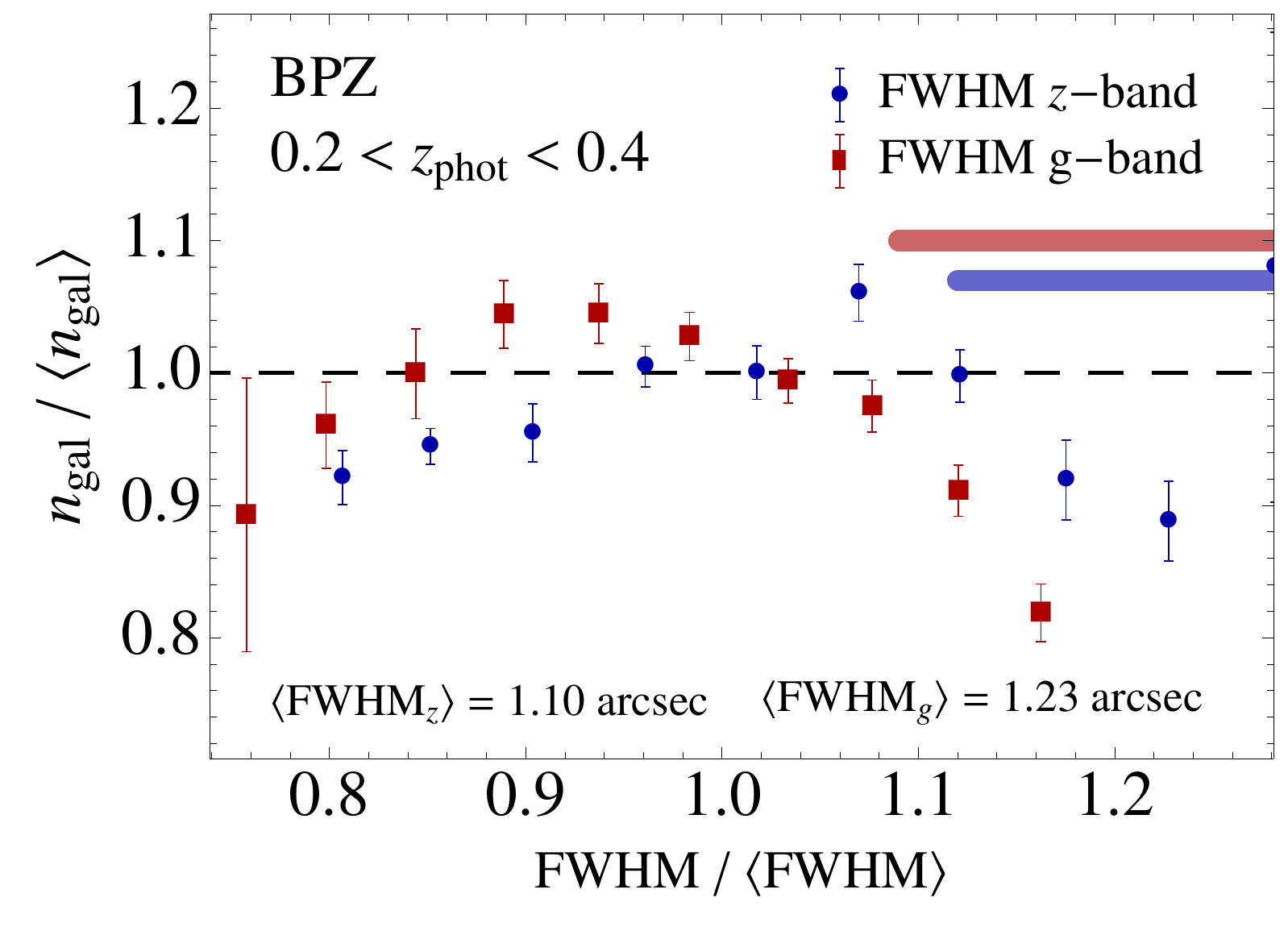}
\includegraphics[width=0.4\textwidth]{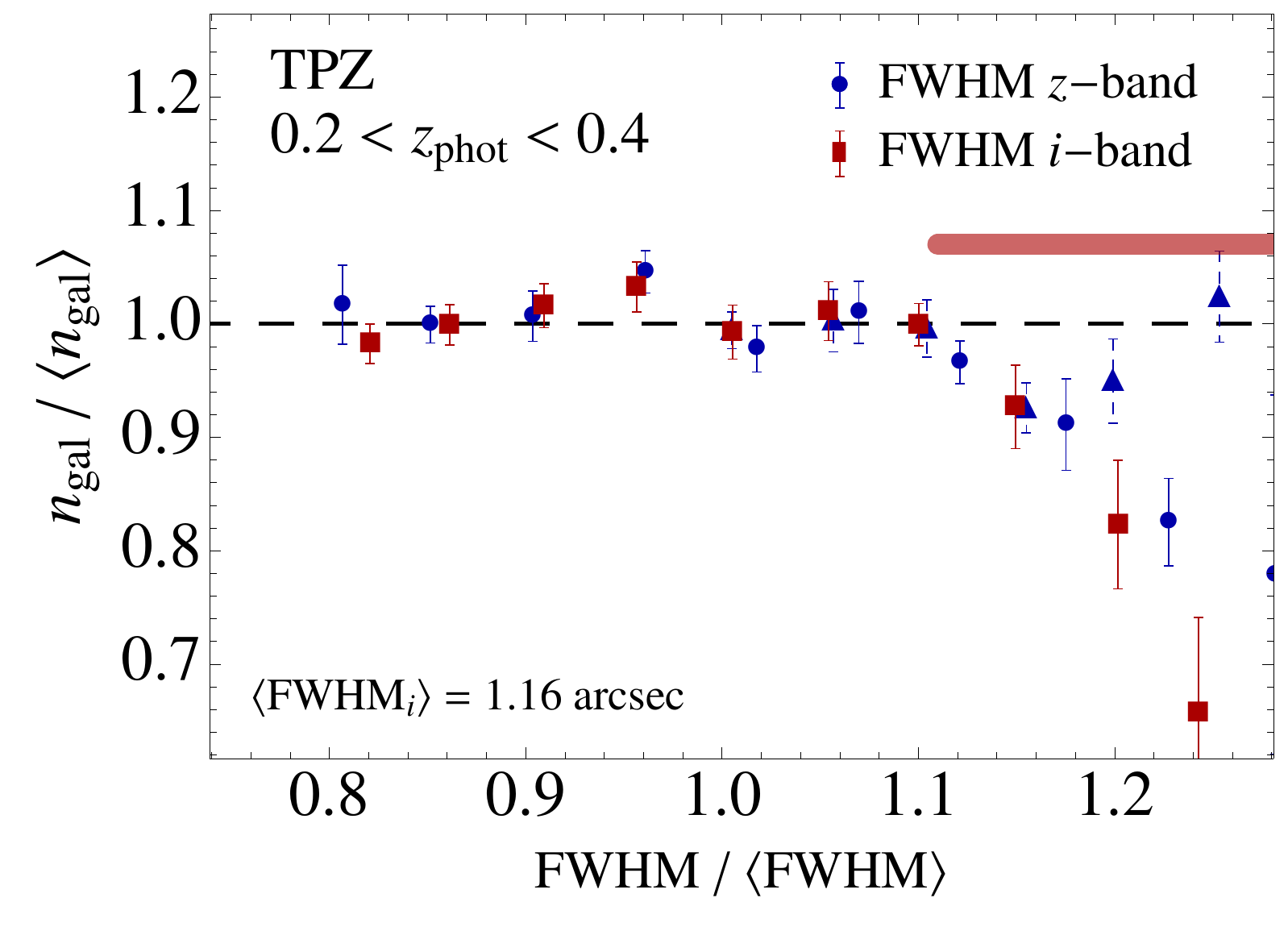}
\includegraphics[width=0.4\textwidth]{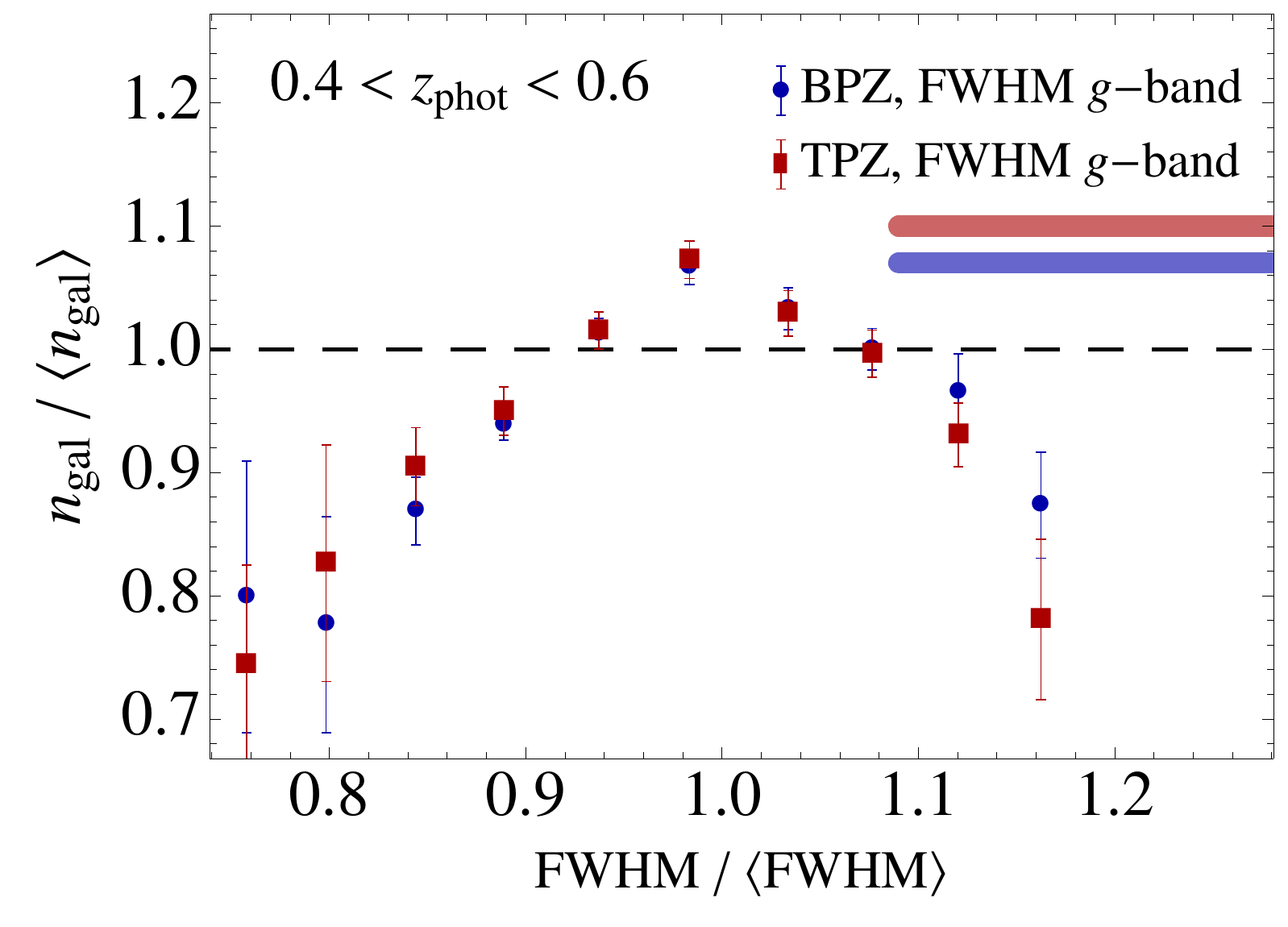}
\includegraphics[width=0.4\textwidth]{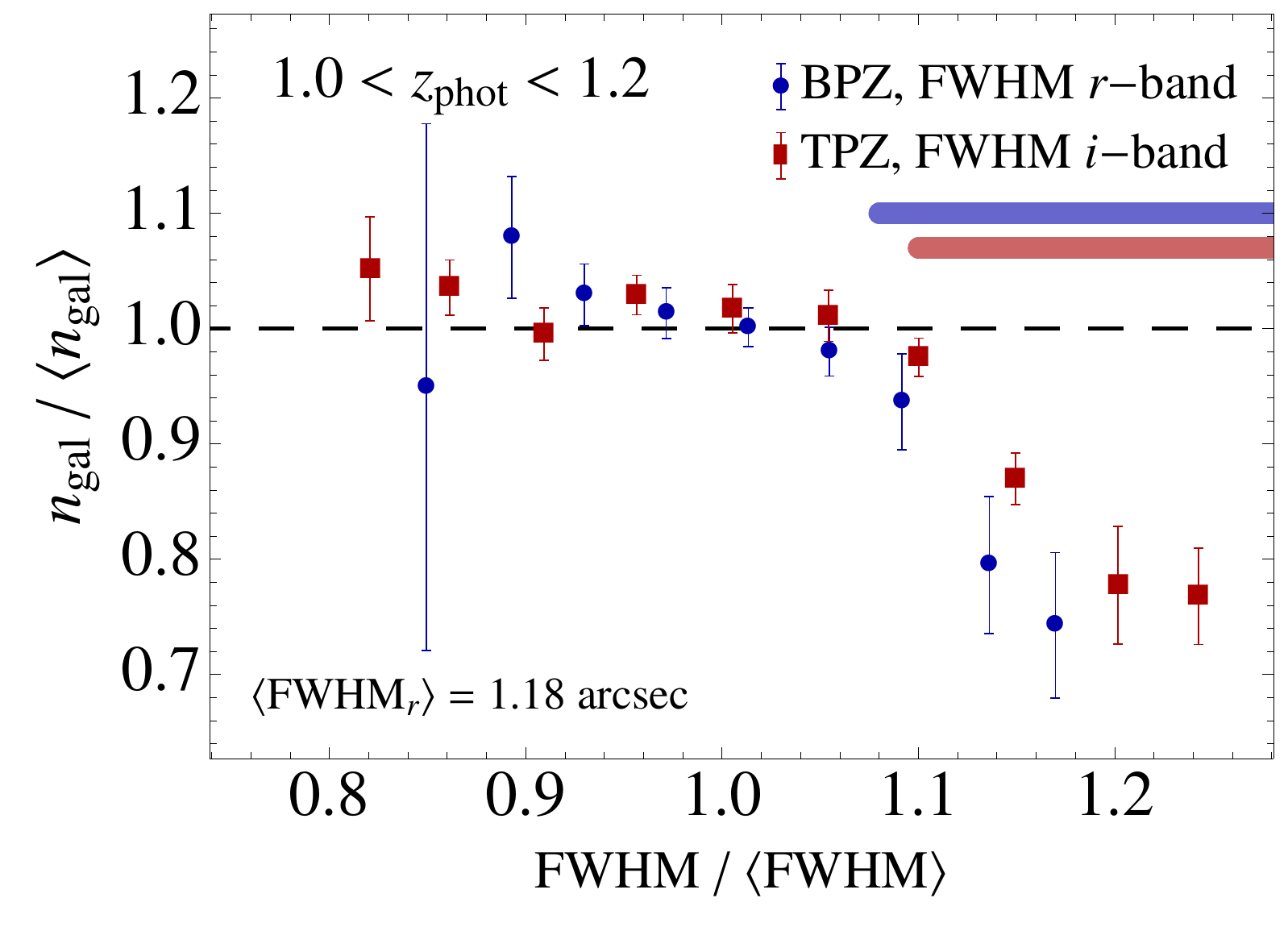}
\caption{Galaxy density (normalized to the mean over the footrpint) as
  a function of the value of different potential sources of systematic
  errors, in regions where those galaxies reside. This is displayed
  for our two photo-$z$ estimators and
  different tomographic bins. 
  We only show those cases where the density of galaxies
  drops steadily after some threshold value indicating regions that
  may lead to systematic effects. To minimize them we mask these regions
  on top of our nominal mask (as a function of photo-$z$ bin). Values
  masked out in each case are indicated by the inset boxes in blue/red. Cases where this relation is roughly linear (e.g. $0.4 < z < 0.6$ after high $g$-band FWHM values are masked) induce a change to the
  clustering measurements that can be corrected for using
  the cross-correlations between the maps for
  galaxy distribution and the ones for systematic effects  (see
  Sec.~\ref{sec:relsys})}. 
\label{fig:dens_plots}
\end{figure*}

\section{Mitigating systematic effects in angular clustering}
\label{sec:systematics}

In this section we give a detailed explanation of our general
procedure to address the impact of potential systematic effects due
to varying observing conditions, stellar contamination, or any other
source of spurious density fluctuations across our footprint. We
assume that all potential sources of systematics are encoded in the maps described 
in Section \ref{sec:sysmaps}.

Let us first describe our step-by-step procedure in broad terms with
more details and results given in subsequent sections,
\newline

[1] {\it Galaxy Density vs. Potential Systematics} : Our first step is
to study the galaxy number density in each tomographic bin as a
function of the value of
each potential systematic variable. If the galaxy density is
independent of the value of the given potential systematic we do not
consider it as impacting our data. Otherwise we try to either mask it or
correct for its impact in the clustering measurement, as shown below.
\newline

[2] {\it Bad Regions Masking} : The relation between galaxy density vs. potential
systematics can take several forms. If this relation is such that the
galaxy density is constant as function of the potential systematic and changes
sharply after some threshold value, we mask
the regions of such anomalous dependence (such as bad seeing, or high airmass) to minimize its
effect. This defines a veto mask that we then use in all subsequent
clustering analysis. The quality of the data is such that
we typically mask only small fractions of the footprint. We will only mask data worse than a given quantity, as when the data quality is worse, we naturally expect photo-$z$s, star/galaxy separation, object detection, etc., to all perform worse and thus be more likely to cause spurious fluctuations in the observed galaxy density.
\newline

[3] {\it Clustering corrections using Cross-correlations} : If 
the relation between galaxy density vs. potential systematic is smooth
and close to linear (after imposing the veto mask discussed above) we use
cross-correlation between galaxies maps and the observational maps to correct our
measurements.
\newline

[4] {\it Stellar Contamination} : As described in Section \ref{sec:stargalaxyseparation}, we estimate 1.5\% of the SV galaxy sample is comprised of mis-classified stars. The clustering of stars over the SV footprint is close to zero, implying that the main effect of the contamination is to proportionally remove power from the $w(\theta)$ measurements. We estimate the stellar contamination in each tomographic bin and derive how to correct our clustering measurements given such contamination in Section \ref{sec:stellcont}.
\newline


\subsection{Galaxy Density vs. Potential Systematics}
\label{sec:sysmask}

Our first step is to study how the galaxy density depends on each
potential systematic variable, if at all.  
The starting point for these measurements is pixelized maps of the
distribution of galaxies and of the potential systematics. 
We divide the range of the values of each potential systematic variable into roughly ten bins.  
For the sky area corresponding to each bin, we calculate the mean galaxy number density 
normalized by the mean across the whole footprint.  In
Fig.~\ref{fig:dens_plots} we show the resulting normalized galaxy density as a
function of the different potential systematic variables. After examining all $z$ bins, we only show the
relevant cases. Error bars are computed using the jack-knife method by
splitting the footprint into 100 regions. Hence errors in Fig.~\ref{fig:dens_plots} 
might be under-estimated due to a limited sampling of large-scale
fluctuations. Furthermore, note that extreme values of
the potential systematics are typically not sampled in most JK regions.
In such cases (i.e. the first and/or last points in each panel of 
Fig.~\ref{fig:dens_plots}) we choose to estimate error bars as Poisson
sampling noise. 

\begin{figure}
\centering
\includegraphics[width=0.22\textwidth]{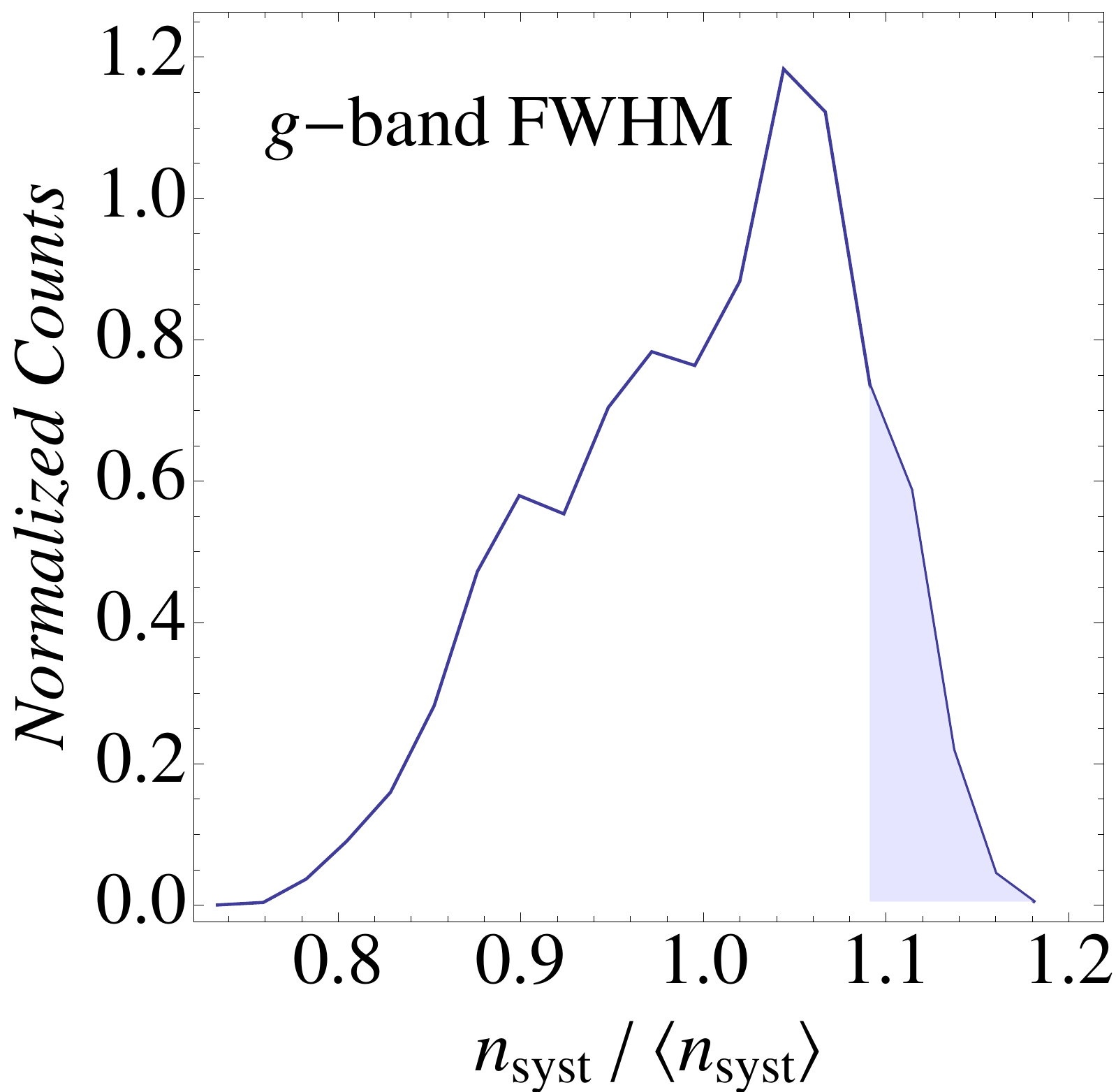}
\includegraphics[width=0.22\textwidth]{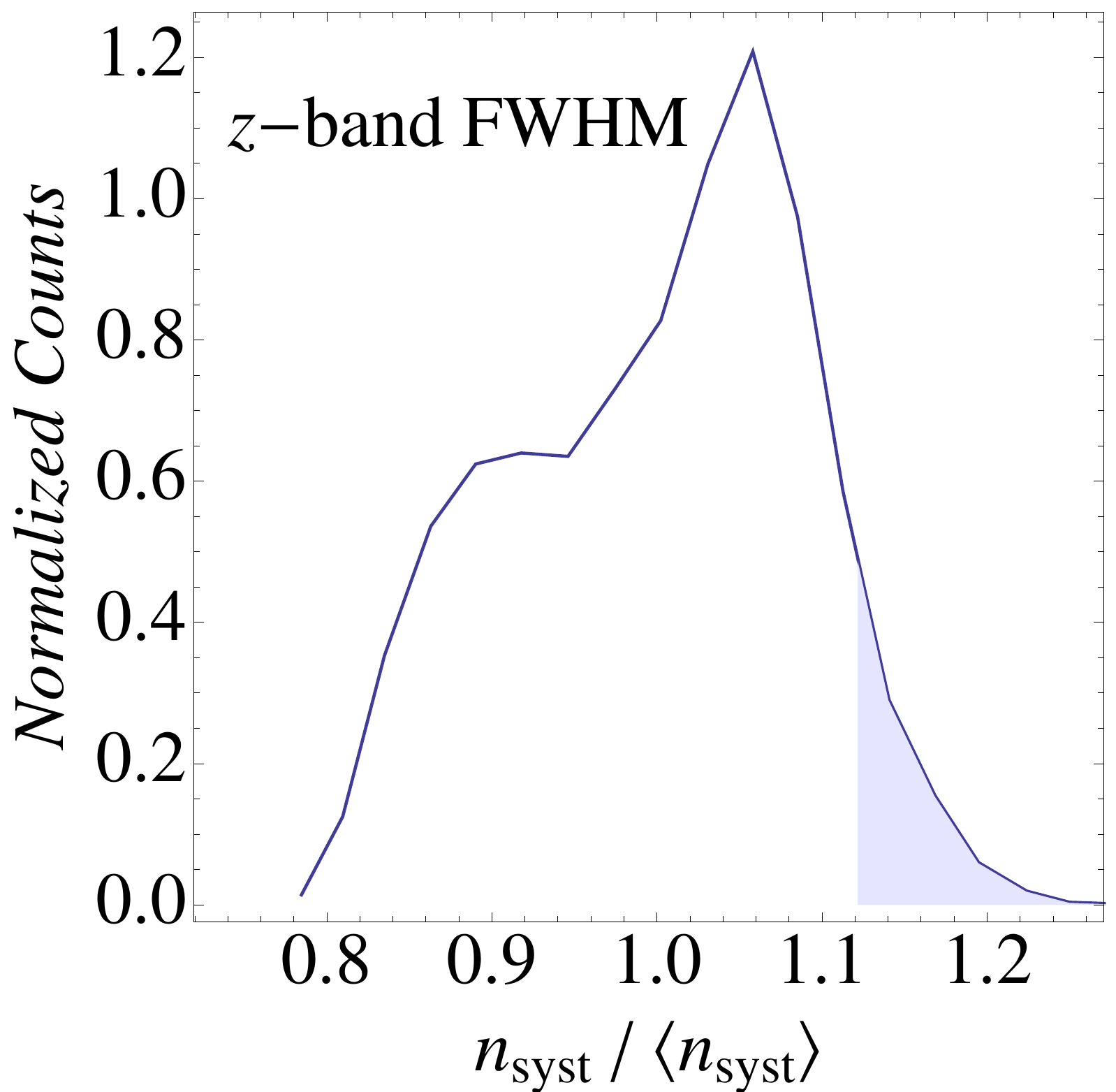} \\
\includegraphics[width=0.22\textwidth]{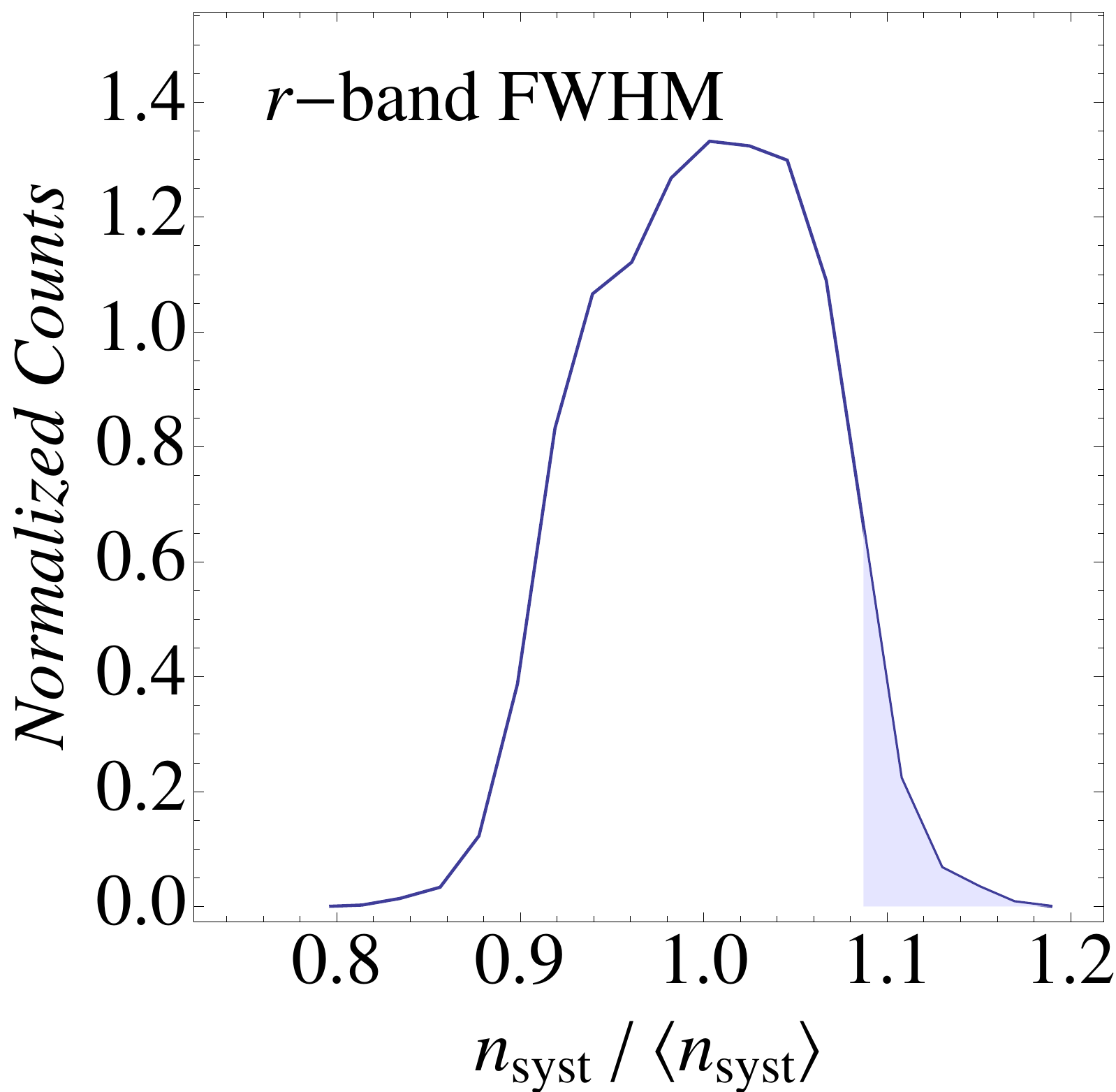}
\includegraphics[width=0.22\textwidth]{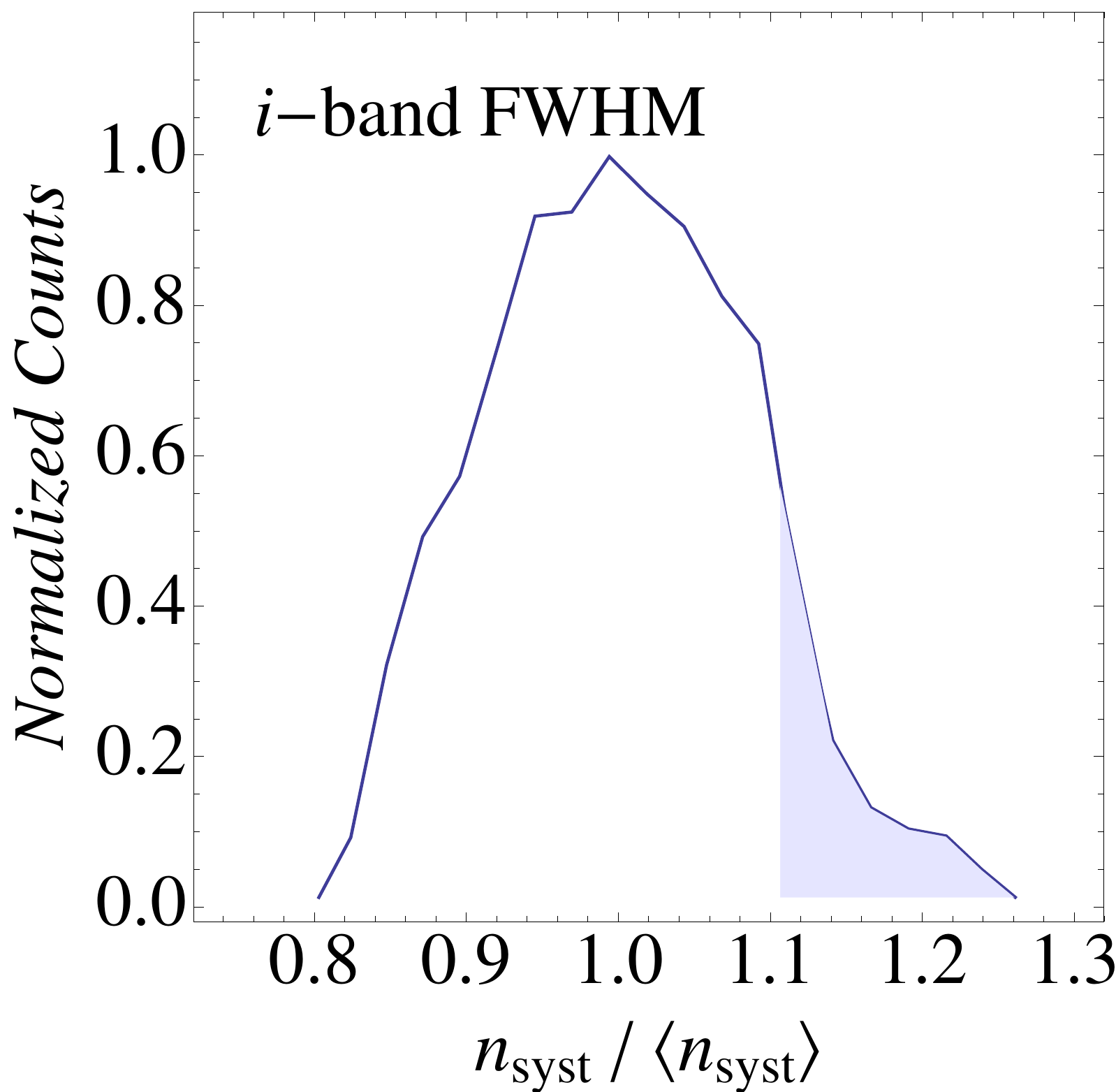}
\caption{Distribution of FWHM weighted co-added values in different bands across our
  footprint. The shaded parts of the histograms correspond to regions
  where large decrement of galaxy density in different photo-$z$ bins is
  found. To avoid a systematic effect they were removed from the
  corresponding clustering analysis. See Sec.~\ref{sec:sysmask} for further details.}
\label{fig:syst_histo}
\end{figure}

We perform this procedure in every redshift bin, and define an angular
mask due to systematics per bin. These are: 

\begin{itemize}

{\item $0.2 < z < 0.4$: The top left panel of Fig.~\ref{fig:dens_plots}
  shows the relation of galaxies selected in this bin with BPZ as a
  function of FWHM in $g$-band and $z$-band. For high values of seeing we
  see a clear drop of galaxy density and therefore choose to remove these
  areas. We cut regions with $g$-band FWHM $> 1.34$ arcsecs ($1.09$
  of the mean $g$-band FWHM)
  which removes $13\%$ of the area. Similarly, we remove area with $z$-band FWHM $> 1.24$
  arcsecs ($1.12$ of the mean $z$-band FWHM) which removes a further $6\%$. 
  In all $19\%$ of our nominal footprint is masked for
  the BPZ sample at this photo-$z$ bin. In the top left and right panels of
  Fig.~\ref{fig:syst_histo} we show histograms of FWHM shading the portion removed by this selection
   (at the same time, Fig.~\ref{fig:syst_maps} shows the maps for FWHM in all bands, providing an idea of
  which regions are removed). The same procedure but using galaxies
  selected with TPZ yields the top right panel of Fig.~\ref{fig:dens_plots}. In this case
  we remove $i$-band FWHM $> 1.28$ arcsecs ($1.11$ of the mean $i$-band FWHM). Note that there is also a
  dependence with $z$-band FWHM. However after the regions with worst $i$-band
  FWHM are removed, this dependence disappears. This is shown by the
  triangular symbols with dashed error-bars . Hence for TPZ we only mask bad
  $i$-band FWHM regions removing $\sim 11\%$ of the original
  footprint. The distribution of $i$-band FWHM is shown in the
  bottom right panel of Fig.~\ref{fig:syst_histo}, where the masked out part has been shaded.} 
\newline

{\item $0.4 < z < 0.6$: For this bin only the $g$-band FWHM affects the
  galaxy density in a way suitable for masking. It does so in a
  very similar manner for
  both photo-$z$ samples, as shown in the bottom left panel of
  Fig.~\ref{fig:dens_plots}. Hence we define a common veto mask for
  BPZ and TPZ
  removing regions with $g$-band FWHM $> 1.34$ arcsecs ($1.09$ of the mean
  $g$-band FWHM). This removes $13\%$ of the area. According to
  Fig.~\ref{fig:syst_histo} the threshold value could be smaller because the
  relation turns around at $\sim 1.28$ arcsecs. We have checked that this extra masking does not
  change the final clustering results, but it implies removing about $30\%$ of the
  area. Hence we cut $g$-band FWHM $> 1.34$ arcsecs what should
  account for the bulk of the effect. Note that after removing this region there is a clear dependence of galaxy density increasing with $g$-band FWHM. This residual
is exactly the type that can be addressed by means
of cross-correlations between galaxies and potential systematics, as detailed in Sec.~\ref{sec:relsys}.}
\newline
{\item $1.0 < z < 1.2$: For the highest redshift bin we again define
  one additional mask for BPZ and one for TPZ based on
  Fig.~\ref{fig:dens_plots} 
  bottom right panel. For BPZ we remove
  regions with $r$-band FWHM $>1.28$ arcsecs ($1.08$ of the mean
  $r-$band FWHM) which amounts to $6\%$ of the original footprint 
  (leaving $109.2$ deg$^2$). For TPZ we cut regions with $i$-band FWHM $>1.28$ arcsec ($1.1$ of the mean
  $i$-band FWHM and $10\%$ of the area). Above these threshold values the density of galaxies
drops sharply compared to the mean.}
\end{itemize}

For the bins at intermediate redshifts ($0.6 < z < 0.8$ and $0.8 < z <
1.0$) we find no dependence of the data with the potential
systematics. Thus, we do not correct the clustering or
introduce additional cuts to the nominal footprint when analyzing these bins.

\subsection{Galaxy-Systematics cross correlations}
\label{sec:relsys}

Once regions with bad observing conditions that translate into an
anomalous drop of the density of galaxies have been identified and masked out
we devote our attention to correcting more subtle dependencies where the density
varies roughly proportionally with the strength of the systematic. 
For this we follow and extend the work of
\cite{Ross11,2012ApJ...761...14H}. We start by assuming that
fluctuations in the potential sources of systematic effects $\delta^i_{\rm sys} $
will induce spurious fluctuations into the observed
galaxy number density $\delta_{\rm obs}=n_g / \bar{n}_g$, such that
\beq
\delta_{\rm obs} = \delta_{\rm true} + \sum_{i} \alpha_i \, \delta^i_{\rm sys} 
\label{eq:two}
\eeq
where $\delta_{\rm true}$ are the true underlying fluctuations we want
to recover and $i$ refers to any of the maps described in Section~\ref{sec:sysmaps}.
By definition we assume that true galaxy fluctuations do not
correlate with the ones induced by the systematics\footnote{This means
  we must be careful in choosing systematics: e.g. photometric redshift
  error is typically smallest for luminous red galaxies, which cluster
  strongly;  we therefore must not use a map of redshift error (or,
  similarly, magnitude error which correlates with galaxy luminosity)
  as a systematic. Otherwise the correlation can be strong, see for example
  \cite{2014MNRAS.437.3490M}.},
\beq
\langle \delta_{\rm obs}\,  \delta^j_{\rm sys}\rangle =  \sum_{i}
\alpha_i \, \langle \delta^i_{\rm sys} \delta^j_{\rm sys} \rangle. 
\label{eq:one}
\eeq
If we define,
\begin{align}
w_{{\rm cross}, j}  &\equiv  \langle \delta_{\rm obs}\,\delta^j_{\rm
  sys}\rangle, \nonumber \\
w_{{\rm auto}, ij}  &\equiv  \langle \delta^i_{\rm sys}\,\delta^j_{\rm sys}\rangle,
\end{align}
then Eq.~(\ref{eq:one}) can be written as, 
\beq
w_{{\rm cross},j} = \sum_i \alpha_i \, w_{{\rm auto},ij},
\eeq
and the solution is a matrix inversion,
\beq
\vec{\alpha} = \vec{w}_{\rm cross} \cdot (w_{\rm auto})^{-1}.
\eeq
Hence if we have 6 systematic maps we have to invert a 6x6 matrix for each
$\theta$-bin (notice that we are a priori assuming that $\alpha$ is
spatially independent to factor it out of the correlations).
From Eq.~(\ref{eq:two}) one can work out the auto correlation of true
fluctuations
\begin{align}
w(\theta)_{\rm true} &= w(\theta)_{\rm obs} - \sum_i \sum_j \alpha_i
\alpha_j \langle \delta^i_{\rm sys} \delta^j_{\rm sys}\rangle \\
&= w(\theta)_{\rm obs} - \vec{\alpha} \cdot \vec{w}_{\rm cross},
\label{correction_equ}
\end{align}
which reduces to the standard case,
\beq
w(\theta)_{\rm true} =
w(\theta)_{\rm obs} - w^2_{\rm cross} / w_{\rm auto}
\label{correction_1sys_equ}
\eeq
for just one systematic.

Note that our starting assumption in Eq.~(\ref{eq:two}) of a 
roughly linear relation between $\delta_{\rm obs}$ and $\delta_{\rm
  sys}$ can be tested and confirmed with the
one-dimensional relation of 
galaxy density as a function of potential systematics as done in
Sec.~\ref{sec:sysmask}. If strong non-linear dependencies are found then Eq.~(\ref{eq:two}) will
not apply and the corrections induced will bias the measurements. For this reason, it is important to mask the regions (e.g., high $g$-band seeing) where the relationships are most clearly non-linear, as is done in the previous section.

In addition, note that our approach does account for the correlation
among the potential systematics themselves. Nevertheless, solving for an arbitrarily large
number of maps might still induce over-corrections and biases due to
the inversion in Eq.~(\ref{correction_equ}) and the noise of the
measurements particularly on large-scales. 

Further below we explain our quantitative criteria based on cross-correlations to
choose which maps impact the galaxy sample and need to be corrected
for. Then we follow a combined approach in which we also
cross check against the density relations as in Sec.~\ref{sec:sysmask}
for those systematics that pass the criteria that a linear relation exists.

To use a quantitative criterion to select the most relevant systematics maps, we  cross-correlate them
with the galaxy distribution maps at each redshift bin after applying the masking described in Section \ref{sec:sysmask}.
For each potential systematic we calculate the correction to the
galaxy correlation according to Eq.~(\ref{correction_1sys_equ}),
$w_{\rm cross}^2/w_{\rm auto}$, ignoring cross-correlations among
systematics to begin with.  This correction is determined for each
angular bin $\theta$.  Using JK resampling over the footprint, we
calculate an error on the correction.  Then, if the correction is
inconsistent with zero at a 1$\sigma$ level\footnote{We use $10$
  logarithmically distributed angular bins in $0.12-4$ deg. The $1\sigma$ limit corresponds to
  $\chi^2 > 11.53$, where the $\chi^2$ is computed using the
  covariance of the $w_{\rm cross}^2/w_{\rm auto}$ estimates in the
  JK regions.}, the systematic is deemed significant and taken
into account in further analysis;  otherwise, it is neglected.

\begin{figure}
\centering
\includegraphics[width=0.44\textwidth]{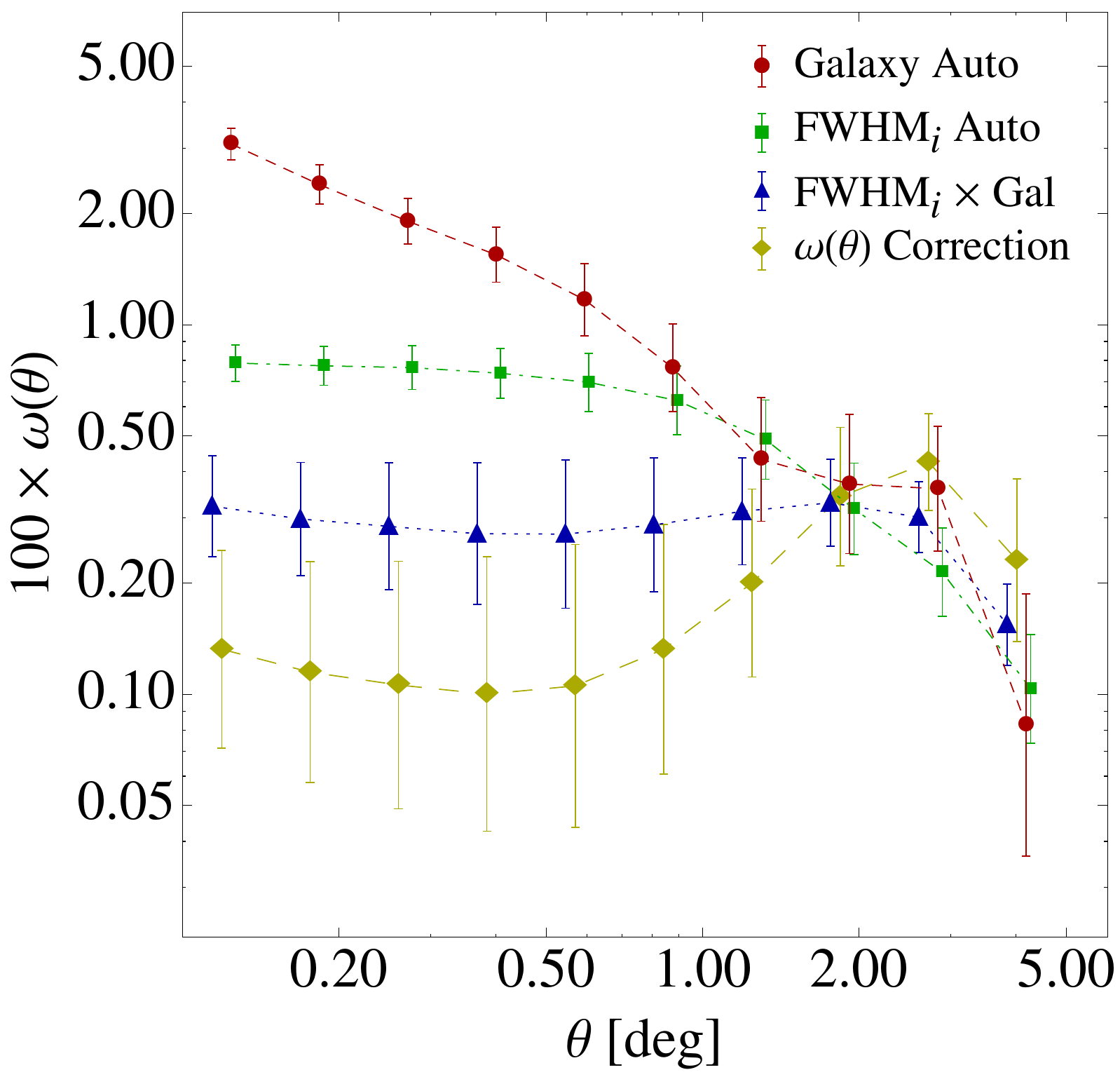}
\caption{Different two-point correlations entering the clustering
  correction in the bin $0.2 < z < 0.4$ for the BPZ sample due to spatial
  variations in $i$-band FWHM over the DES footprint. See
  Sec.~\ref{sec:relsys} for details.}
\label{fig:sys_correction_example}
\end{figure}

Figure \ref{fig:sys_correction_example} shows as an example the
correlation between the central redshift bin, $ 0.2 < z < 0.4$ for the
BPZ sample and
the FWHM map in the $i$ band.  The figure shows the auto-correlation
of the galaxy sample, the auto-correlation of the FWHM map, the
cross-correlation between the data and the systematic, and the
corresponding correction that should be applied to the data (ignoring
covariance with any other systematic effects), see Eq.~(\ref{correction_1sys_equ}).  The correction is significantly non-zero, and therefore
we consider the $i$-band FWHM map to be a relevant systematic in our
analysis.

For the BPZ photo-$z$ catalog the systematics that correlate significantly with the data set are: 
\begin{itemize}
{\item $0.2 < z < 0.4$ : $i$-band $r$-band and $z$-band FWHM, $r$-band Skybrite
  and dust extinction (5 maps)}
{\item $0.4 < z < 0.6$ : $g$-band and $r$-band FWHM (2 maps)}
{\item $0.6 < z < 0.8$ : None}
{\item $0.8 < z < 1.0$ : None}
{\item $1.0 < z < 1.2$ : None after $r$-band FWHM masking}
\end{itemize}

For the TPZ photo-$z$ catalog the systematics that correlate significantly with the data set are: 
\begin{itemize}
{\item $0.2 < z < 0.4$ : $r$-band FWHM, $r$-band and $z$-band Skybrite (3 maps)}
{\item $0.4 < z < 0.6$ : $r$-band FWHM (1 map)}
{\item $0.6 < z < 0.8$ : None}
{\item $0.8 < z < 1.0$ : None}
{\item $1.0 < z < 1.2$ : None after $i$-band FWHM masking}
\end{itemize}

Despite the fact that the limiting depth is the same in all of the
redshift bins we use, we find slight variations in the type and degree of
systematic contamination as a function of redshift. The corrections
arise mainly at low redshifts, hence they might be due to a correlation between the
observing conditions and the determination of photometric redshifts
(the DES filter systems does not contain $u$-band, which degrades the
low redshift photo-$z$). Nonetheless, the
significance of the corrections is never beyond 2$\sigma$ at any given
angular scale which
signals that the data is not very impacted by systematics and our
results not dominated by these corrections. This is discussed in more
detail in Sec.~\ref{sec:results} and Tables \ref{tab:bpz_syst_bias}
and \ref{tab:tpz_syst_bias}.

\begin{figure}
\centering
\includegraphics[trim=0 0 1cm 0, width=0.43\textwidth]{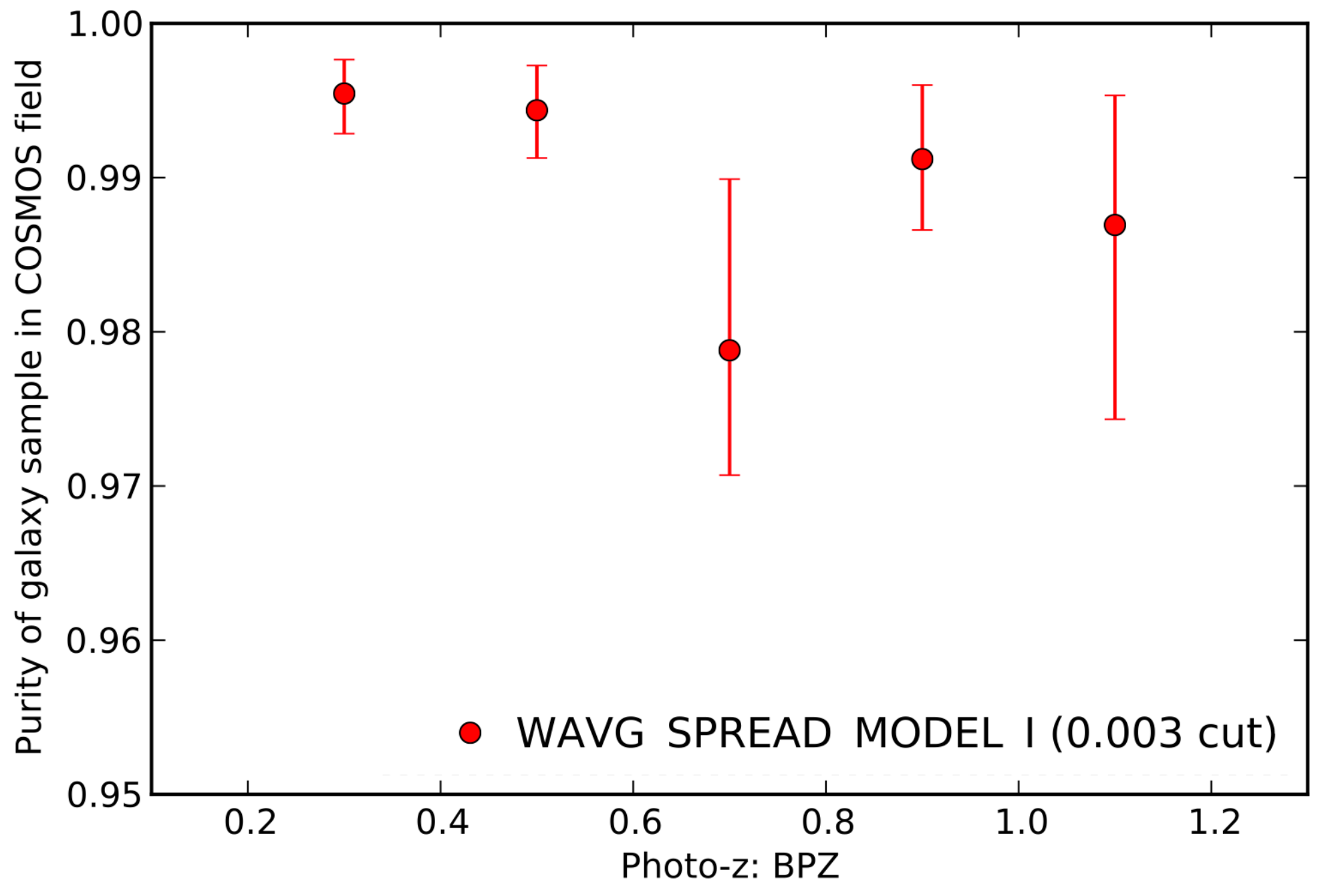}
\caption{The purity of the DES galaxy sample in the COSMOS field in each of the BPZ tomographic redshift bins we use, determined using the default cut on \texttt{WAVG\_SPREAD\_MODEL}  $> 0.003$. The stellar contamination is at most $2\%$.}
\label{fig:stellar_photoz}
\end{figure}

\subsection{Stellar Contamination}
\label{sec:stellcont}

Stellar contamination will affect the measured clustering signal, even if the cross-correlation between the galaxies and stars is negligible. This is due to the fact that rather than modulate the selection function of galaxies, stellar contamination introduces a separate population. In the limit where the stars are un-clustered and the stellar contamination is constant the observed galaxy density is, $\delta_{o} = (1-f_{star})\delta_{gal}$ and thus
\begin{equation}
w_{\rm gal} = \frac{w_{\rm o}}{(1-f_{\rm star})^2},
\end{equation}
i.e., the measured clustering is diluted by the square of the purity of the galaxy sample.

We derive a general expression for the clustering of a galaxy
population, given some stellar contamination (full details are  shown
in Appendix \ref{sec:appendix_stellarcontam})
\begin{equation}
w_{gal} = (1+f_{\rm star})^2\left(w_{\rm o}-f^2_{\rm star}w_{{\rm star},S}-\frac{f_{\rm star}^4}{(1+f_{\rm star})^2}\right),
\label{eq:wgst}
\end{equation}
where in this formalism, the number density of contaminating stars is allowed to be a function of not just the total density of stars, but also the survey observing conditions.

The cross-correlation tests presented in previous sections did not find stellar contamination to be a significant systematic contaminant. This suggests that the term in Eq. \ref{eq:wgst} proportional to $w_{{\rm star},S}$ is consistent with zero. However, the terms $(1+f_{\rm star})^2w_{\rm o}-f_{\rm star}^4$ remain, and thus any stellar contamination must be accounted for.

We estimate the stellar contamination in each photometric redshift bin
by using DES observations in the COSMOS field. This region contains
space-based observations which identify matching DES sources as
point-like or not (as detailed in Section
\ref{sec:stargalaxyseparation}). Figure \ref{fig:stellar_photoz} shows
the purity of the DES sample in the COSMOS field and its uncertainty,
that we estimate in each redshift bin for the classifier discussed
in Sec.~\ref{sec:stargalaxyseparation}. The uncertainty includes
the statistical Poisson variance, the error due to sample variance
fluctuations of the galaxy sample, which impacts the purity too, and
the fluctuations of the stellar sample in the SPTE area. These are
large due to the presence of the Large Magellanic Cloud, and are
included when extrapolating to the SPTE field. This extrapolation to
the DES SPTE area is done by finding the density of highly
probable stars in the COSMOS field with a $19 < i < 21$ cut and
$|\texttt{WAVG\_SPREAD\_MODEL})| < 0.001$ and we find that the average
density is the same as that found in the DES SPTE area when applying
the same cuts, to within 10\% (below our error bars due to intrinsic
fluctuations of the galaxy and stellar samples).  We apply the
correction given by Eq.~(\ref{eq:wgst}) to  the $w(\theta)$ measurements
and propagate this uncertainty into measurements of the galaxy
bias. The correction to the bias is largest in the $0.6 < z < 0.8$ bin, but is at most 2$\%$. 

%% file: Sec5.tex
\section{Results}
\label{sec:results}

\begin{figure*}
\begin{flushleft}
\includegraphics[width=0.32\textwidth]{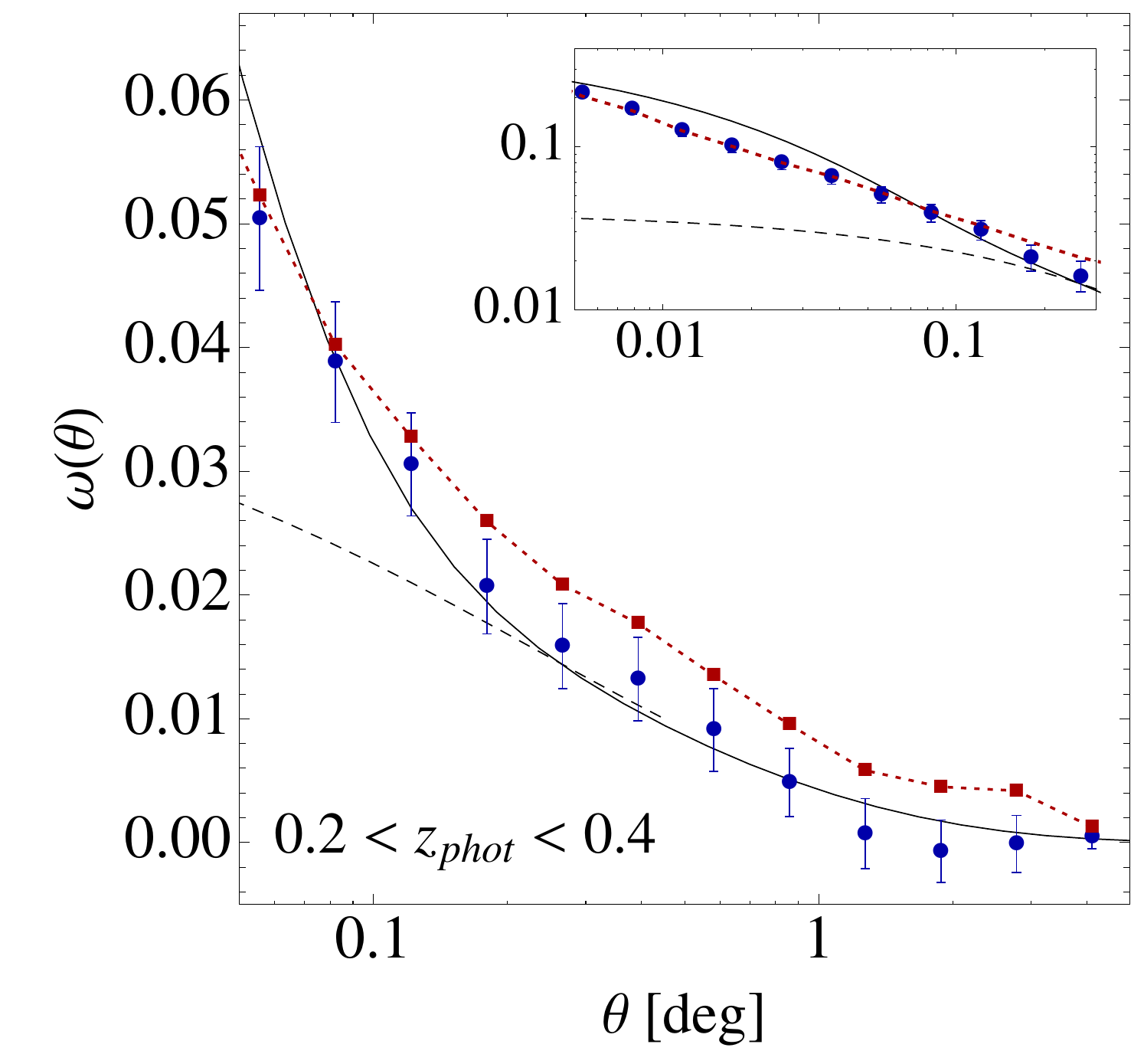} 
\includegraphics[width=0.32\textwidth]{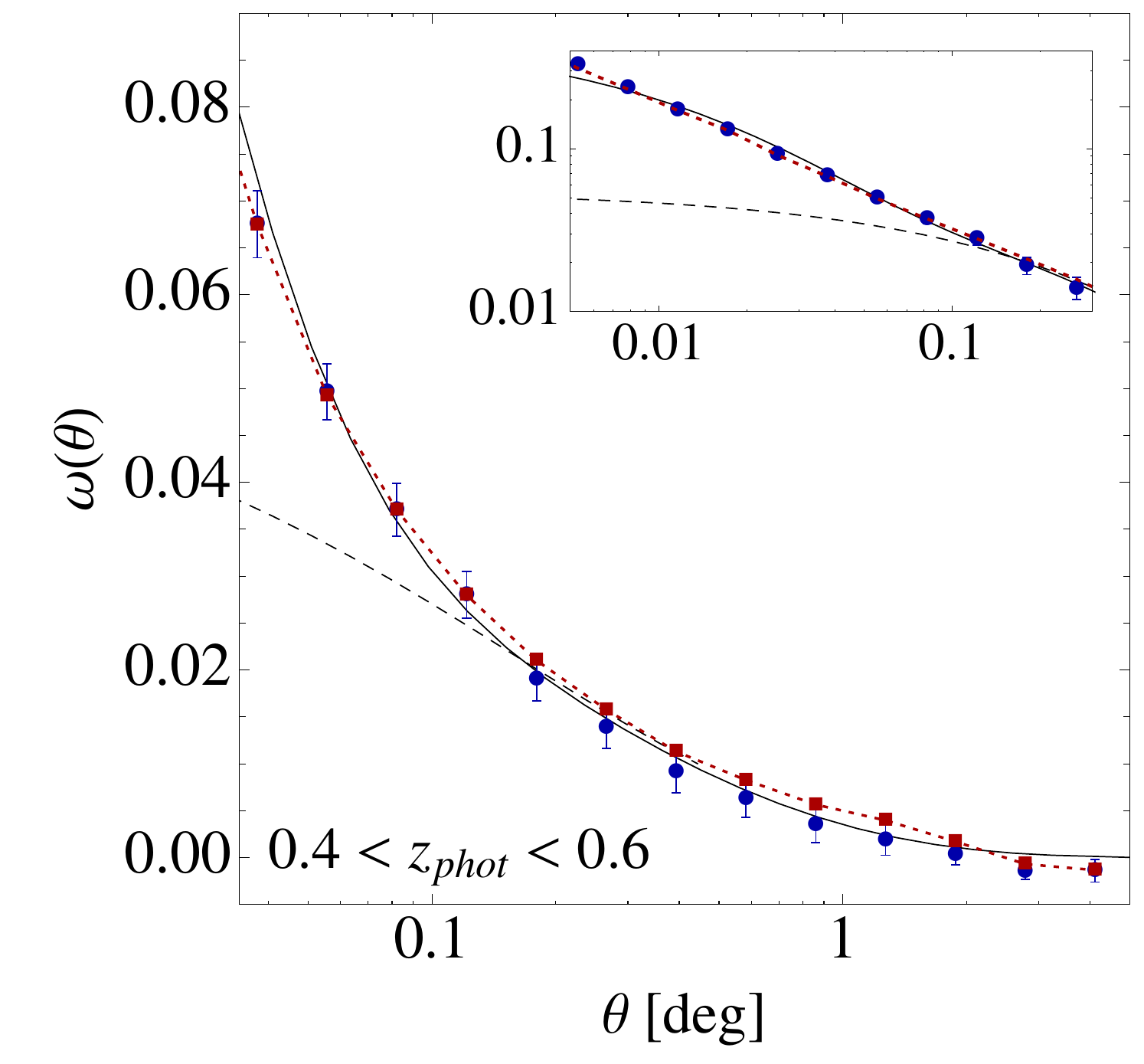} 
\includegraphics[width=0.32\textwidth]{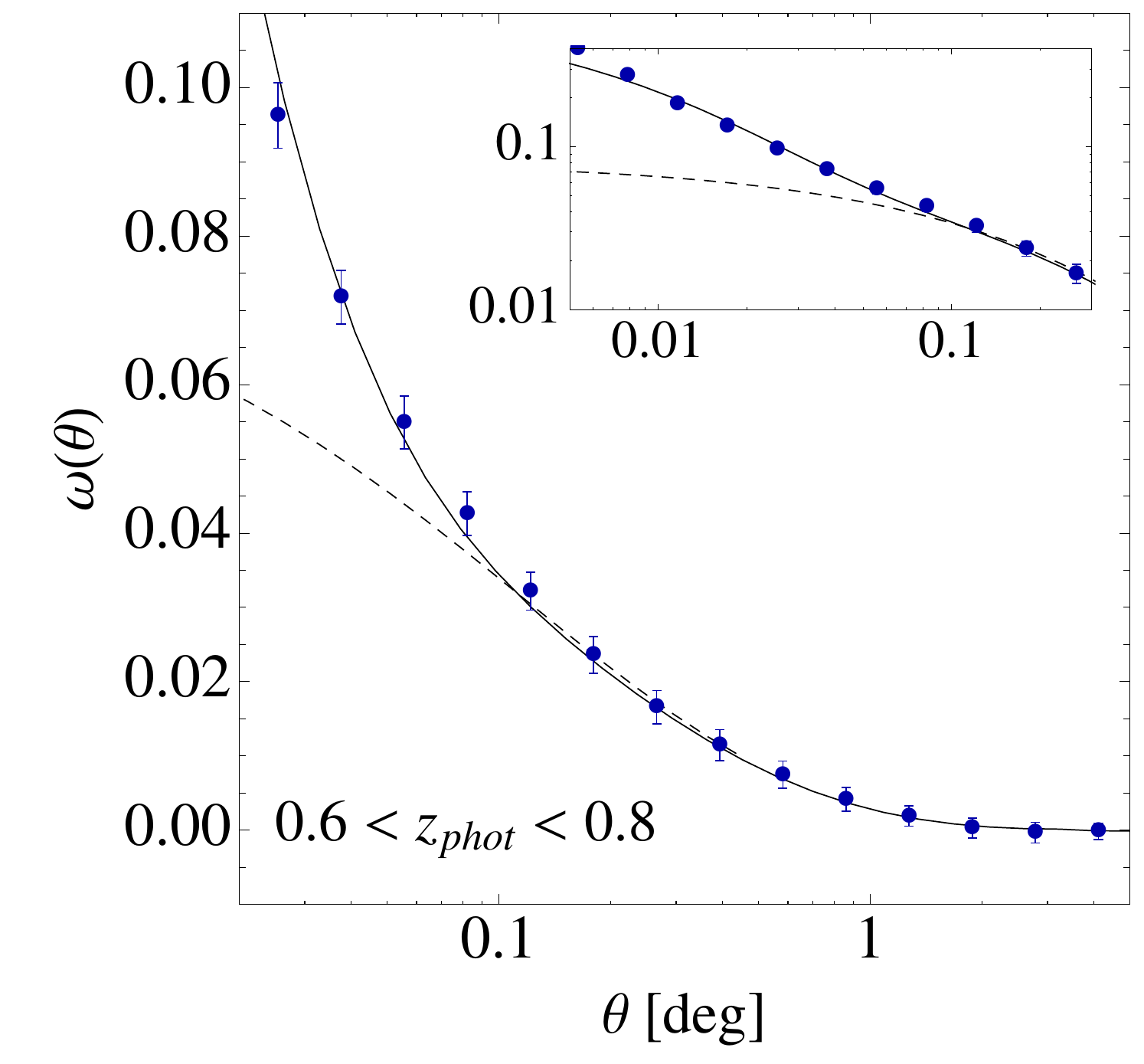} 
\includegraphics[width=0.32\textwidth]{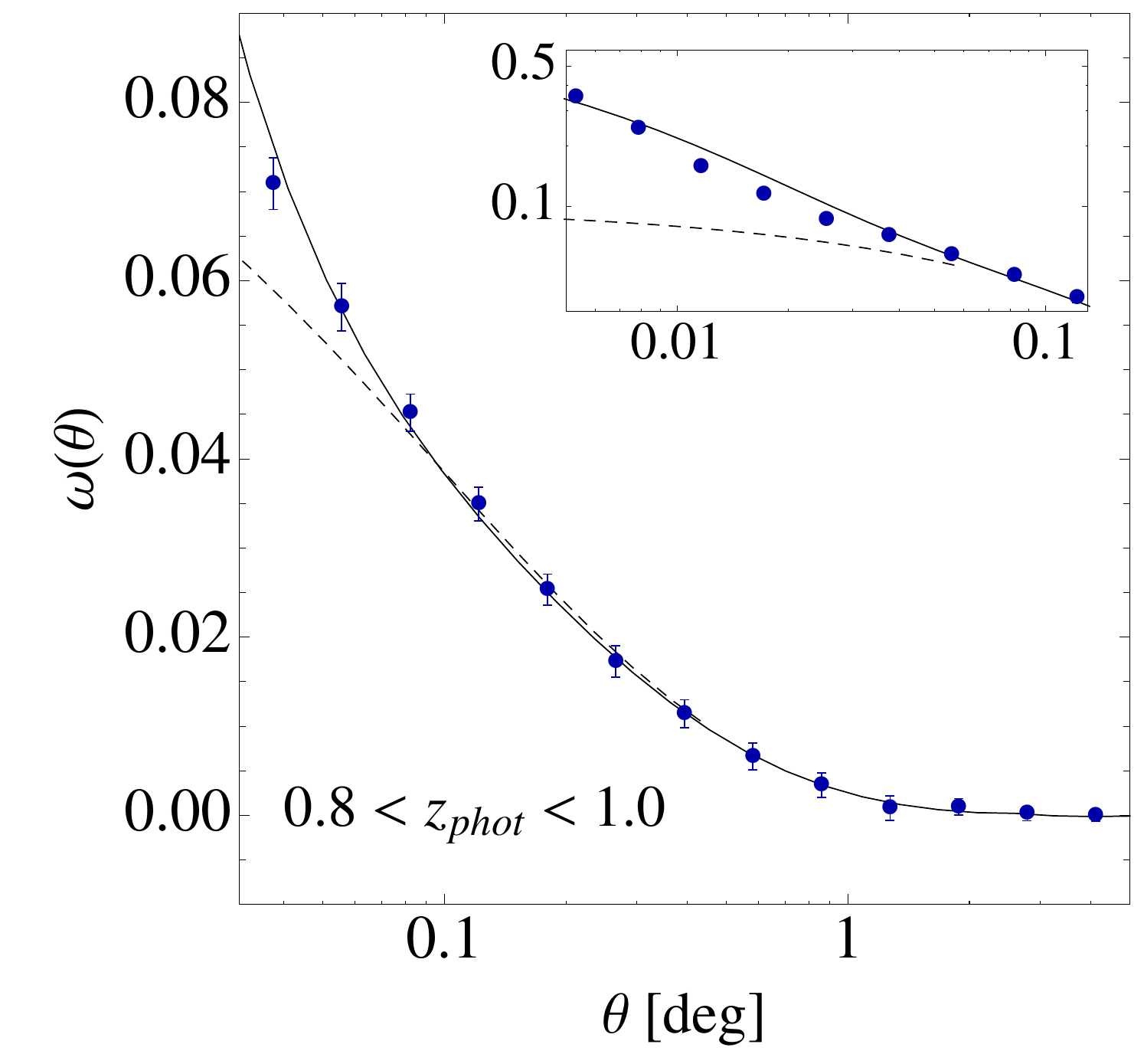} 
\includegraphics[width=0.32\textwidth]{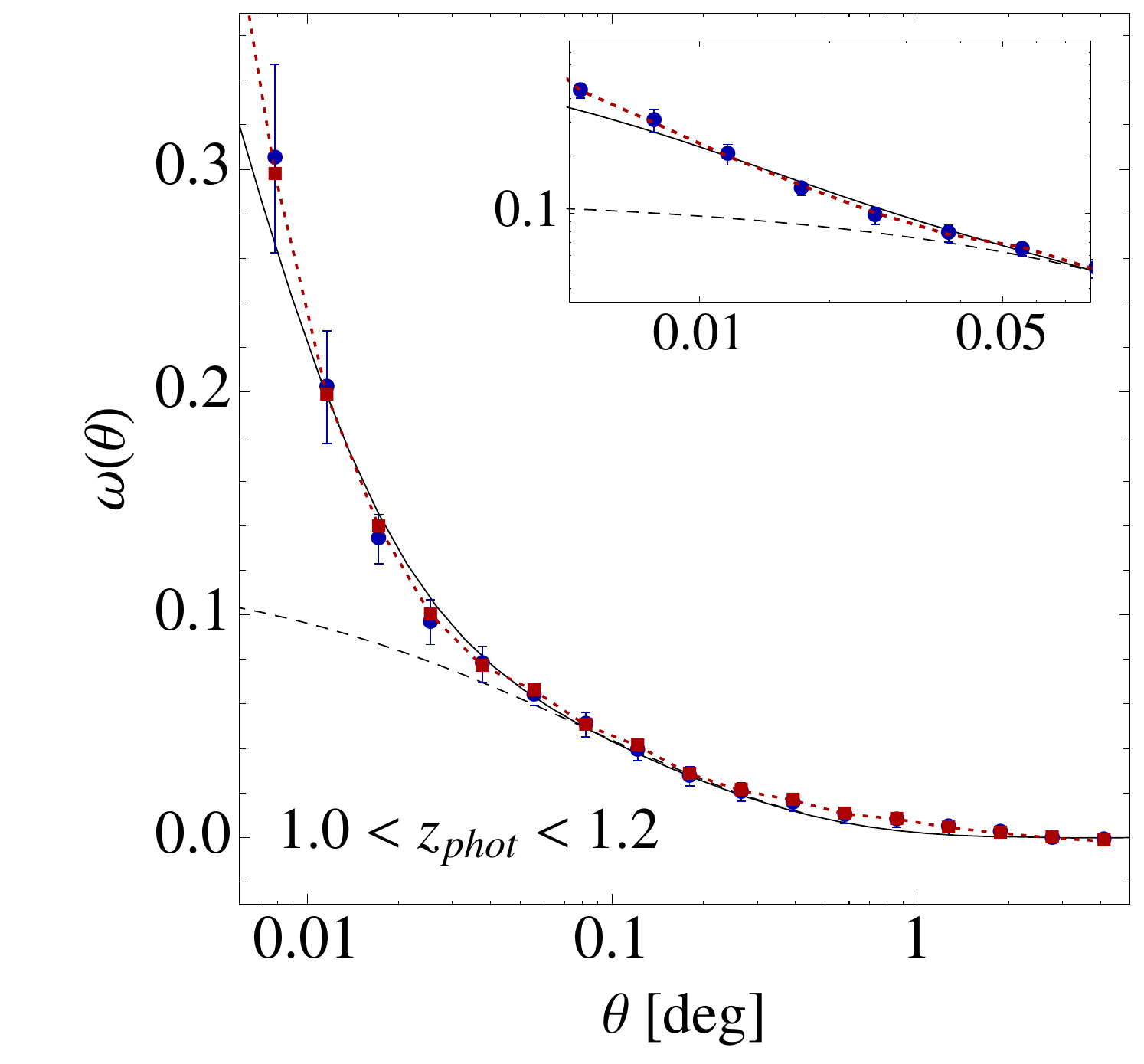} 
\caption{Angular auto-correlation functions, $w(\theta)$, in photo-$z$ bins of the BPZ estimator for our
  flux limited sample ($i < 22.5$) spanning a broad range in
  redshift, from $z \sim 0.2$ to $1.2$. Blue circles correspond to the measurements after correcting our sample for systematics as discussed in Sec.~\ref{sec:systematics}, while red squares is before such corrections (only shown when relevant). Solid lines correspond to a linear bias model applied to the non-linear matter $w(\theta)$ computed with {\tt Halofit}. The best-fit bias displayed was obtained from fitting the range of scales shown in each case (main panels). Dashed lines correspond to linear theory, with the same value of the bias. Note how the simple ``linear bias'' model describes the clustering towards scales considerably smaller than the linear regime shown by dashed lines. The inset panels show the performance of the best-fit ``linear bias'' model towards smaller scales than the ones used in the fit (main panels), see text for a detailed discussion.}
\label{fig:clustering_bins}
\end{flushleft}
\end{figure*}

We now discuss different aspects of our angular clustering measurements in the tomographic bins, in particular the corrections for systematics discussed above.
Let us first note that in what follows we define the best-fit bias $b$ and 1$\sigma$ error $\sigma_b$ in the standard way, i.e., given 
\begin{equation}
\chi^2(b) = \sum_{\theta,\theta^\prime} (\hat{w}(\theta)-b^2 \, w_T (\theta)) {\rm Cov}^{-1}(\theta,\theta^\prime) (\hat{w}(\theta^\prime)-b^2 \, w_T (\theta^\prime))
\label{eq:chi2note}
\end{equation}
we first find its minimum to define $b$ and then vary it around this value such that $\Delta \chi^2 = 1$ with $\Delta \chi^2 \equiv \chi^2(b+\sigma_b)-\chi^2(b)$. {\color{black} The sum runs through a range of scales that will be specified accordingly while the covariance has been defined in Sec.~\ref{sec:covariance} as the ``mixed'' approach, see Eq.~(\ref{eq:mixcov}). In Appendix \ref{sec:appendix_b} we discuss the best-fit bias results when different ways of estimating the covariance are considered. Note that we do not perform joint fits including multiple redshift bins, hence Eq.~(\ref{eq:chi2note}) does not include the covariance between different redshift bins.}
The $b^2\,w_{T}$ refers to the modeling of Eq.~(\ref{eq:wtheta1}), where $w_T$ is the 
underlying dark-matter clustering multiplied by a linear bias $b$ to yield the model galaxy-galaxy correlation function.
If $w_T$ is computed starting from the linear theory power spectrum, we call this model ``linear growth''. If instead this is computed with the nonlinear prescription {\tt Halofit} \citep{Takahashi2012}, but still maintaining a linear bias, we refer to it as the ``linear bias'' model. On large scales {\tt Halofit} reduces to linear theory, while on small scales the effects of non-linear growth increase the model correlation above its linear value.  The smallest scale at which $w_{T}$ from {\tt Halofit} equals $w_T$ from linear theory will be referred to as ``scale of linear growth''. These are reported in the caption of Table~\ref{tab:bpz_syst_bias} for each of the five photo-$z$ bins.

Tables \ref{tab:bpz_syst_bias} and \ref{tab:tpz_syst_bias} show the result of fitting a ``linear growth'' model 
 to the clustering measurements before any systematic correction is applied (second column, labeled ``baseline mask''), after a subsequent masking of regions with the worst observing conditions, as discussed in Sec. \ref{sec:sysmask} (third column, labeled ``+ bad regions masking'') and lastly, after the remaining correlation with relevant maps is corrected for as detailed in Sec.~\ref{sec:relsys} (fourth column, ``Gal-Syst cross correlations''). Table \ref{tab:bpz_syst_bias} corresponds to the BPZ photo-$z$ catalog, while Table \ref{tab:tpz_syst_bias} to the TPZ one. 

The corrections for systematic effects lower the clustering amplitudes and hence the derived best-fit biases. They are mainly important for low redshift, in particular the $0.2 < z < 0.4$ bin. Here the BPZ best-fit bias goes down by $\delta b / \sigma_b \sim 3$, while for TPZ the change is $\delta b / \sigma_b \sim 1.6$. For the $0.4 < z < 0.6$ bin the change in bias is $\mathcal{O}(\sigma_b)$. These two redshift ranges are important for the combination of galaxy-galaxy clustering with weak lensing observables, which typically use lens samples at low-$z$.
Hence these corrections are likely to be necessary to produce an unbiased cosmological analysis using such a combination of probes, e.g., a spuriously large clustering amplitude could translate to a spuriously high measurement of $\sigma_8$. At higher redshifts, which we expect to be more suitable for future BAO studies due to larger volumes and better photo-$z$ performance, we do not find significant observational systematic biases affecting our measurements, at least for the range of angular scales probed in with our current data.

\begin{table}
\begin{center}
\begin{tabular}{|c|c|c|c|c|} 
 \hline
 \multicolumn{5}{|c|}{BPZ (template method): best-fit bias and 1$\sigma$ error} \\
 \hline
 Photo-z & Baseline &  + Bad Area  & + Gal-Syst  & $\chi^2/$ \\
 Bin     & Mask     &  Masking        &  Cross-Corr & d.o.f. \\
 \hline
 $\! \! 0.2 < z < 0.4$ & $\! 1.28 \pm 0.07$ & $\! 1.20 \pm 0.07$ & $\! 1.05 \pm 0.07$  & $\!\! 2.5 / 7$   \\ 
 $\! \! 0.4 < z < 0.6$ & $\! 1.25 \pm 0.05$ & $\! 1.26 \pm 0.05$ & $\! 1.23 \pm 0.05$  & $\!\! 8.3 / 8$   \\ 
 $\! \! 0.6 < z < 0.8$ & $\! 1.35 \pm 0.04$ & $\! -$             & $\! - $             & $\!\! 2.3 / 9$   \\
 $\! \! 0.8 < z < 1.0$ & $\! 1.54 \pm 0.02$ & $\! -$             & $\! - $             & $\!\! 10.3 / 10$ \\ 
 $\! \! 1.0 < z < 1.2$ & $\! 2.20 \pm 0.07$ & $\! 2.17 \pm 0.09$ & $\! - $             & $\!\! 3.6 / 10$  \\ 
 \hline
\end{tabular}
\end{center}
\caption{Impact of the different corrections for observational systematic effects on the derived best-fit bias $b$ for the tomographic bins selected with our template method. The baseline mask is described in Sec.~\ref{sec:completeness} and corresponds to all regions where our sample is complete (i.e. $10\sigma$ depth  $i>=22.5$). The third column corresponds to an additional masking of regions with high values of potential systematic variables such as seeing, where we observe large decrements in galaxy density (as described in Sec.~\ref{sec:sysmask}). The fourth column refers to further corrections to $w(\theta)$ after these masks, in cases where the data still correlates with maps for potential systematics (as discussed in Sec.~\ref{sec:relsys}). The fifth column reports the $\chi^2/{\rm d.o.f}$ after all corrections applied. Empty entries refer to cases where such corrections were not necessary.  These values were obtained after fitting the ``linear growth'' model for scales $\theta >\theta_{min}=(0.26, 0.18, 0.12, 0.08, 0.06)$ deg., from first to last $z$ -bin.}
\label{tab:bpz_syst_bias}
\end{table}

\begin{table}
\begin{center}
\begin{tabular}{|c|c|c|c|c|} 
 \hline
 \multicolumn{5}{|c|}{TPZ (machine learning method)} \\
 \hline
 Photo-z & Baseline &  + Bad Area  & + Gal-Syst  & $\chi^2/$ \\
 Bin     & Mask     &  Masking     &  Cross-Corr & d.o.f. \\
 \hline
 $\! \! 0.2 < z < 0.4$ & $\! 1.18 \pm 0.07$ & $\! 1.13 \pm 0.08$ & $1.07 \pm 0.08$ & $\!\! 2.1 / 7$   \\ 
 $\! \! 0.4 < z < 0.6$ & $\! 1.29 \pm 0.04$ & $\! 1.30 \pm 0.04$ & $1.24 \pm 0.04$ & $\!\! 6.7 / 8$   \\ 
 $\! \! 0.6 < z < 0.8$ & $\! 1.34 \pm 0.05$ & $\! -$             & $\! -$          & $\!\! 14.5 / 9$   \\
 $\! \! 0.8 < z < 1.0$ & $\! 1.56 \pm 0.03$ & $\! -$             & $\! -$          & $\!\! 3.7 / 10$   \\ 
 $\! \! 1.0 < z < 1.2$ & $\! 1.97 \pm 0.09$ & $\! 1.96 \pm 0.06$ & $\! -$          & $\!\! 4.5 / 10$   \\ 
 \hline
\end{tabular}
\end{center}
\caption{Same as Table \ref{tab:bpz_syst_bias} but for the tomographic bins selected with the machine learning method. Different algorithms for photometric redshift estimation use different data quantities (most notably, template based ones use magnitude errors
while most machine learning do not). Thus one expects a different response to potential systematics.}
\label{tab:tpz_syst_bias}
\end{table}

\begin{figure*}
\begin{center}
\includegraphics[width=0.6\textwidth]{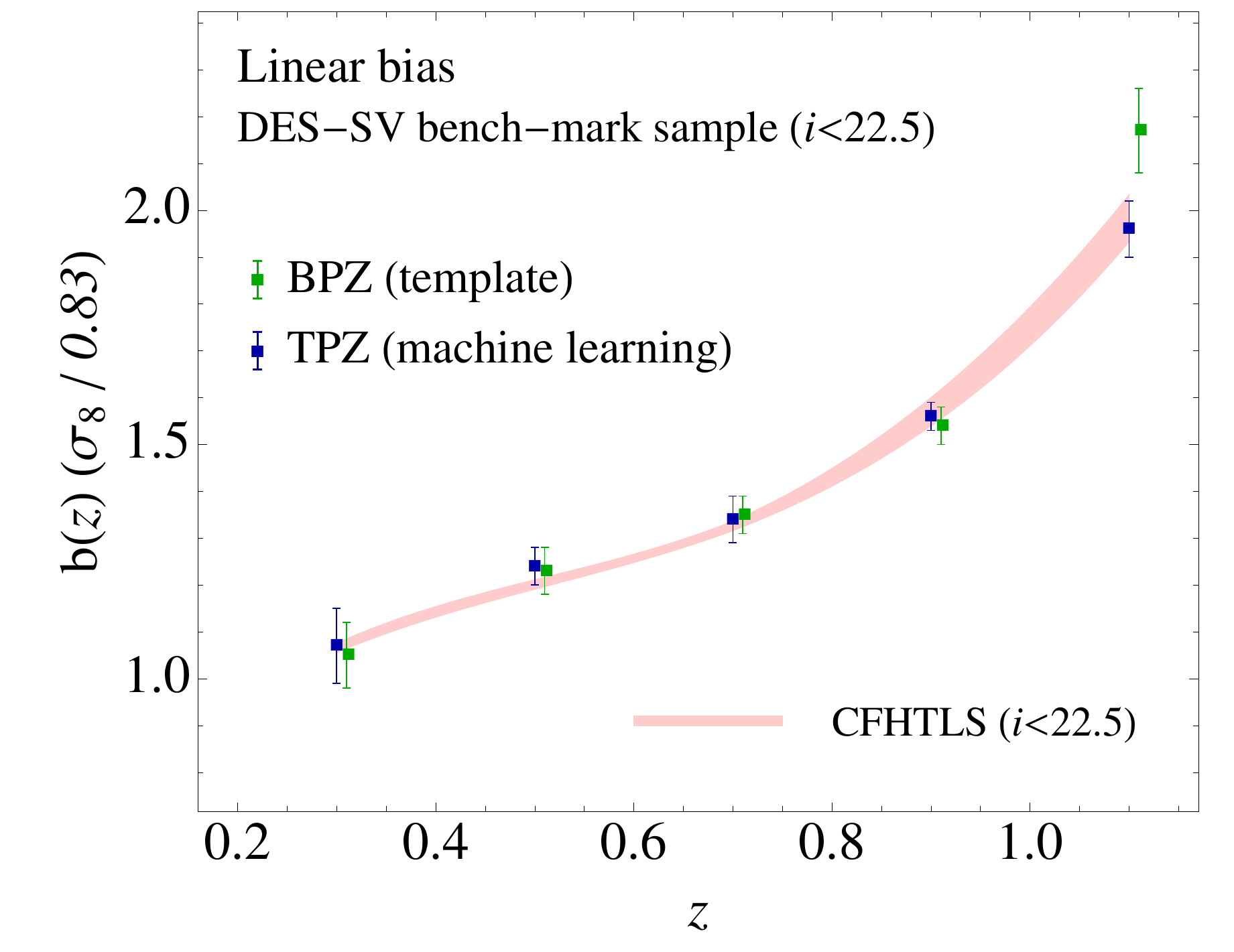} 
\caption{Comparison of the large-scale bias measured in a DES-SV flux limited sample ($i<22.5$) to equivalent measurements from CFHTLS derived from Coupon et al. (2012). We present DES results for two different photometric redshift catalogs, one obtained using a template method (BPZ), another with a machine learning approach (TPZ). 
The overall agreement between the two DES samples as a function of redshift is better that 2 per cent for $z<1$. At $z>1$ is difference is not statistically significant ($\sim 2 \sigma$). This represents a non-trivial test for DES-SV photometric redshift estimation. Our results are also in good agreement with those from CFHTLS, with $\chi^2/d.o.f=4/5$ for TPZ and $8.7/5$ for BPZ,  representing a cross-validation of data quality and sample selection.}
\label{fig:bias_vs_CFHTLS} 
\end{center}
\end{figure*}

Figure \ref{fig:clustering_bins} shows the angular correlation function measurements in the five
consecutive photo-$z$ bins in which we have split our sample (for the BPZ sample). Blue circles show our measurements after applying all of the corrections for observational systematic effects that we have detailed, while the square red symbols show the measurements before any correction. As is clear from Fig.~\ref{fig:clustering_bins} these corrections do not affect the small scale clustering, and are most important at low redshifts. The dashed curve in each panel corresponds to a linear theory model (``linear growth''), while the solid curve uses the non-linear {\tt Halofit} prescription (``linear bias''). Each model curve is for the bias given in Table \ref{tab:bpz_syst_bias}, fit over the range of angular scales indicated in its the caption. In Sec.~\ref{sec:linearbias} we discuss the angular range of validity of a linear bias prescription based on {\tt Halofit}, therefore in Fig.~\ref{fig:clustering_bins} we have chosen to display the clustering measurements to scales that go beyond the linear regime. 

There is good qualitative agreement between the linear bias model and the observed clustering at the scales shown in the main panels of Fig. \ref{fig:clustering_bins} (we discuss this more in quantitative terms below in Sec.~\ref{sec:linearbias}). This result is interesting because it implies that, at least for projected clustering in angular coordinates, the scale of non-linear biasing is different from the one of non-linear dark matter clustering. The latter is currently better understood, so in general terms this result is relevant. We will come back to it in Sec.~\ref{sec:linearbias}.

We further note that at large scales ($\theta \gtrsim 2^\circ$) all the correlations tend to zero. This is a signal that systematic effects are under control, as systematic variations tend to introduce spurious large-scale power \citep{Ross11,2012ApJ...761...14H,Leistedt13}). 

In the inset panels of Fig. \ref{fig:clustering_bins}, we show the clustering measurements at smaller angular scales and using a {\rm log} scaling for $w(\theta)$. The model curves are merely extrapolations using the best-fit bias recovered on larger-scales (main panels). Qualitatively, this simple model does not depart strongly from the measurements. The data at these scales will be used to constrain the halo occupation distribution of DES SV galaxies in Sobreira et al. (in prep.). However, the general agreement signals that a more elaborate non-linear bias prescription based on perturbation theory \citep{1993ApJ...413..447F,2006PhRvD..74j3512M,2012PhRvD..85h3509C}
might be adequate to describe the data in this high signal-to-noise regime. Such study is left for future work. Instead, in Sec.~\ref{sec:linearbias} we will show that the current data size and quality of DES-SV is able to distinguish the breakdown of a ``linear bias'' prescription in detail. 

Lastly we note that we have chosen not to show the clustering for the TPZ sample in a manner analogous to Fig.~\ref{fig:clustering_bins}, as the results look almost identical at the qualitative level.

\subsection{Bias Evolution and Comparison to CFHTLS}

The DES science verification data used in this paper (and in a series of papers accompanying it) is in several regards similar to that collected by the CFHTLS collaboration, for example in depth, photometry and area.
CFHTLS has, to this point in time, been regarded as the state of the art for deep wide area photometric data. Hence in this section
we compare our clustering measurements to those presented by \cite{2012A&A...542A...5C}.

Galaxies in \cite{2012A&A...542A...5C} are selected according to SEXtractor $\magauto$ magnitudes of $i <22.5$, and thus comprise a very similar sample to the one presented in this paper. 
To facilitate the comparison further we have chosen our photo-$z$ bins similarly to those in the clustering study of \cite{2012A&A...542A...5C}. Although photometric redshifts in CFHTLS were estimated using a different template method, LePhare (\cite{2006A&A...457..841I}, \cite{2009A&A...500..981C}) we have used a very similar set of templates to obtain our DES BPZ data \citep{2014MNRAS.445.1482S}. The comparison to a neural network photo-$z$ catalog such as TPZ is novel in this regard.

The apparent magnitude sample $i<22.5$ in \cite{2012A&A...542A...5C} was then split and reported as several volume limited absolute magnitude ``threshold" samples in each photo-$z$ bin. Therefore in each redshift bin, our sample (also selected by $i<22.5$) corresponds to the faintest absolute magnitude sample in \cite{2012A&A...542A...5C}, except for the fact that \cite{2012A&A...542A...5C} imposed cuts to make the sample volume limited, where we have not. For example, for $0.4 < z < 0.6$ there is a group of galaxies brighter than $i = 22.5$ but fainter than $M_r-5 \log h=-18.8$ (see their Fig. 4). These galaxies are in our sample but not in the $M_r - 5 \log_{10} h < -18.8$ sample of 
\cite{2012A&A...542A...5C}. In Appendix \ref{sec:appendix_c} we estimate the impact of these differences in selection when comparing $b(z)$ results. We find the differences are most relevant at high $z$, where the galaxies in the samples have luminosity thresholds, $L$, such that $db/dL$ is large.

The bias evolution reported by \cite{2012A&A...542A...5C}, accounting for the differences in sample selection as described in Appendix \ref{sec:appendix_c}, is shown by a filled light-red region in Fig.~\ref{fig:bias_vs_CFHTLS} (where the width corresponds to the statistical errors bars reported by CFHTLS) \footnote{We have re-scaled the CFHTLS measurements that assumed at $z=0$ $\sigma_8=0.8$ to our cosmology $\sigma_8 = 0.83$}. The green squares with error bars correspond to our bias measurements for the BPZ sample presented in the previous section and in Table \ref{tab:bpz_syst_bias}, while the blue squares correspond to our TPZ sample as reported in Table \ref{tab:tpz_syst_bias} (both for the cases where all corrections for observational systematics have been applied). 

Overall, the match in the recovered bias as a function of redshift between our apparent magnitude sample and the corresponding one by CFHTLS is good, with a slight tension at the highest bin for BPZ.
Note however that for CHFTLS the error bars depicted in Fig.~\ref{fig:bias_vs_CFHTLS} are statistical only (as reported by \cite{2012A&A...542A...5C}), while in our case we include effects such as 
the reduction of effective area due to further masking of bad observing regions, or 
the impact of stellar contamination.
If we define 
\begin{equation}
\chi^2 = \sum_{{\rm bin}\,j=1}^{5} (b^{(j)}_{\rm DES}-b^{(j)}_{\rm CFHTLS})^2/(\sigma^{(j)}_b)^2
\end{equation}
and add the errors in quadrature, $\sigma_b^2 = \sigma_{b,\,{\rm DES}}^2 + \sigma_{b,\,{\rm CFHTLS}}^2$ we find that $\chi^2(BPZ) = 8.7$ while $\chi^2(TPZ) = 4$ for five degrees of freedom in each case. Assuming that the DES and CFHTLS data are drawn from the same underlying Gaussian distribution, greater $\chi^2$ would be expected in 55 per cent of cases for the TPZ comparison and 12 per cent of cases for the BPZ comparison. This means that the agreement is better than 1$\sigma$ for TPZ and better than 2$\sigma$ for BPZ. 

\begin{figure}
\begin{center}
\includegraphics[width=0.49\textwidth]{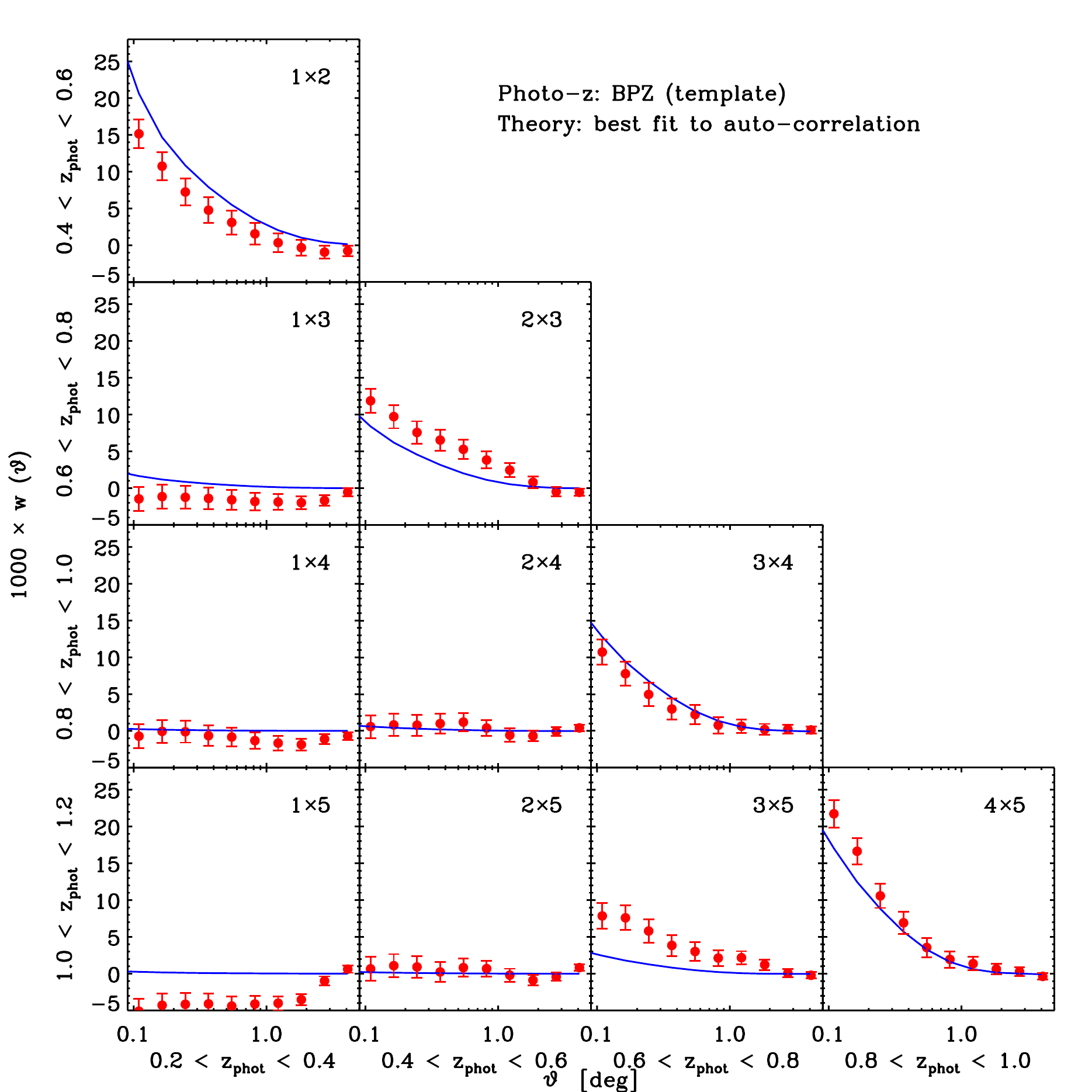}  
\caption{Full set of two-point cross-correlation functions between all of the BPZ redshift bins we use. The theory curves represent a Planck fiducial model, with linear bias values fitted to the auto-correlations displayed in Fig. \ref{fig:clustering_bins} over scales $\theta > 0.2$ deg. Cases where this produces a match to the observed cross-correlations demonstrates the robustness of the $dn/dz$ estimation.}
\label{fig:crosscorr_bpz} 
\end{center}

\end{figure}
\begin{figure}
\begin{center}
\includegraphics[width=0.49\textwidth]{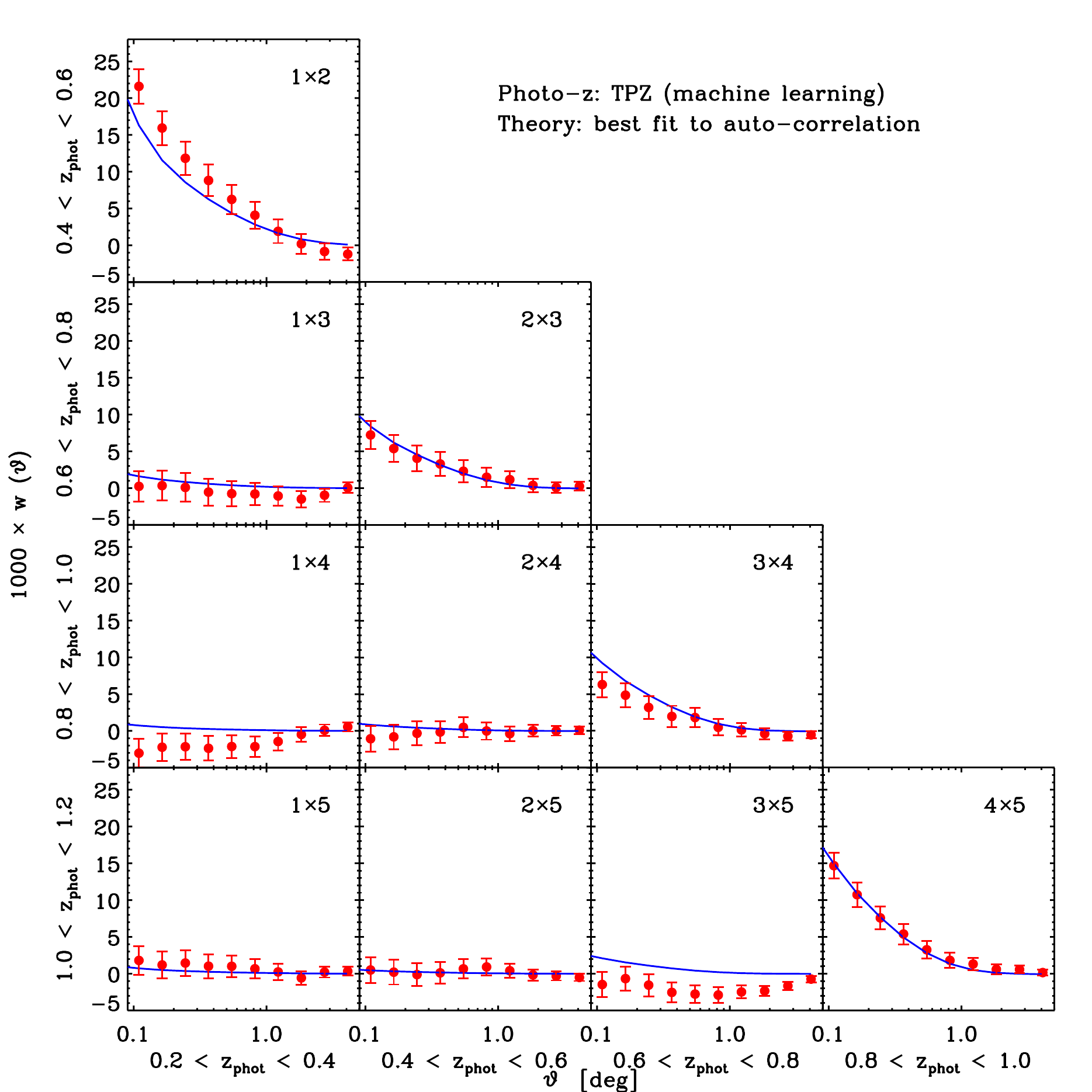}  
\caption{Same as Fig.~\ref{fig:crosscorr_bpz} for the photometric catalog derived with TPZ.}
\label{fig:crosscorr_tpz} 
\end{center}
\end{figure}

We stress that the agreement for the bias evolution between the two photo-$z$ catalogs is non-obvious because different photo-$z$ codes could, in principle, select different types of galaxies in different redshift bins (note that the algorithms implemented in BPZ and TPZ are very different). This might explain the difference at the last $z$-bin if BPZ
were selecting a rarer population with a higher bias. However we note that if the two samples were independent, the disagreement would be 2.1$\sigma$, and thus not remarkable. For $z < 1.0$, the two techniques return samples with bias values that agree to better than 2 per cent, suggesting that at these redshifts, systematic uncertainties associated with the photo-$z$s in this redshift range are well-controlled. 

The increase in bias with redshift found in Fig.~\ref{fig:bias_vs_CFHTLS} is primarily a consequence of the effective luminosity threshold, as this threshold naturally increases with redshift for a flux limited sample. Only the high luminosity galaxies will be observed at high $z$, and they are more biased. This effect accentuates for $z \gtrsim 0.8$ as the majority of galaxies come from the exponential tail of the luminosity function. The $b(z)$ shown in Fig.~\ref{fig:bias_vs_CFHTLS} is therefore not physical in the sense that the sample is changing with redshift. In Appendix \ref{sec:appendix_a} we show that this evolution of galaxy bias as a function of redshift is also recovered in simulations based on Halo Occupation Distribution techniques to populate halos with galaxies.

\subsection{Consistency from cross-correlations of photo-$z$ bins}

As a further test of potential systematics and photo-$z$ we show in Figs.~\ref{fig:crosscorr_bpz} and \ref{fig:crosscorr_tpz} the full set of cross-correlation measurements between photo-$z$ bins for the BPZ and TPZ samples, respectively (see \cite{2012MNRAS.425.1527C} for a related analysis using SDSS data). These measurements were done using a cumulative mask, for each photo-$z$ estimator, combining the ones derived for each photo-$z$ bin in Sec. \ref{sec:sysmask}.

For bins that are well-separated in redshift, we observe that their cross-correlations are consistent with zero, as expected from the redshift distributions in Fig.~\ref{fig:dndz}. One notable exception is the measured cross-correlation between bins $0.2 < z < 0.4$ and $1.0 < z < 1.2$ for BPZ, which is significantly negative. Understanding the cause of this is left for future work. 

Solid lines in Figs.~\ref{fig:crosscorr_bpz} and \ref{fig:crosscorr_tpz} correspond to our fiducial model, given by Eq.~(\ref{eq:wtheta1}), for $w_{ij}$ provided with the redshift distributions in photo-$z$ bin $i$ and $j$. They use a bias $b=\sqrt{b_i b_j}$ where $b_i$ and $b_j$ are the 
the best fit ``linear growth'' model bias from the auto-correlations. This is therefore a fully predictive model given the auto-correlations, and a test for outliers and the reconstructed $dn/dz$.

In Tables \ref{tab:bpz_cross_bias} and \ref{tab:tpz_cross_bias} we compare the predicted $\sqrt{b_i b_j}$ with bias values determined by directly fitting the cross-correlation functions over the same range of scales ($\theta > 0.2$ deg), which we call $b_{ij}$\footnote{Note that we use the same range of scales here for all redshift bins, so that a fair comparison is possible for all cross-correlations. The biases thus differ slightly than those that would be inferred from Table \ref{tab:bpz_syst_bias}.}. 
We estimate the error on the bias prediction $\sqrt{b_ib_j}$ by a direct propagation of the error on the auto-correlation biases, $\sigma_{b_i}$ and $\sigma_{b_j}$,
\begin{equation}
\sigma_{\sqrt{b_i b_j}} =(\sqrt{b_i b_j}/2) (\sigma_{b_i} / b_i + \sigma_{b_j} / b_j).
\end{equation}
The fourth column displays the absolute difference between $\sqrt{b_i b_j}$ and $b_{ij}$ divided by their error combined in quadrature. In the 20 total cases tested (considering both photo-$z$ methods), we find only one case where $\Delta b/\sigma_b$ is greater than 2.21$\sigma$ (BPZ bin 3$\times$5) and only two cases where they are greater than 1.45$\sigma$ (the additional case is TPZ bin 1$\times$2). Considered in total, the cross-correlation results agree with the auto-correlation results at roughly the expected statistical level.

The picture recovered from Figs.~\ref{fig:crosscorr_bpz} and \ref{fig:crosscorr_tpz} and Tables  \ref{tab:bpz_cross_bias} and \ref{tab:tpz_cross_bias} is that outliers are not causing obvious inconsistencies in our analysis, and that to a large extent our estimated redshift distributions are correct. However, future DES data will require more work to assess the level of systematic uncertainty associated with photo$-z$ estimation.

\subsection{On the Scale of Linear Bias}
\label{sec:linearbias}

In Sec.~\ref{sec:results} and Fig.~\ref{fig:clustering_bins} we showed that on large angular scales the clustering recovered in our data agrees well with the standard linear theory prescription: $b^2 \,w_{\rm DM,\,Lin}(\theta)$ for all photo-$z$ bins. We further noted that at smaller scales the assumption of linear bias appeared to still hold provided the description of the underlying dark matter clustering accounted for non-linear growth of structure (for which we used {\tt Halofit} from \citealt{Takahashi2012}). In this section we explore in quantitative terms the scale at which this assumption leads to
a significant data-theory discrepancy\footnote{In this section, we focus on the BPZ data for concreteness but the conclusions reached in this section do not significantly depend on this choice. }. 

\begin{table}
\begin{center}
\begin{tabular}{|c|c|c|c|}
 \hline
 \multicolumn{4}{|c|}{BPZ\,cross-correlations} \\
 \hline
 Bin i $\times$ j & $b_{ij}$  & $\sqrt{b_i b_j}$ & $\Delta b / \sigma_b$ \\
 \hline
 1$\times$2 & $0.98 \pm 0.07$ & $1.10 \pm 0.05$ & $1.45$ \\
 1$\times$3 & $0.00 \pm 1.23$ & $1.12 \pm 0.05$ & $0.91$ \\
 1$\times$4 & $0.01 \pm 4.86$ & $1.19 \pm 0.05$ & $0.24$ \\
 1$\times$5 & $0.00 \pm 3.99$ & $1.44 \pm 0.07$ & $0.36$ \\ \\

 2$\times$3 & $1.25 \pm 0.14$ & $1.26 \pm 0.05$ & $0.04$ \\
 2$\times$4 & $0.00 \pm 1.64$ & $1.33 \pm 0.05$ & $0.81$ \\
 2$\times$5 & $0.04 \pm 5.74$ & $1.61 \pm 0.08$ & $0.27$ \\ \\

 3$\times$4 & $1.22 \pm 0.13$ & $1.36 \pm 0.06$ & $0.97$ \\
 3$\times$5 & $3.27 \pm 0.43$ & $1.65 \pm 0.08$ & $3.74$ \\ \\

 4$\times$5 & $1.96 \pm 0.12$ & $1.75 \pm 0.08$ & $1.39$ \\
\hline
\end{tabular}
\end{center}
\caption{Comparison of best-fit biases $b_{ij}$ obtained from fitting the two-point cross-correlation functions between photo-$z$ bins $i$ and $j$ (displayed in Fig.~\ref{fig:crosscorr_bpz}) vs. the prediction for such signal from the best-fit to auto-correlations in $z$-bin $i$ and $z$-bin $j$, given by $\sqrt{b_i b_j}$. 
The fourth column shows the absolute difference $\Delta b = |b_{ij}-\sqrt{b_i b_j}|$ divided by their error combined in quadrature.}
\label{tab:bpz_cross_bias}
\end{table}

\begin{table}
\begin{center}
\begin{tabular}{|c|c|c|c|}
 \hline
 \multicolumn{4}{|c|}{TPZ\,cross-correlations} \\
 \hline
 Bin i $\times$ j & $b_{ij}$  & $\sqrt{b_i b_j}$ & $\Delta b / \sigma_b$ \\
 \hline
 1$\times$2 & $1.35 \pm 0.08$ & $1.13 \pm 0.05$ & $2.21$ \\
 1$\times$3 & $0.00 \pm 1.91$ & $1.12 \pm 0.05$ & $0.58$ \\
 1$\times$4 & $0.00 \pm 1.79$ & $1.24 \pm 0.06$ & $0.69$ \\
 1$\times$5 & $0.00 \pm 3.28$ & $1.36 \pm 0.07$ & $0.41$ \\ \\

 2$\times$3 & $1.01 \pm 0.21$ & $1.25 \pm 0.05$ & $1.11$ \\
 2$\times$4 & $0.00 \pm 1.40$ & $1.39 \pm 0.06$ & $0.99$ \\
 2$\times$5 & $0.00 \pm 2.35$ & $1.53 \pm 0.08$ & $0.64$ \\ \\

 3$\times$4 & $1.11 \pm 0.20$ & $1.37 \pm 0.06$ & $1.27$ \\
 3$\times$5 & $0.00 \pm 2.14$ & $1.51 \pm 0.07$ & $0.70$ \\ \\

 4$\times$5 & $1.61 \pm 0.15$ & $1.68 \pm 0.08$ & $0.39$ \\
\hline
\end{tabular}
\end{center}
\caption{Same as Table \ref{tab:bpz_cross_bias}, but for the TPZ cross correlations displayed in Fig.~\ref{fig:crosscorr_tpz}.}
\label{tab:tpz_cross_bias}
\end{table}

We define the transition from linear to non-linear dark matter clustering as the minimum scale such that $w_{\rm DM,\,Halofit} = w_{\rm DM,\,Lin}$.
Given our photo-$z$ bins this occurs at $\theta_{\rm Linear\,Growth} \sim (0.26, 0.16, 0.11, 0.09, 0.08)$ deg. for bins one to five. This corresponds to roughly $4\,{\rm Mpc}\,h^{-1}$ of comoving separation across redshift given our cosmological model. The best-fit biases and $\chi^2$ on linear scales ($\theta > \theta_{\rm Linear\,Growth}$) are provided in Table~\ref{tab:bpz_syst_bias}. 

To state at what point the ``linear bias'' model becomes notably worse than the ``linear growth'' model we perform the following test. We first find the $\chi^2$ using the data vector and covariance matrix for the angular bins where $\theta > \theta_{\rm Linear\,Growth}$ and the ``linear growth'' model. Then we add new data to the fit by decreasing the minimum angular scale considered, $\theta_{\rm min}<\theta_{\rm Linear\,Growth}$, and using a ``linear bias'' model we determine the difference in $\chi^2$ values between the extended data vector and the ``linear growth'' one.
A  large $\Delta \chi^2$ (depending on the extra data-points, $\Delta ({\rm d.o.f.})$) implies that a model with a single bias parameter is no longer a good description of the extended data vector. In quantitative terms, we define $\theta_{\rm min}$ such that,
\begin{equation}
\Delta \chi^2 -\Delta({\rm d. o .f}) = 4
\end{equation}
as the point where a $2\sigma$ preference for an improved (but unknown) model for the bias as a function of scale exists. We thus define the largest angular scale at which this condition is met as our measurement of $\theta_{\rm Linear\,Bias}$. In what follows, we refer to the results of this test as the results of the `$\chi^2$ test'.

\begin{figure}
\begin{center}
\includegraphics[trim= 0 1.2cm 0 0, width=0.45\textwidth]{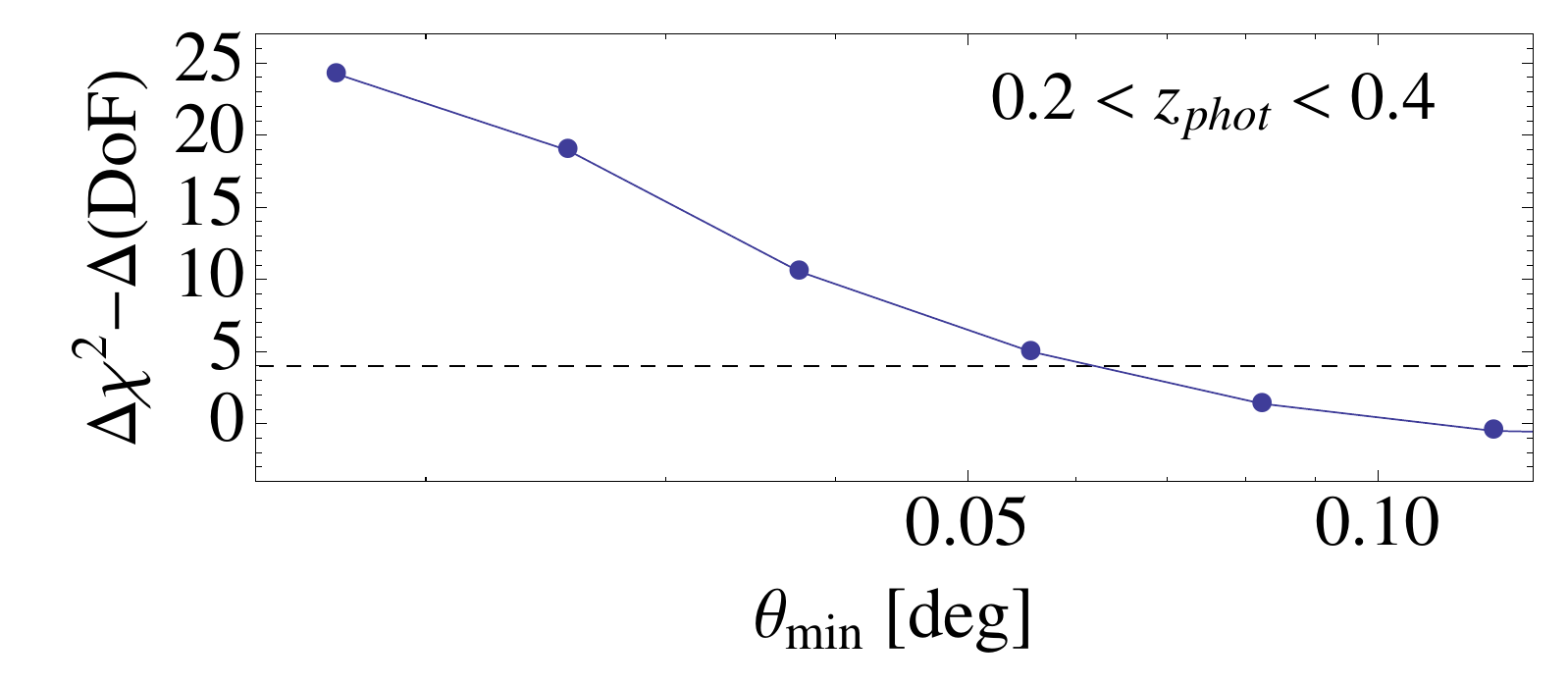} 
\includegraphics[trim= 0 1.2cm 0 0, width=0.45\textwidth]{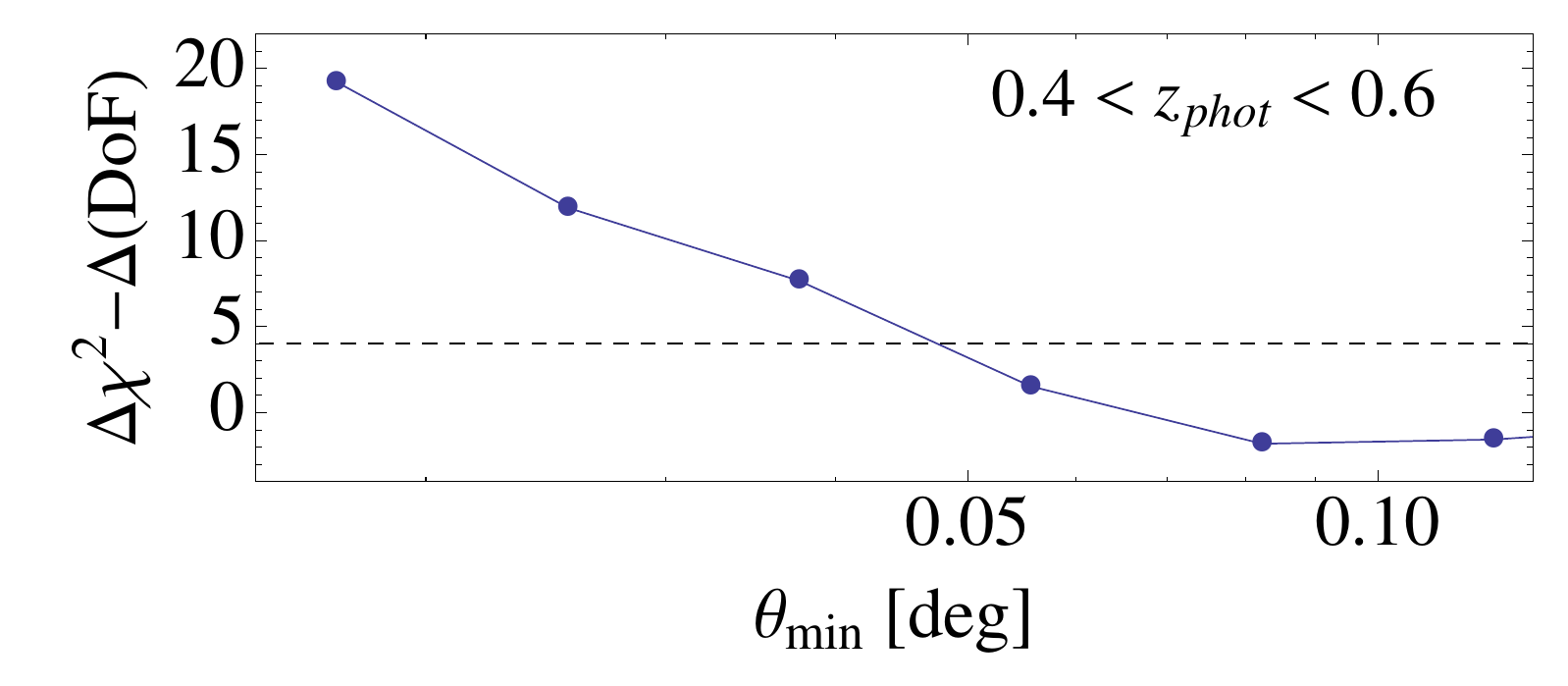} 
\includegraphics[trim= 0 1.2cm 0 0, width=0.45\textwidth]{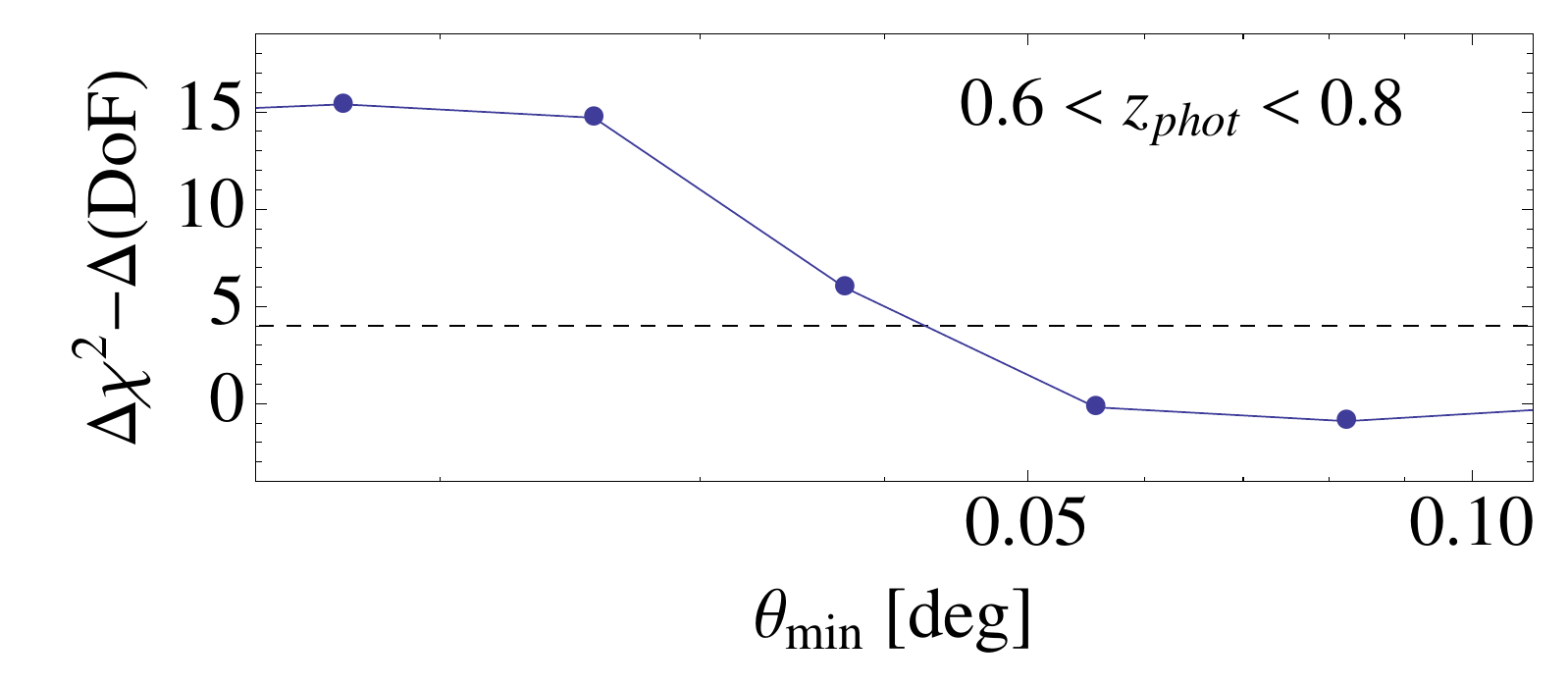} 
\includegraphics[trim= 0 1.2cm 0 0, width=0.45\textwidth]{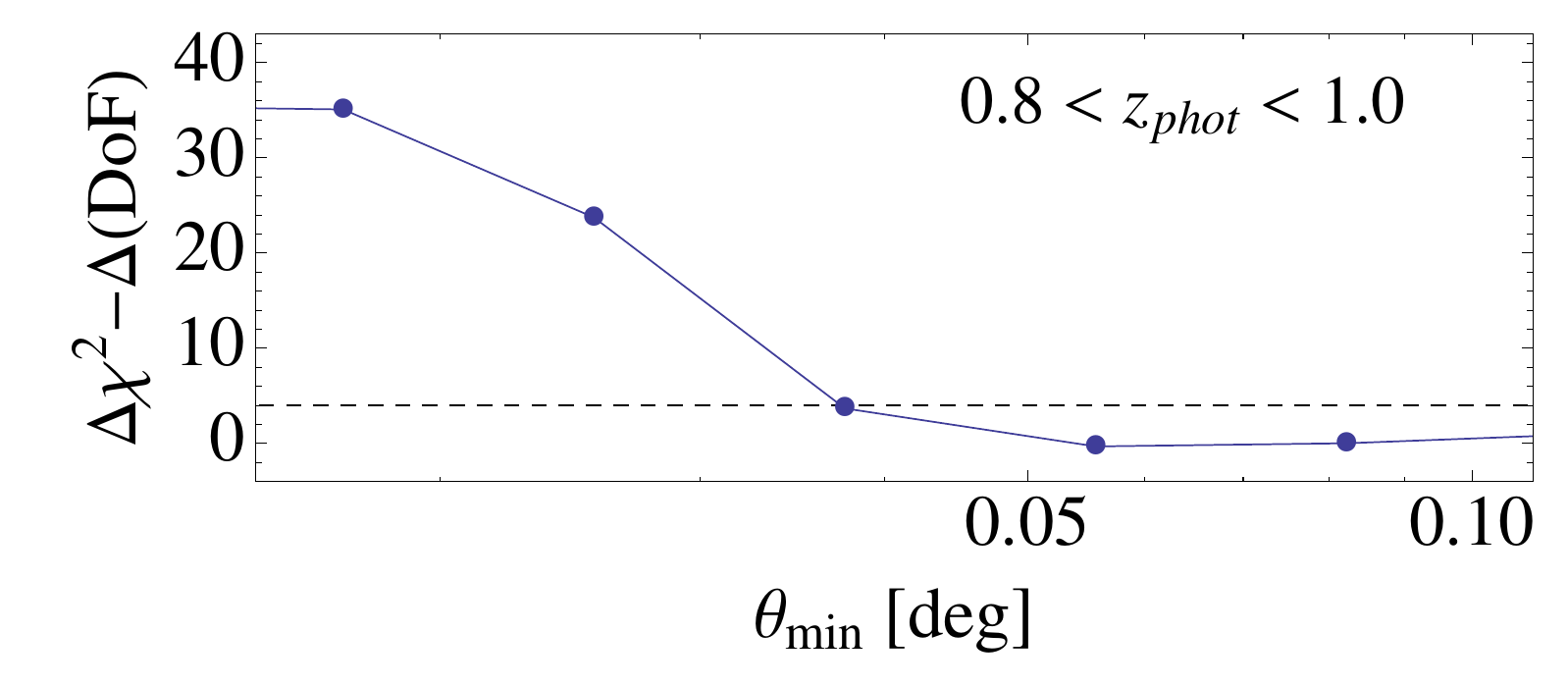} 
\includegraphics[trim= 0 0cm   0 0, width=0.45\textwidth]{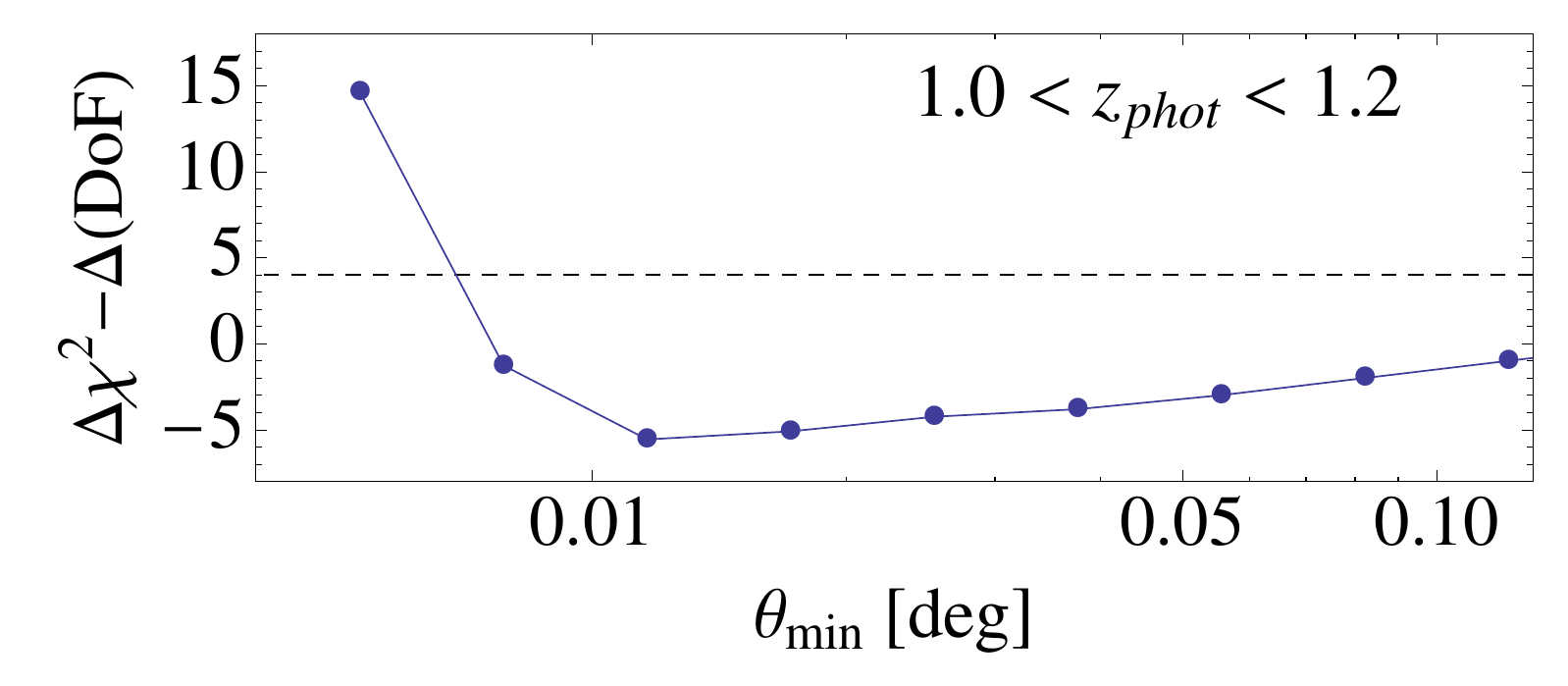}
\caption{We estimate the smallest scale $\theta_{\rm min}$ at which a linear bias model is still a good description of the clustering by computing the $\chi^2$ difference between a fit extending to $\theta_{\rm min}$ and a fit only on large linear scales. A $\Delta \chi^2 \sim 4 + \Delta (dof)$ roughly corresponds to $2\sigma$ evidence that the linear bias model fails to describe the data.}
\label{fig:bias_chi2} 
\end{center}
\end{figure}

\begin{figure}
\begin{center}
\includegraphics[trim=0 1cm 0 0, width=0.4\textwidth]{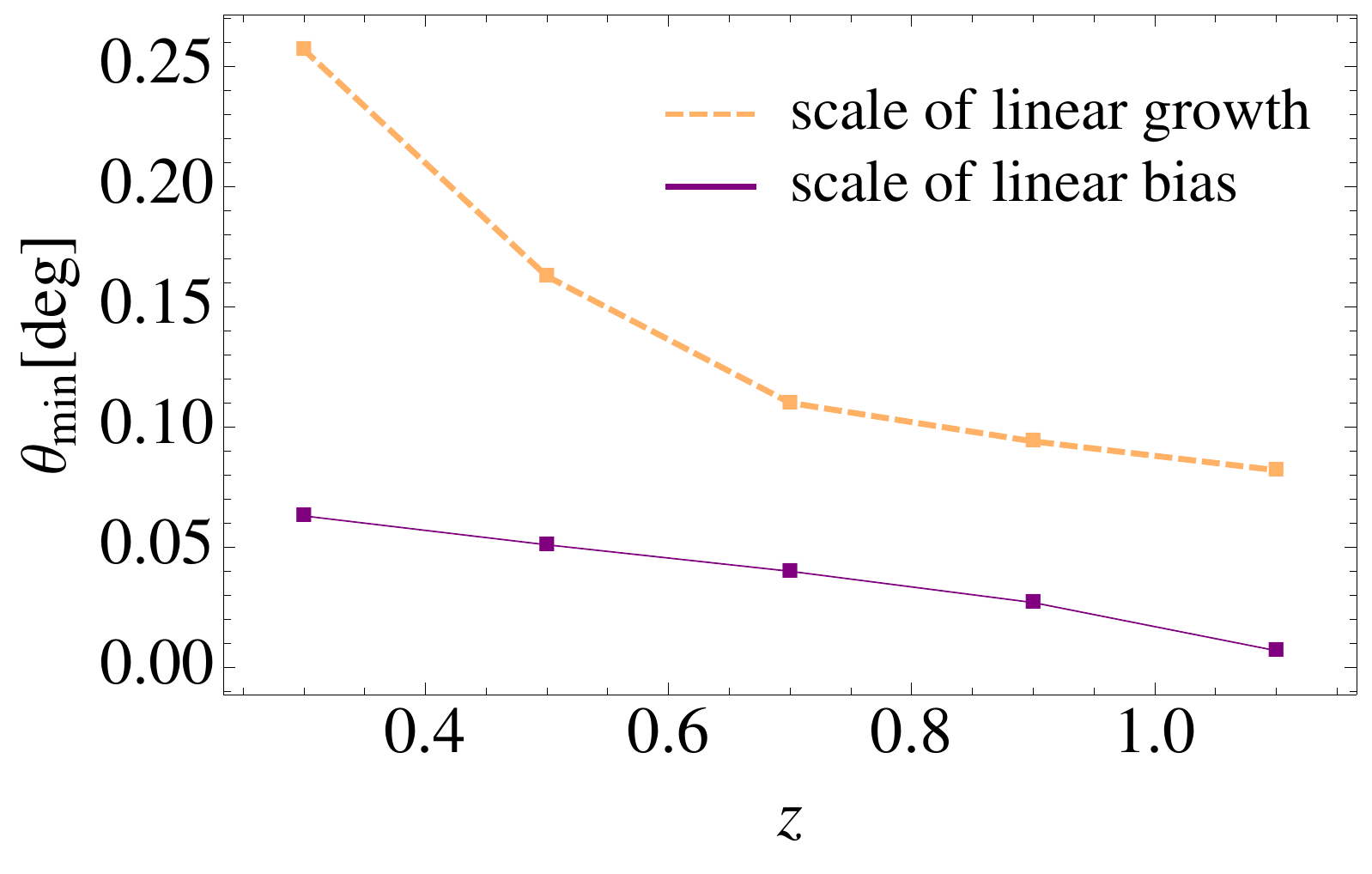} 
\includegraphics[width=0.4\textwidth]{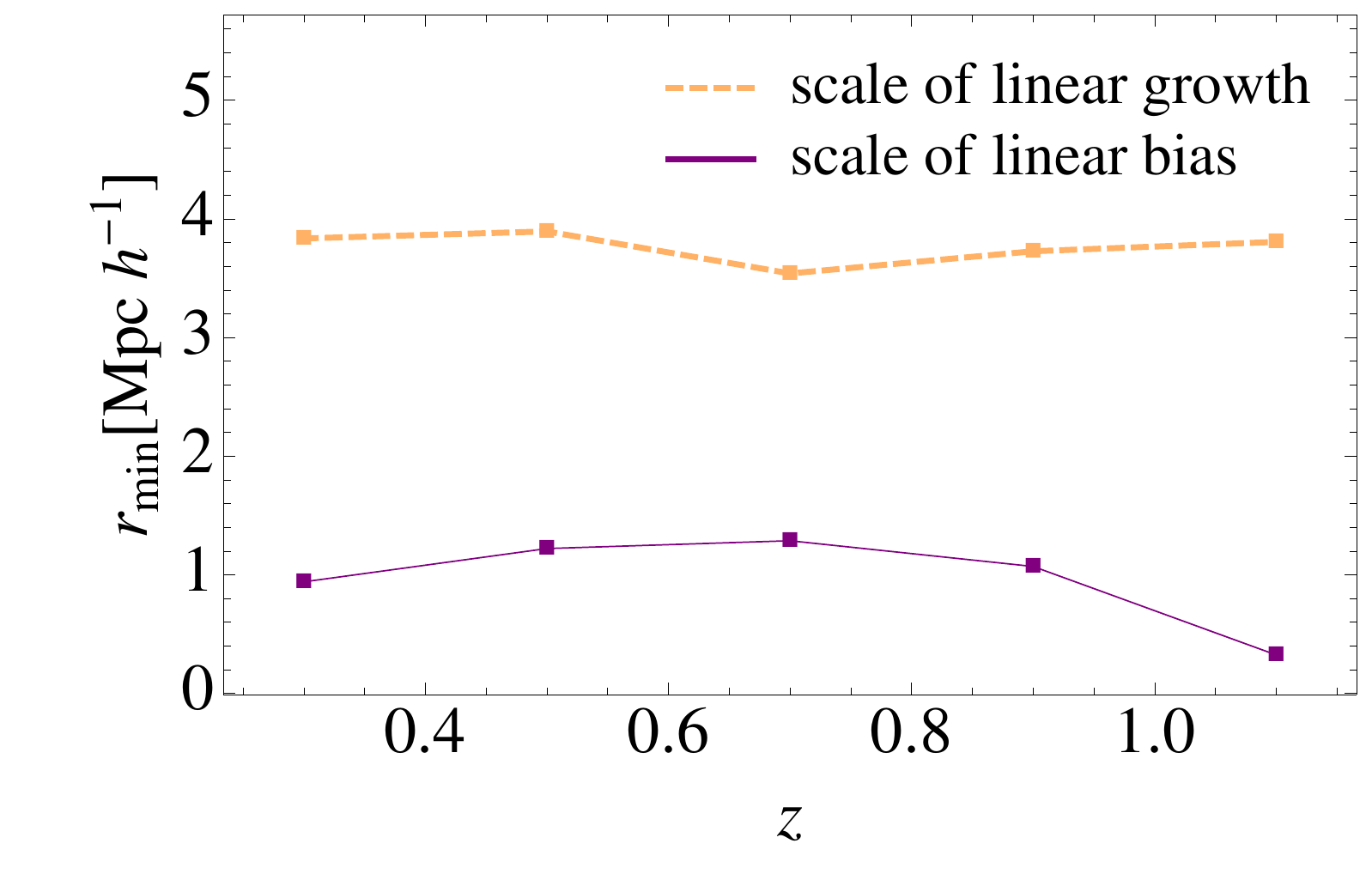}\caption{Top Panel: the evolution of the scale down to which the linear bias model (with non-linear dark matter) reproduces our angular clustering measurements, compared to the evolution of the linear bias + linear dark matter clustering. The former is valid to considerably smaller scales. Bottom Panel: Same information but translated into co-moving distances at the given redshift.}
\label{fig:bias_vs_nonlinscale} 
\end{center}
\end{figure}

The results of the $\chi^2$ test are shown in Fig.~\ref{fig:bias_chi2} for all $z$ -bins. We find $\theta_{\rm Linear\,Bias} \sim (0.06, 0.05$, $0.04, 0.037, 0.007)$ deg. for bins one to five.
We consider the accuracy at which we detect the breakdown of the ``linear bias'' scheme as half\footnote{Because $w(\theta) \propto b^2$} the precision in the measurement of the correlation function at $\theta_{\rm Linear\,Bias}$. For our current SV sample this is $\sim 2.5\%$ for the central photo-$z$ bins, and $\sim 4\%$ for bins one and five.

In Fig.~\ref{fig:bias_vs_nonlinscale} (top panel) we plot the resulting $\theta_{\rm Linear\,Bias}$ as a function of redshift (solid purple curve) and compare it to $\theta_{\rm Linear\,Growth}$ (dashed orange curve). The scale where the ``linear bias'' model fails is approximately three times smaller than the scale where non-linear growth becomes important,
except for the $1.0 < z < 1.2$ bin, where we find a factor of $10$.
The bottom panel of Fig.~\ref{fig:bias_vs_nonlinscale} shows the same information but translated to co-moving separations at the centre of each redshift bin, assuming our fiducial cosmological model. The scale of linear growth is roughly constant with redshift, $\sim 4\Mpc$, while the scale of linear bias is $\sim 1\Mpc$ (except in the last bin). 

\begin{figure}
\begin{center}
\includegraphics[trim= 0 1.2cm 0 0, width=0.45\textwidth]{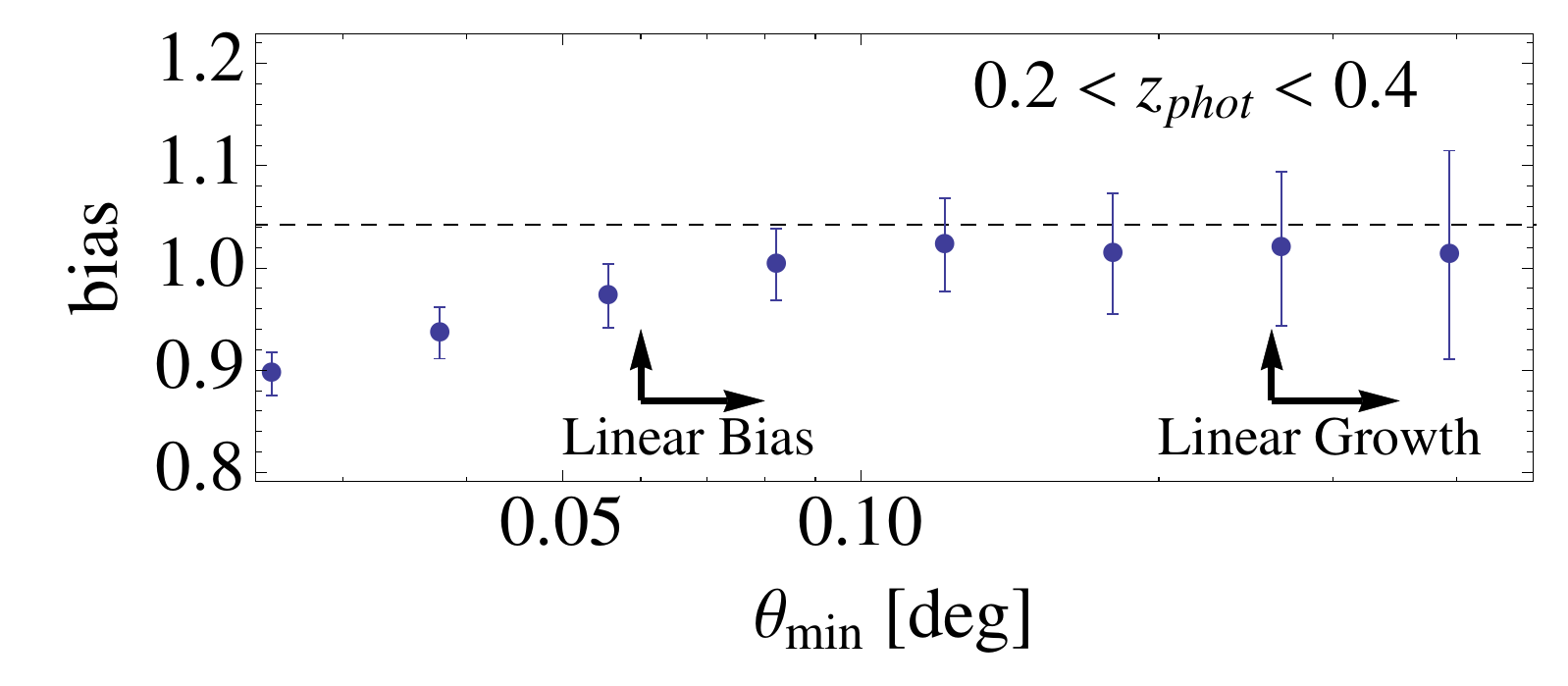} 
\includegraphics[trim= 0 1.2cm 0 0, width=0.45\textwidth]{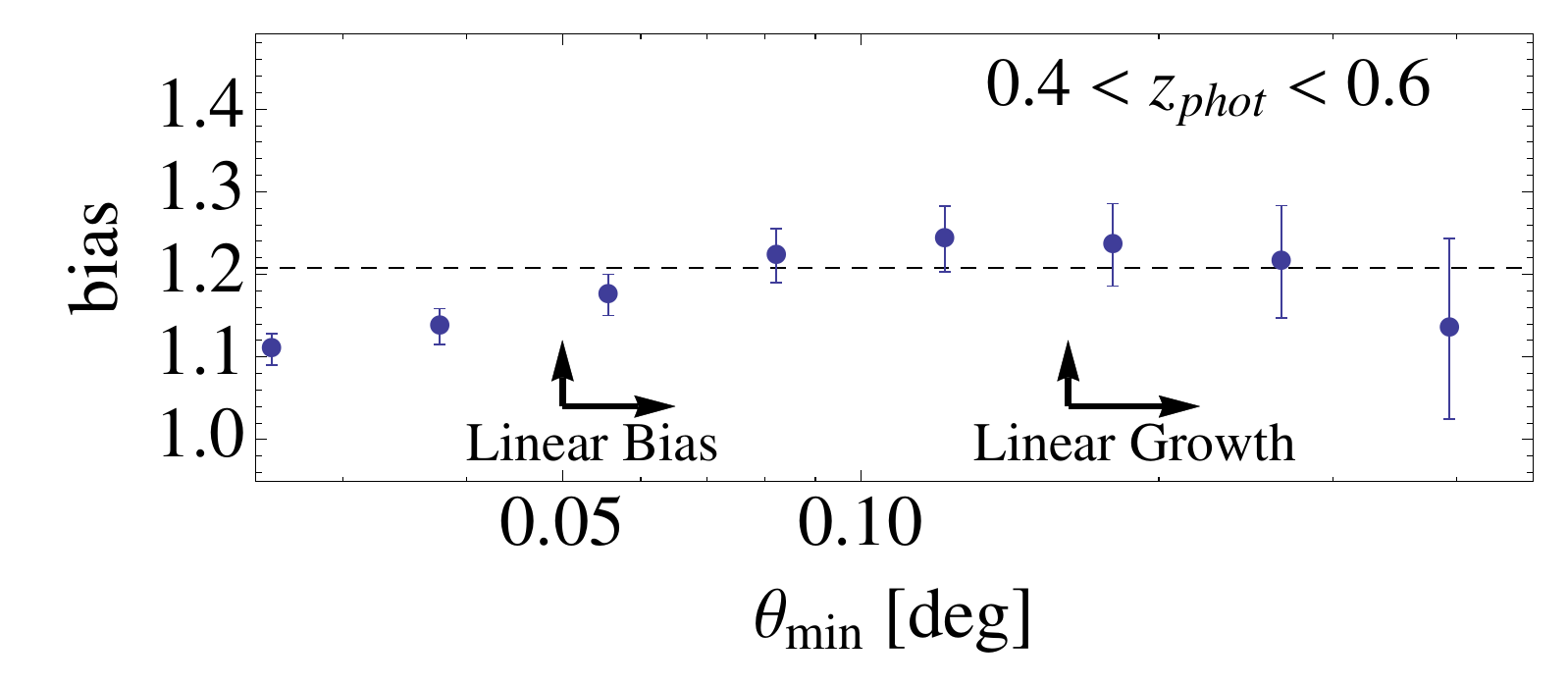} 
\includegraphics[trim= 0 1.2cm 0 0, width=0.45\textwidth]{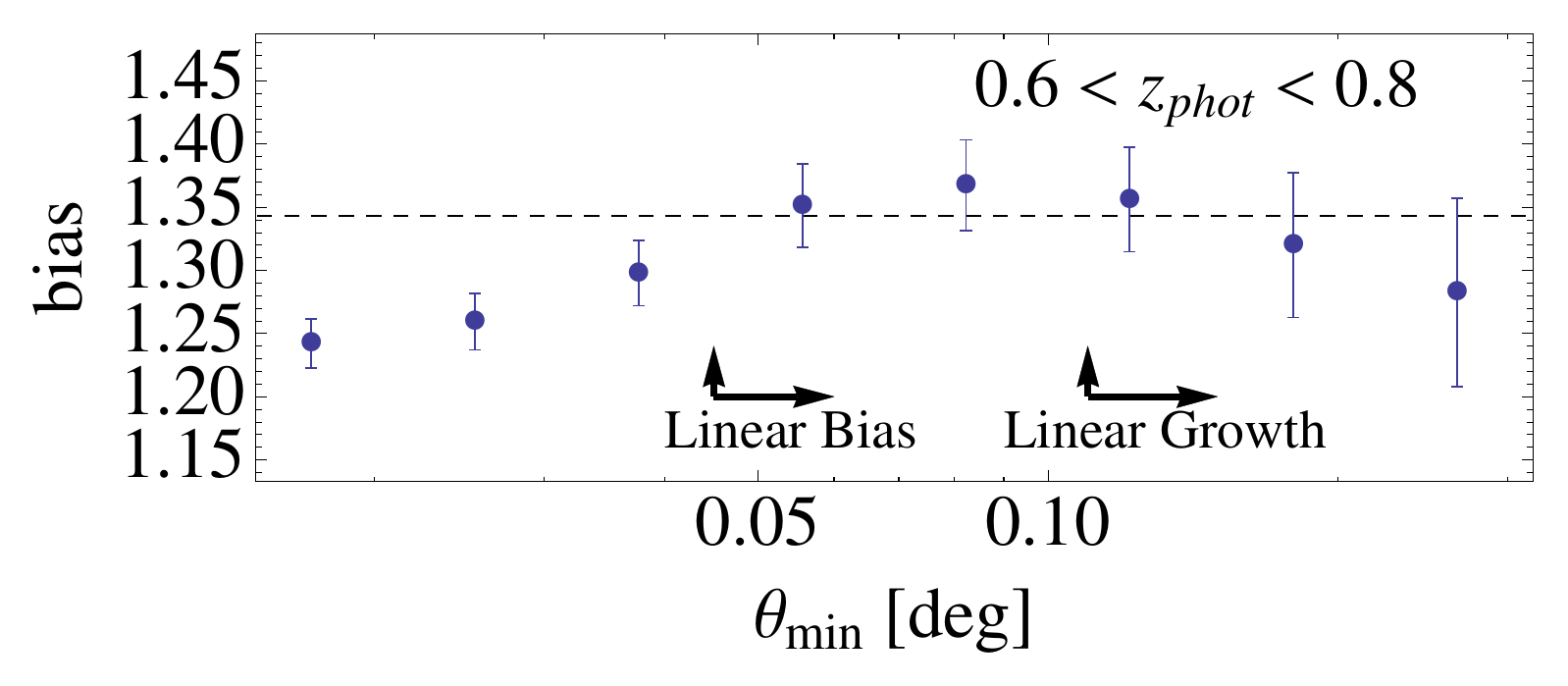} 
\includegraphics[trim= 0 1.2cm 0 0, width=0.45\textwidth]{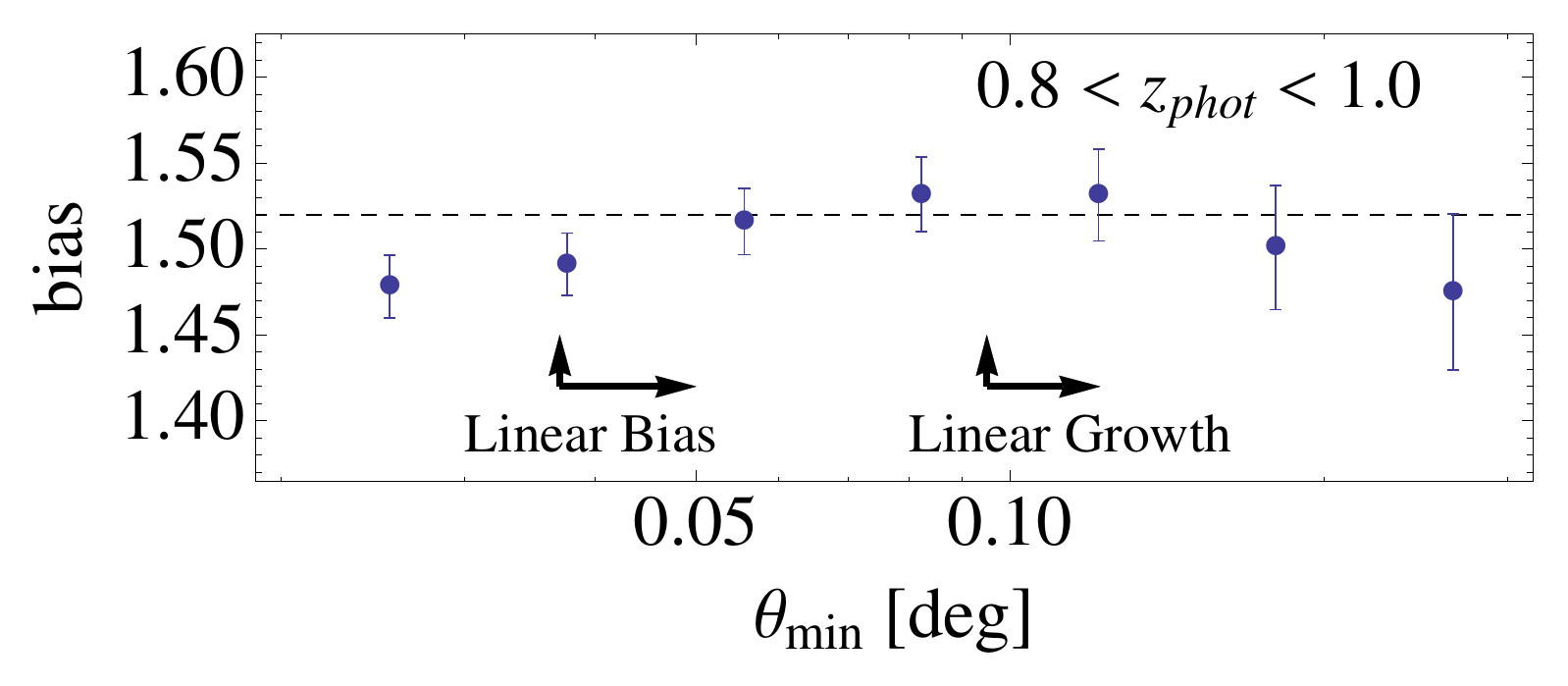}
\includegraphics[trim= 0 0cm   0 0, width=0.45\textwidth]{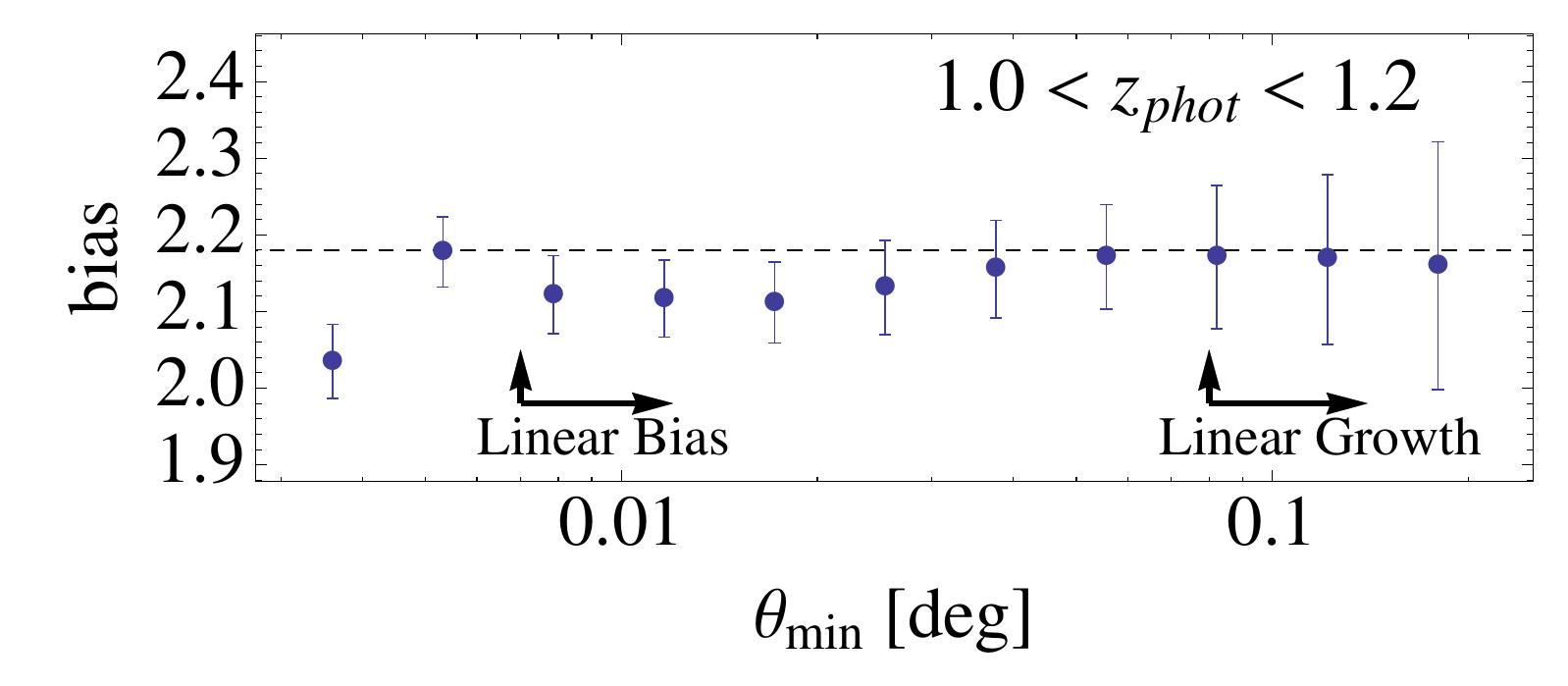}  
\caption{Best fit bias as a function of the minimum angular scale included in the fit. The inset label ``Linear Growth'' indicates the range of scales where a model with linear bias and linear matter clustering (hence ``linear growth'')  applies in our data. The ``Linear Bias'' label corresponds to a model using nonlinear matter clustering ({\tt Halofit} by Takahashi et al. (2012)) but keeping the bias linear.}
\label{fig:bias_thetamin} 
\end{center}
\end{figure}

In Fig~\ref{fig:bias_thetamin} we present, as a cross-check, the resulting best-fit bias as a function of the $\theta_{min}$ tested when obtaining the $\chi^2$ test results.
The dashed line in each panel shows the ``linear growth'' bias from Tables and 1 and 2. As we extend the fit down to smaller $\theta_{\rm min}$ we recover smaller error-bars, as expected. Further, the measured bias values remain consistent with the dashed line, until approximately the scale of Linear Bias determined using the $\chi^2$ test. We find that in all redshift bins, the measured bias begins to be clearly under-estimated compared to the asymptotic large-scale value (this would be consistent with a negative 2nd-order bias parameter in the local Eulerian bias model) at close to the $\theta_{min}$ determined using the $\chi^2$ test.


These results are interesting because they imply that forthcoming DES studies might be able to incorporate higher signal-to-noise data into cosmological analyses that use the full shape of the correlation function information with a simple bias model. We do stress that a more thorough analysis would involve allowing for a more complex model with non-linear and/or non-local bias terms and then investigate how the best-fit value for the linear order term in such scheme compares to the linear model results we present. This is however beyond the scope of the present paper and we postpone it to future work.

%% file: conclusions.tex
\section{Conclusions}
\label{sec:conclusions}

We present the first large-scale clustering analysis of a galaxy sample selected from the Science Verification Data of the Dark Energy Survey. The sample is selected as an apparent magnitude limited sample 
$i < 22.5$, with no color selection except for very conservative cuts to remove color outliers. The sample extends from $z_{\rm phot} =0.2$ to $z_{\rm phot}=1.2$. We performed our analysis in five tomographic bins of width $\Delta z_{\rm phot} = 0.2$. 

This paper has three main foci:

\begin{enumerate}

\itemindent1cm\item{We perform a detailed analysis and amelioration of potential observational systematic effects. We use a set of maps for different variables that can modulate galaxy detection efficiency, which include varying observing conditions, stellar density, and Galactic dust. We analyze which of them are affecting our sample by means of angular cross correlations and a measurement of galaxy density as a function of the value of the potential systematic variable across the footprint. We then apply cuts to minimize these systematic correlations,
with a small loss of statistical power. We then show how these two approaches complement and validate each other. These methods can be widely applied to future clustering studies, including our detailed accounting of the effects of stellar contamination.}
\newline

\itemindent1cm\item{We evaluate the clustering in the five redshift bins comparing it against simple linear theory models. We find good agreement with these models and between the results we obtain from two separate photo-$z$ methods (one template based, the other machine learning) of determining DES photometric redshifts. Both sets of measurements are also consistent with bias measurements, determined at the same redshifts for the same $i< 22.5$ flux limit, obtained by the CHFTLS survey. We measure the cross-correlation between these redshift bins as a test of the photo-$z$ reconstruction, finding the cross-correlation matching predictions from the autocorrelation signal at the expected statistical level.}
\newline

\itemindent1cm\item{We explore the regime of validity of our simple models that involve either linear / non-linear dark matter clustering, and a linear bias term. We find that, in angular clustering, the scale at which a linear bias model is not consistent with our data is considerably smaller than the scale at which non-linear growth in the clustering of dark matter (as predicted by {\tt Halofit} 2012) becomes important compared to the linear theory predictions. These results are relevant for probes that aim to combine weak lensing with large scale structure, as they suggest a linear bias model could be sufficient to scales as small as 1$h^{-1}$Mpc. }
\end{enumerate}

The data we analyze is approximately 1/30th the size of the of final DES data set\footnote{The SV SPTE region of observations is 150 deg$^{2}$ and the final DES footprint will be 5000 deg$^{2}$}. In the near future the methods presented here will be used and extended to larger DES samples suitable to measure the Baryon Acoustic Oscillation scale, constrain the shape of the matter power spectrum and measure the growth of structure. As part of this future analysis, we will use additional tools to robustly assess statistical and systematic uncertainties. We will make greater use of the {\tt Balrog} tool presented in \cite{balrog}, which will enable both a larger data sample and a more comprehensive assessment of systematic variations. Most importantly, we will use detailed mock catalogues and simulated data to provide robust and accurate covariance matrix estimates. These simulations will also be used to assess the systematic uncertainty introduced by the various systematic mitigation techniques and uncertainties in the photo-$z$s. Further, we will consider the impact of time-dependent variations in the data quality comprising the co-add. The work presented in this paper will help inform the creation of these simulations and we will continue to refine and apply the tools we have presented throughout this paper. In this sense, the methods and results we present represents a first step towards the goal of constraining cosmology with DES LSS measurements.

%% file: acknowledgments.tex
\section*{Acknowledgments}

We are grateful for the extraordinary contributions of our CTIO colleagues and the DECam Construction, Commissioning and Science Verification
teams in achieving the excellent instrument and telescope conditions that have made this work possible.  The success of this project also 
relies critically on the expertise and dedication of the DES Data Management group.

Funding for the DES Projects has been provided by the U.S. Department of Energy, the U.S. National Science Foundation, the Ministry of Science and Education of Spain, 
the Science and Technology Facilities Council of the United Kingdom, the Higher Education Funding Council for England, the National Center for Supercomputing 
Applications at the University of Illinois at Urbana-Champaign, the Kavli Institute of Cosmological Physics at the University of Chicago, 
the Center for Cosmology and Astro-Particle Physics at the Ohio State University,
the Mitchell Institute for Fundamental Physics and Astronomy at Texas A\&M University, Financiadora de Estudos e Projetos, 
Funda{\c c}{\~a}o Carlos Chagas Filho de Amparo {\`a} Pesquisa do Estado do Rio de Janeiro, Conselho Nacional de Desenvolvimento Cient{\'i}fico e Tecnol{\'o}gico and 
the Minist{\'e}rio da Ci{\^e}ncia, Tecnologia e Inova{\c c}{\~a}o, the Deutsche Forschungsgemeinschaft and the Collaborating Institutions in the Dark Energy Survey. 
The DES data management system is supported by the National Science Foundation under Grant Number AST-1138766.

The Collaborating Institutions are Argonne National Laboratory, the University of California at Santa Cruz, the University of Cambridge, Centro de Investigaciones En{\'e}rgeticas, 
Medioambientales y Tecnol{\'o}gicas-Madrid, the University of Chicago, University College London, the DES-Brazil Consortium, the University of Edinburgh, 
the Eidgen{\"o}ssische Technische Hochschule (ETH) Z{\"u}rich, 
Fermi National Accelerator Laboratory, the University of Illinois at Urbana-Champaign, the Institut de Ci{\`e}ncies de l'Espai (IEEC/CSIC), 
the Institut de F{\'i}sica d'Altes Energies, Lawrence Berkeley National Laboratory, the Ludwig-Maximilians Universit{\"a}t M{\"u}nchen and the associated Excellence Cluster Universe, 
the University of Michigan, the National Optical Astronomy Observatory, the University of Nottingham, The Ohio State University, the University of Pennsylvania, the University of Portsmouth, 
SLAC National Accelerator Laboratory, Stanford University, the University of Sussex, and Texas A\&M University.

The DES participants from Spanish institutions are partially supported by MINECO under grants AYA2012-39559, ESP2013-48274, FPA2013-47986, and Centro de Excelencia Severo Ochoa SEV-2012-0234.
Research leading to these results has received funding from the European Research Council under the European Union Seventh Framework Programme (FP7/2007-2013) including ERC grant agreements 
 240672, 291329, and 306478. MC has been partially funded by AYA2013-44327. FS acknowledges financial support provided by CAPES under contract No. 3171-13-2. We thank Jean Coupon and Martin Kilbinger for useful discussions and help at different stages of this work.

This paper has gone through internal review by the DES collaboration.
The DES publication number for this article is DES-2015-0055. The Fermilab pre-print number is FERMILAB-PUB-15-305.

%% file: appendix.tex
\subsection{Testing the bias evolution against MICE bench-sample}
\label{sec:appendix_a}

Here we test the evolution of bias for our flux-limited sample using the MICECATv2.0, simulation 
which is an updated version from
the MICECATv1.0 catalog (see \cite{2015MNRAS.448.2987F}, \cite{2013arXiv1312.2013C}, \cite{2015MNRAS.447.1319F}, \cite{2015MNRAS.447..646C}) 
extended to lower mass halos and therefore less luminous galaxies.
This version is complete to $i \simeq 24$ at all redshifts. 
Magnitudes are assigned to galaxies following a combination of Halo Occupation Distribution (HOD) and 
subhalo abundance matching (SHAM) prescriptions tuned to
fit the local ($z \sim 0$) galaxy luminosity function and clustering as a function of luminosity
and color (see  \cite{2015MNRAS.447..646C}). 
To account for evolution in galaxy luminosity on top of the one induced by halo mass evolution,
 we transform the galaxy magnitudes using:
\begin{equation}
i_{evol} = i -0.8[atan(1.5z) - 0.1489],
\end{equation}
which gives 
similar counts to those in COSMOS and other
galaxy surveys (Castander et al, in prep.). 
We then restrict the sample to $i_{evol}<22.5$
to compare to the DES SV bench-mark sample.
\begin{figure}
\centering
\includegraphics[width=0.45\textwidth]{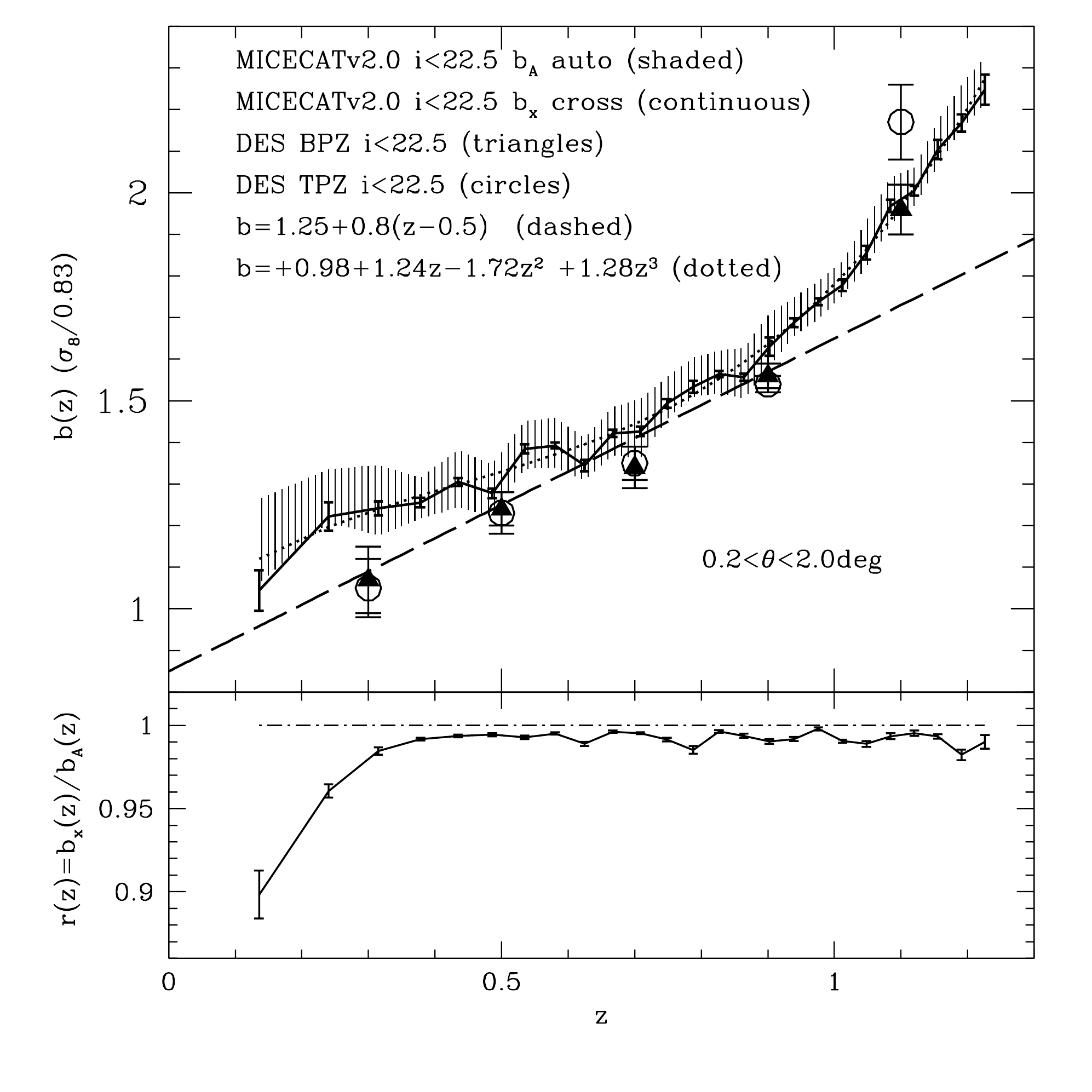}
\caption{Top panel shows the evolution of galaxy bias with redshift in a flux limited sample ($i<22.5$) similar to the one defined in this paper but selected from our MICECATv2.0 mock galaxy catalog. We display bias derived from galaxy auto-correlations $b_A=\sqrt{w_{gg}/w_{mm}}$ and galaxy-matter cross correlation $b_X=w_{gm}/w_{mm}$. The evolution is similar to the one in the DES-SV data. Bottom panel shows the ratio of the two estimates, i.e. the stochasticity coefficient $r=b_A/b_X$, which we found close to
 unity in the redshift range relevant for DES data.}
\label{fig:bz_mice}
\end{figure}

The top panel of Fig.~\ref{fig:bz_mice} shows a comparison of the linear bias measured in
MICECATv2.0 against the one in our BPZ and TPZ samples shown in Fig.~\ref{fig:bias_vs_CFHTLS}.
In the case of MICECATv2.0 we estimate the bias  as $b_A=\sqrt{w_{gg}/w_{mm}}$, where $w_{gg}$
is the measured galaxy angular correlation function and $w_{mm}$
the corresponding measured matter-matter correlation in the same redshift bin. This is done using true redshifts to avoid redshift space distortions, which can have a large effect when using narrow redshift bins as in Fig.~\ref{fig:bz_mice} \citep{2014arXiv1412.2208E} but are negligible (given scales and errors) for $\Delta z = 0.2$
bins used in this paper \citep{2011MNRAS.414..329C,2011MNRAS.415.2193R}. The MICECATv2.0 bias is fitted from the ratio in the range $0.2<\theta<2.0$ deg. The solid  line is the bias measured from cross-correlating galaxies and mass directly, $b_X=w_{gm}/w_{mm}$. The shaded region tries to reproduce the expected sampling and shot-noise errors for DES SV in $b_A$.
It corresponds to  $\Delta b_A=(\Delta w_{gg}/w_{mm})/(2 b_A)$, where matter is kept fixed and
the error in the galaxy-galaxy correlation $\Delta w_{gg}$ is scaled to 116 deg$^2$ 
and $\Delta z \simeq 0.2$,  from the scatter 
in 24 JK regions over a larger 30x30 deg$^2$ simulation patch. The smaller error bars
include sample variance cancelation between matter and galaxy fluctuations (i.e. the scatter in $b_X$ from
different regions). The dotted line shows a cubic fit in simulations to $b_X$ and errors.
The evolution with redshift is very similar to the one in the LSS bench-mark sample, in particular the fast rise at high redshift. At $z <0.5$ DES-SV data seems to be $1\sigma$ lower than simulations. Larger areas are needed 
to decide if this is a significant discrepancy.
The bottom panel of Fig.~\ref{fig:bz_mice} shows the stochasticity coefficient:
$r \equiv w_{gm}/\sqrt{w_{gg} w_{mm}} = b_X/b_A$. The error bars are scaled from the dispersion in $r$ from JK regions: 
thus, they include sample variance cancelation and correspond to the 30x30 deg. patch used 
for this test. There seems to be some significant detection of $r$ at the smallest redshifts, but this 
is most likely  a reflection of the poor performance of JK errors for the smallest volumes. 
At $z>0.3$ we find $r$ to be in the range $r \simeq  0.98-1.00$,
which is consistent with unity when errors are scaled to the SV area.
 This is relevant for analyses using probes that cross correlate the bench-mark galaxy sample defined here with other data sets such as CMB lensing \citep{GiannantonioCMBLens} or weak lensing maps.

Note that the bias is displayed in units of $(\sigma_8/0.83)$ to account for the difference in the MICE value of $\sigma_8=0.80$ and the one used in this paper ($\sigma=0.83$).
There is quite a good agreement in the values of $b(z)$ without need of any other adjustment. This 
illustrates the fact that the steep rise of bias at $z>1$ is in fact expected 
using simple HOD and luminosity evolution models. 

\subsection{Statistical error estimation}
\label{sec:appendix_b}

In Sec.~\ref{sec:covariance} we introduced our approach to estimate the covariance matrix between angular clustering measurements, see Eq.~(\ref{eq:mixcov}).
This combined a theory modeling for the off-diagonal elements with a Jackknife (JK) approach for the variance. The JK approach is intrinsically limited by the fact that typically (given the size of our footprint and data) is not possible to extract many statistically independent samples to construct a reliable full covariance. However it does account better than theory estimations for issues such as small scale mask effects, boundaries, non-linerities and/or contamination or systematics in the data. On the other hand, the theory estimation yields off diagonal elements that have no noise compared to their JK counter-parts. In this appendix we explore this choice in more detail.

\begin{figure}
\centering
\includegraphics[trim=0 1.3cm 0 0, width=0.4\textwidth]{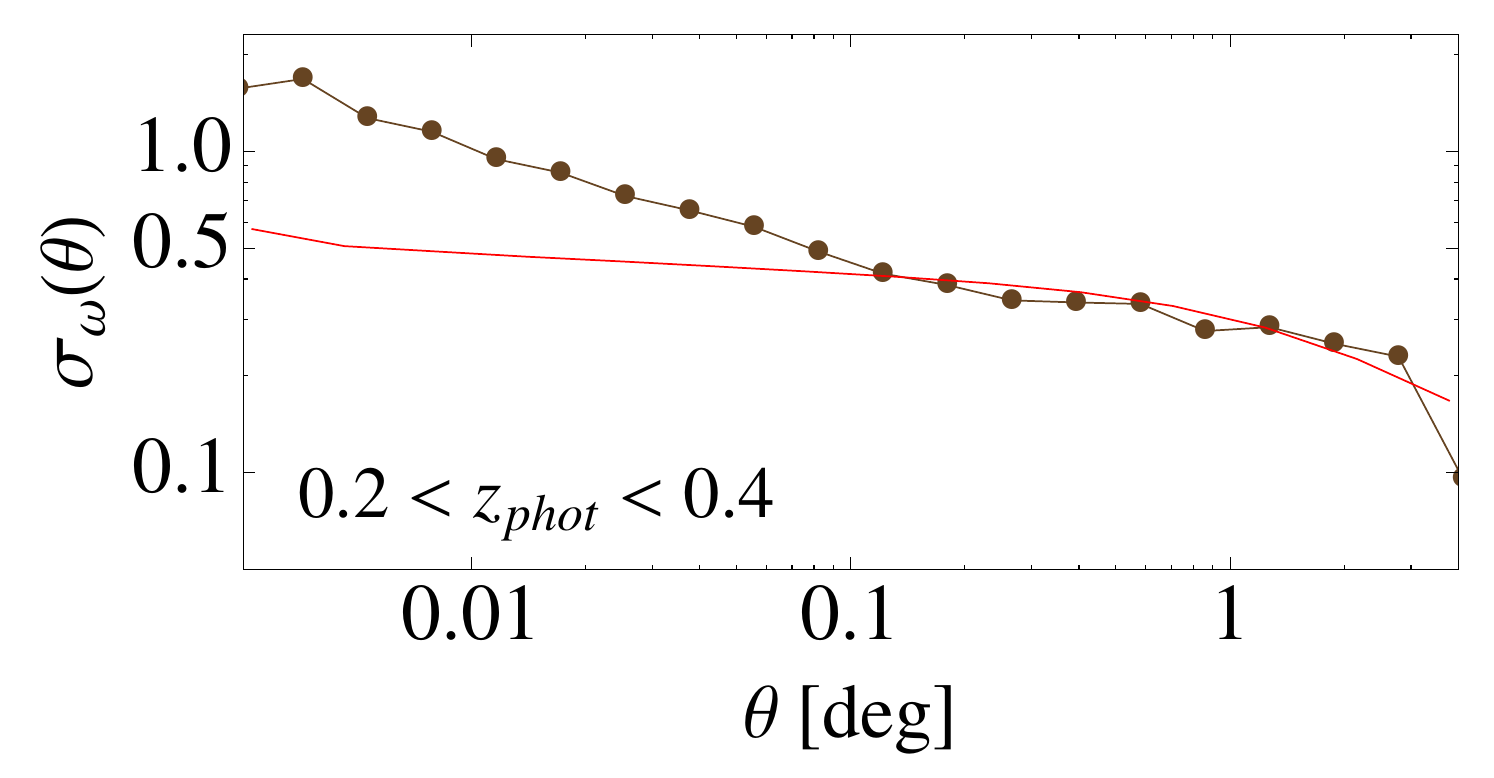}
\includegraphics[trim=0 1.3cm 0 0, width=0.4\textwidth]{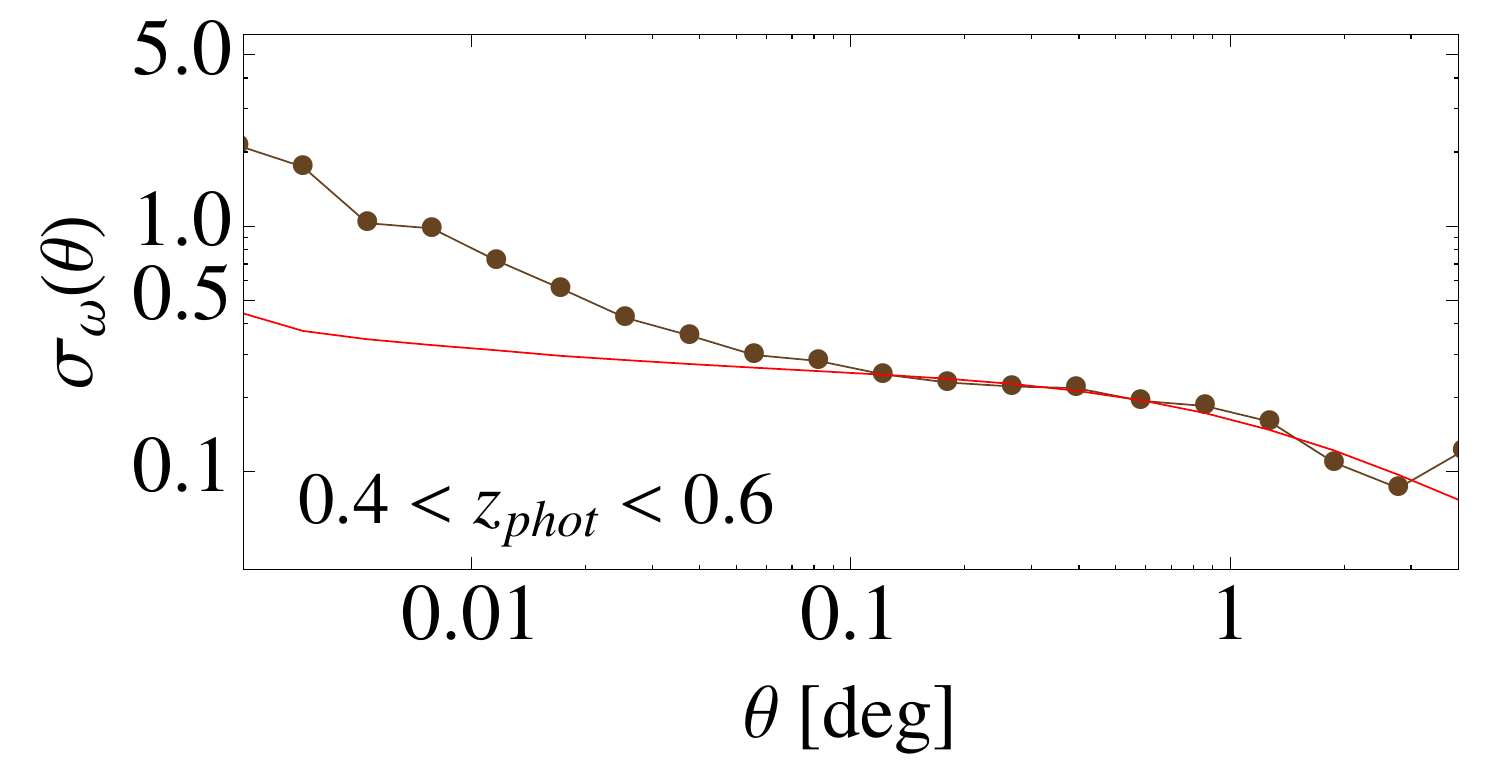}
\includegraphics[trim=0 1.3cm 0 0, width=0.4\textwidth]{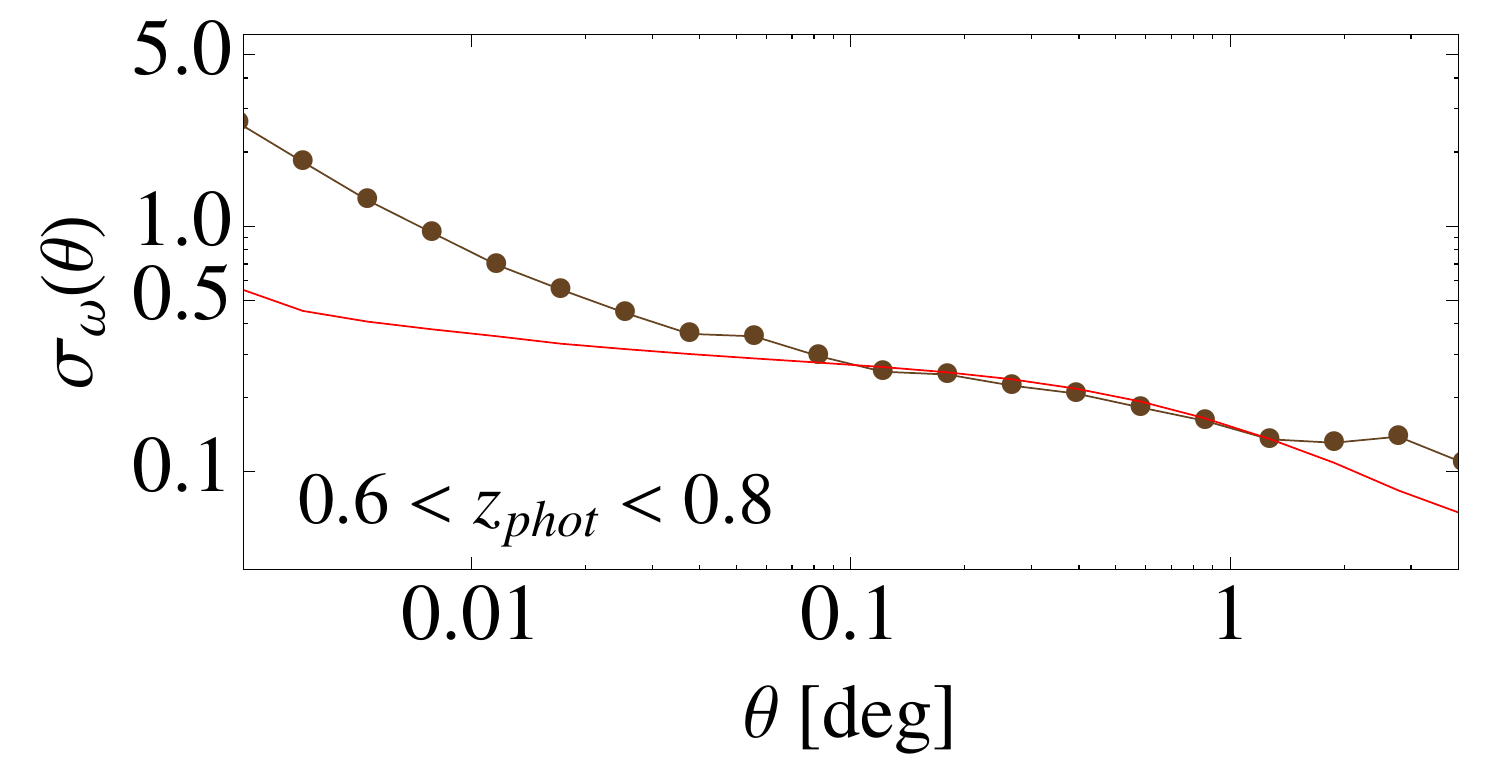}
\includegraphics[trim=0 1.3cm 0 0, width=0.4\textwidth]{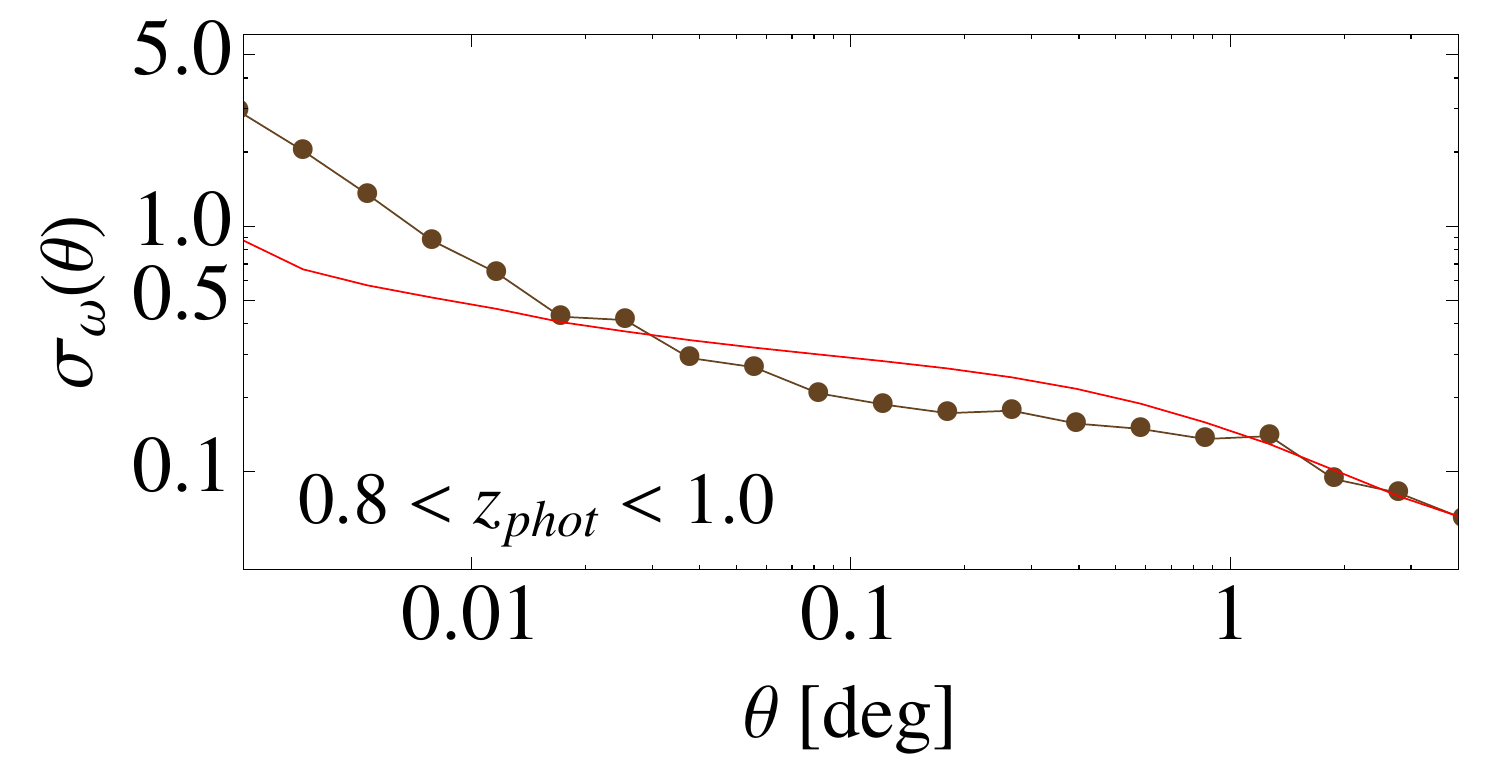}
\includegraphics[width=0.4\textwidth]{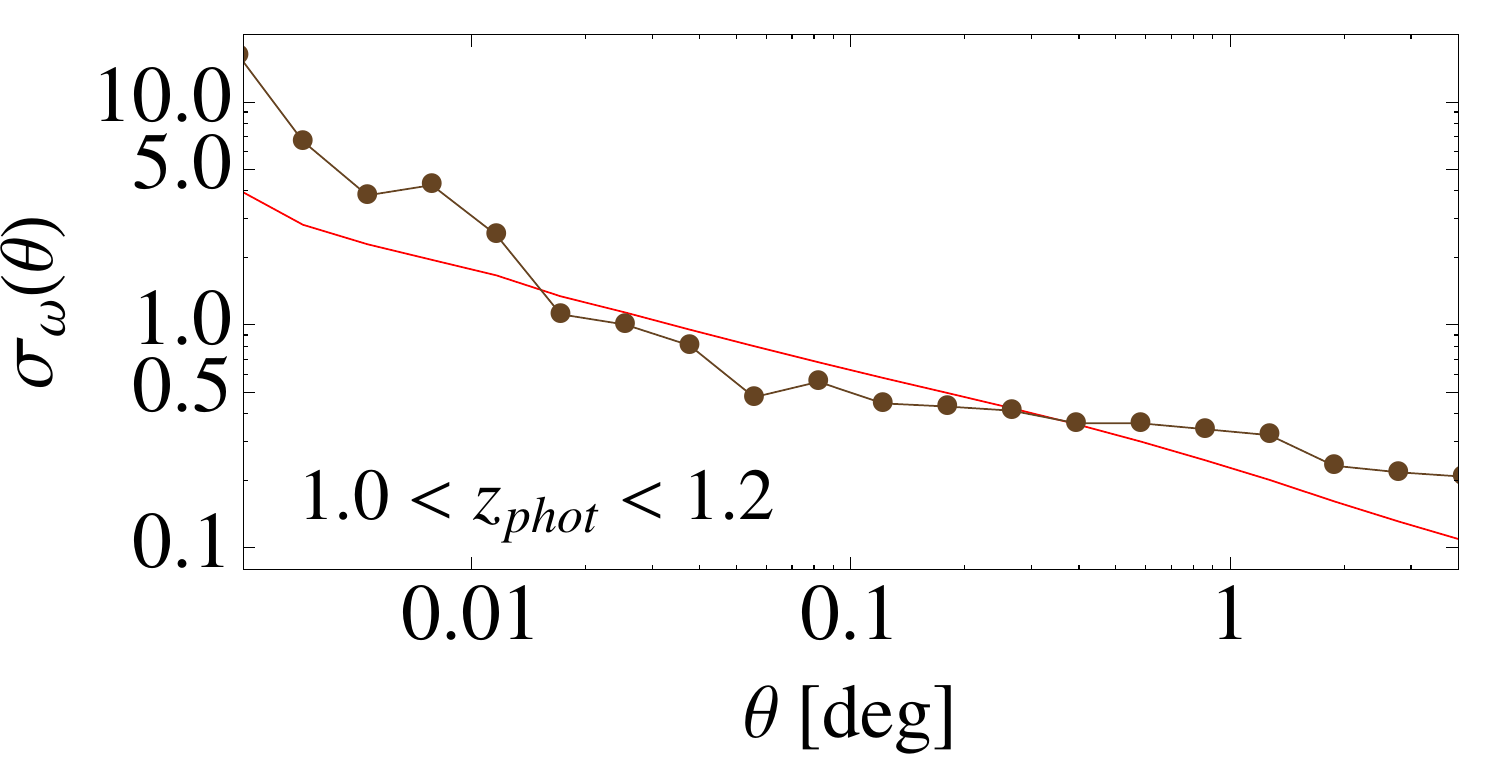}
\caption{Comparison of Jackknife resampling errors (points) and theoretically derived ones (lines) for our five tomographic bins (BPZ photo-$z$). For the former we used 40 equal size regions. The theory estimation uses the best-fit bias, the projected number density and the area of the footprint (and its based on the nonlinear matter power spectra). The agreement is very good on scales ($\theta \gtrsim 1\,{\rm deg}$ at low-$z$ and $\theta \gtrsim 0.1\,{\rm deg}$ at high-$z$), while on smaller scales the structure of the data as captured by the JK resampling yields larger variance.}
\label{fig:jk_vs_theory}
\end{figure}

A first test is to compare the error estimates on $w(\theta)$ measurements (i.e. the variance) that we obtain using the JK approach with the theory ones.  For the theory estimate we used the best-fit bias over linear-scales, the exact footprint size and the exact shot-noise measured in the data (i.e. number density of objects in the photo-$z$ bin after masking). The JK one employs $40$ equal size JK regions, and we tested that the results that we now present do not depend on variations in this set-up (we have explored $n_{JK}$ from 20 to 100). In Fig.~\ref{fig:jk_vs_theory} we show how these two estimates compare to each other. On large-scales ($\theta \gtrsim 0.1$ deg at low $z$, and $\theta \gtrsim 0.01$ deg at high $z$) they match each other well, except for the $0.8 < z < 1.0$ where the theory yields $\sim 20\%-25\%$ higher errors in the range $0.1\,{\rm deg}\,< \theta < 1\,{\rm deg}\,$. 

A second test is to derive the best-fit bias using three possible methods for constructing the covariance matrix, a full theory approach,
the JK method, or the combined one that we use in Sec.~\ref{sec:covariance}. This is shown in Table \ref{tab:cov_bias}. The best-fit biases are not affected by the choice of covariance, while the 1$\sigma$ errors vary, but not so much as to affect any of our conclusions. At smaller scales, the JK uncertainties are significantly larger than the theoretical ones; this is likely due to both non-linear and small-scale mask effects.

\begin{table}
\centering
\begin{tabular}{ |c|c|c|c| } 
 \hline
 Photo-$z$ Bin & Jack-knife Cov. & Theory Cov. & Mix Cov. \\
 \hline
 $0.2 < z < 0.4$ & $1.05 \pm 0.07$ & $1.04 \pm 0.08$ & $1.05 \pm 0.07$ \\ 
 $0.4 < z < 0.6$ & $1.22 \pm 0.04$ & $1.23 \pm 0.06$ & $1.23 \pm 0.05$ \\ 
 $0.6 < z < 0.8$ & $1.37 \pm 0.03$ & $1.36 \pm 0.04$ & $1.36 \pm 0.04$ \\
 $0.8 < z < 1.0$ & $1.55 \pm 0.02$ & $1.54 \pm 0.04$ & $1.54 \pm 0.02$ \\ 
 $1.0 < z < 1.2$ & $2.16 \pm 0.12$ & $2.16 \pm 0.12$ & $2.17 \pm 0.09$ \\ 
 \hline
\end{tabular}
\caption{Dependence of best-fit bias on different methods to estimate the covariance matrix (see text for details). The bias is
obtained from fitting the auto-correlations in Fig.~\ref{fig:bias_vs_CFHTLS} over a  linear range of scales 
This paper uses the mixed approach for its main results.}
\label{tab:cov_bias}
\end{table}

\subsection{Galaxy Bias in CFHTLS}
\label{sec:appendix_c}

Galaxies in \cite{2012A&A...542A...5C} are first selected according to \texttt{SExtractor} \magauto magnitudes of $i <22.5$, similar to the sample  presented in this paper. Their flux limited sample was then split in several volume limited absolute magnitude "threshold" samples in each photo-$z$ bin. Therefore in each redshift bin, our sample corresponds to the faintest absolute magnitude sample in \cite{2012A&A...542A...5C}, except for the fact that our sample is not volume limited. 

For example over the redshift range $0.4 < z < 0.6$ the faintest volume limited sample is defined as brighter than 
$M_r-5 \log h=-18.8$ because this is the absolute magnitude corresponding to apparent magnitude $i < 22.5$ for the farthest galaxies at $z=0.6$. At $z=0.4$ the absolute magnitude corresponding to $i<22.5$ is one magnitude fainter, $M_r-5 \log h=-17.8$. Therefore as one spans the whole redshift range of the sample from $z=0.4$ to $z=0.6$ there is a set of galaxies brighter than $M_r-5 \log h=-17.8$ but fainter than $M_r-5 \log h=-18.8$, that do not enter the volume limited sample in $0.4 < z < 0.6$ but do enter the flux limited one. And this holds for all photo-$z$ bins: there is always one magnitude difference between the faintest absolute magnitude from the lowest to the highest redshift of each photo-$z$ bin \citep{2012A&A...542A...5C}.

To directly compare our bias measurements to those in \cite{2012A&A...542A...5C} we need to know how bias depends on luminosity to then estimate the contribution to the bias of this ``extra'' set of galaxies, as
\beq
b_{\rm flux\,lim, \,bin \,N} = \int_0^1 b_{\rm vol\,lim}(<M_r + x) \,dx
\label{eq:bias_translation}
\eeq
where $M_r$ is the magnitude threshold defining the faintest sample for the photo-$z$ bin N. Here $x=1$ would correspond to the lowest redshift bound of such bin, while $x=0$ the highest.

At low redshift the difference discussed above does not impact the derived bias strongly because these galaxies are faint enough that have biases close to one and are in a regime of very weak dependence with luminosity. However as we move to higher redshifts, the faintest samples move to $M_r - 5 \log_{10} h \lesssim -21$ where bias evolves strongly with luminosity (e.g. see Fig. 18 in \cite{2012A&A...542A...5C}).

\begin{figure}
\begin{center}
\includegraphics[width=0.4\textwidth]{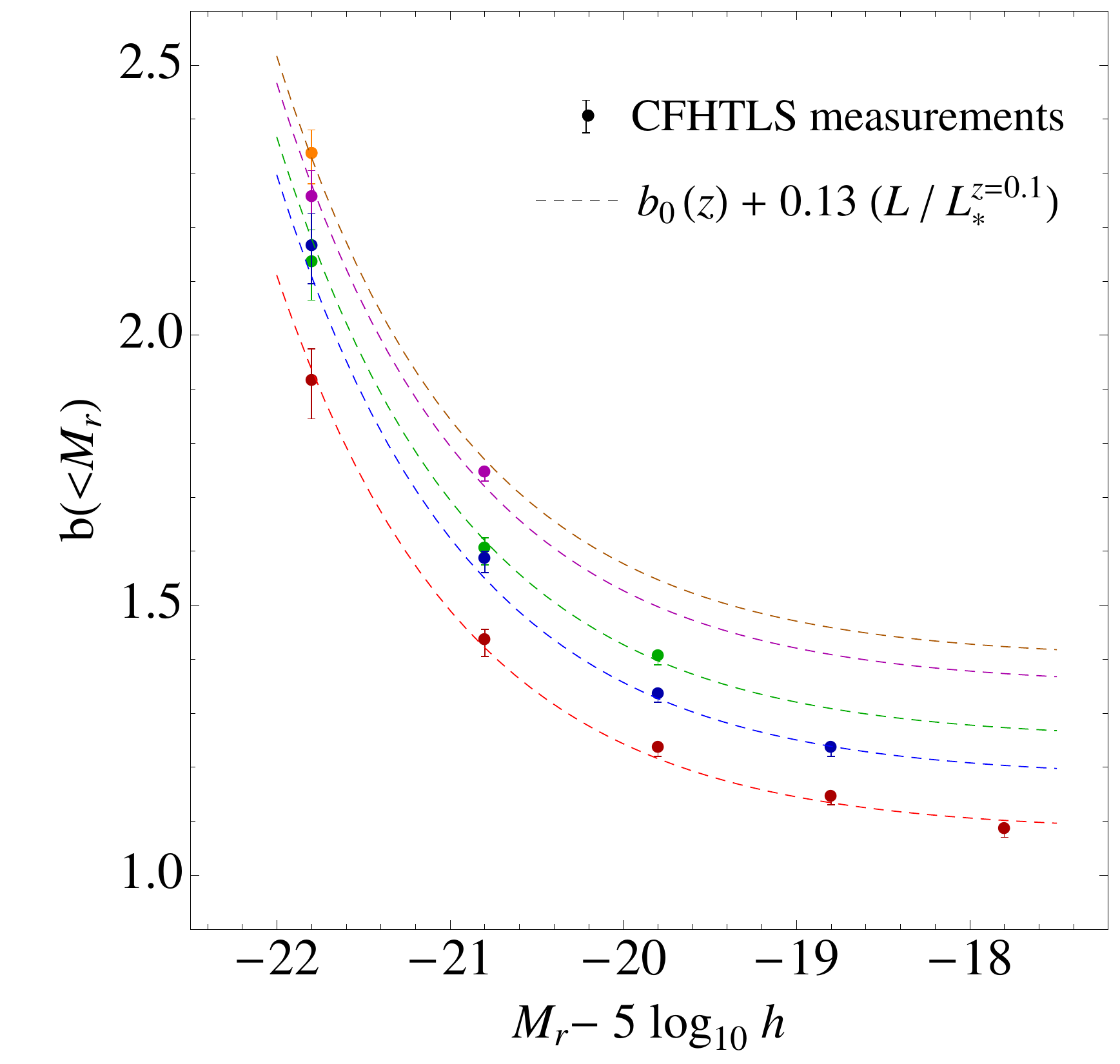} 
\caption{Fit to the bias as a function of luminosity presented by Coupon et al. (2012) for the 5 bins also used throughout our paper (starting from $0.2 < z < 0.4$ at the bottom). The evolution can be simply described as depending linearly with luminosity (see inset label).}
\label{fig:bias_mr} 
\end{center}
\end{figure}

To account for this effect so as to properly compare our bias results to those of CFHTLS we did as follows. We collected all the best-fit biases reported by \cite{2012A&A...542A...5C} in their Table  B.1. This is shown in Fig.~\ref{fig:bias_mr}. We have found that these measurements are very well fit by a simple linear relation with r-band luminosity $b(M_r)=b_0(z) + a_0 L_r$ with constant $a_0 = 0.13 / L_{\star}^{z=0.1}$. This result is in fact very similar to the one discussed by \cite{2012A&A...542A...5C}, although we do not normalize by the characteristic luminosity at the given redshift. We then fit $b_0$ at every photo-$z$ bin, including $1.0 < z < 1.2$ where only one data-point is given by \cite{2012A&A...542A...5C}. We find $b_0=(1.08, 1.18, 1.25, 1.35, 1.4)$ at photo-$z$ bins centered at $(0.3, 0.5, 0.7, 0.9, 1.1)$. These results are displayed in Fig.~\ref{fig:bias_mr}.

\begin{table}
\centering
\begin{tabular}{ |c|c|c|c| } 
 \hline
 Photo-$z$ Bin & $M_r-5 \log h$ & $b_{\rm vol\,lim}( < M_r)$ & $b_{\rm flux\,lim}(< M_r)$ \\
 \hline
 $0.2 < z < 0.4$ & -17.8 & $1.08 \pm 0.01$ & $1.09$ \\ 
 $0.4 < z < 0.6$ & -18.8 & $1.23 \pm 0.01$ & $1.22$ \\ 
 $0.6 < z < 0.8$ & -19.8 & $1.40 \pm 0.01$ & $1.35$ \\
 $0.8 < z < 1.0$ & -20.8 & $1.74 \pm 0.03$ & $1.59$ \\ 
 $1.0 < z < 1.2$ & -21.8 & $2.33^{+0.05}_{-0.06}$ & $2.00$ \\ 
 \hline
\end{tabular}
\caption{Translation of the biases defined by Coupon et al. (2012) where samples are defined in terms
of volume limited quantities to our case where we have a flux-limited sample across the full redshift range (the parent catalog in both cases is similar, $i-band\,\magauto<22.5$). Here $M_r$ is the absolute magnitude corresponding to the faintest sample in each photo-$z$ bin as defined by Coupon et al. (2012), $b_{\rm vol\,lim}$ are the bias values reported in their Table B.1., and $b_{\rm flux\,lim}$ the ones directly comparable to our results.}
\label{tab:bias_CFHTLS}
\end{table}

Once provided with a fit for $b(<M_r)$ we computed the bias conversion as in Eq.~(\ref{eq:bias_translation}). Results are detailed in Table \ref{tab:bias_CFHTLS}. The fourth column displays the values that we use in Sec.~\ref{sec:results} and Fig.~\ref{fig:bias_vs_CFHTLS}. Because the effect is not strong and the bias range not large, the results from an integration as in Eq.~(\ref{eq:bias_translation}) are equivalent to simply computing the bias for samples shifted half by a magnitude (i.e. $b_{\rm flux\,lim}(<-17.8+0.5)$ for $0.2 < z < 0.4$, then $b_{\rm flux\,lim}(<-18.8+0.5)$ for $0.4 < z < 0.6$ and so on). 

\subsection{Stellar Contamination}
\label{sec:appendix_stellarcontam}
Building from the results of, e.g., \cite{Myers06,Ross11,2012ApJ...761...14H}, we derive the effect that stellar contamination has on the measured clustering of a galaxy sample. We assume the observed number of galaxies in some location, $N_o$, is a function of the true number of galaxies, the number of stars, $N_{st}$, that could be mis-classified as a galaxy, and some function of the survey conditions, $S$. If the only source that modulates the observed number of galaxies is stellar contamination, then
\begin{equation}
N_o = N_{gal} + N_{st}F(S),
\end{equation}
where $N_{st}F(S)$ is the number of stars mis-classified as galaxies at a particular location. The fractional stellar contamination, $f_{st}$, over the footprint is
\begin{equation}
f_{st} = \frac{\langle N_{st}f(S) \rangle}{\langle N_o\rangle}
\end{equation}
and
\begin{equation}
\langle N_o\rangle = (1+f_{st})\langle N_{gal} \rangle.
\end{equation}
Given the over-density
\begin{equation}
\delta = N/\langle N \rangle -1,
\end{equation}
we have
\begin{equation}
\delta_o = \frac{1}{1+f_{st}}\frac{N_{gal}}{\langle N_{gal}\rangle}+\frac{1}{1+f_{st}}\frac{N_{st}F(S)}{\langle N_{gal}\rangle}-1,
\end{equation}
which can be re-written in terms of the galaxy over-density
\begin{equation}
\delta_o = \frac{\delta_{gal}-f_{st}}{1+f_{st}}+\frac{1}{1+f_{st}}\frac{N_{st}F(S)}{\langle N_{gal}\rangle}.
\end{equation}
and then the over-density of the stellar contamination, $\delta_{st,S} = \frac{N_{st}F(S)}{\langle N_{st}F(S)\rangle}-1$
\begin{equation}
\delta_o = \frac{\delta_{gal}-f_{st}}{1+f_{st}}+f_{st}(\delta_{st,S}+1) = \frac{\delta_{gal}}{1+f_{st}} +f_{st}\delta_{st,S}+\frac{f^2_{st}}{1+f_{st}},
\end{equation}
after recognizing that $\frac{\langle N_{st,S}F(S)\rangle}{(1+f_{st})\langle N_{gal}\rangle} = f_{st}$.

The measured angular correlation function, $w_o = \langle \delta_o\delta_o\rangle$, is
\begin{equation}
w_o = \frac{w_{gal}}{(1+f_{st})^2} +f_{st}^2w_{st,S}+\frac{f_{st}^4}{(1+f_{st})^2},
\end{equation}
(assuming no true correlation between galaxies and stars). Therefore
\begin{equation}
w_{gal} = (1+f_{st})^2\left(w_o-f^2_{st}w_{st,S}-\frac{f_{st}^4}{(1+f_{st})^2}\right)
\end{equation}

The above requires no approximations, but is limited to the specific case of contamination. One must be able to estimate $w_{st,S}$, but we expect this can be approximated as $w_{st}$ in many cases. Further, estimating the stellar contamination as a function of survey conditions can be done empirically and using forward-modeling tools (the potential exists when using similar tools as, e.g., \cite{LeistedtMAPS,balrog}).

%% file: affiliations.tex
\section*{Affiliations}
\textit{
$^{1}$Institut de Ci\`encies de l'Espai, IEEC-CSIC, Campus UAB, Carrer de Can Magrans, s/n,  08193 Bellaterra, Barcelona, Spain\\
$^{2}$Institut de F\'{\i}sica d'Altes Energies, Universitat Aut\`onoma de Barcelona, E-08193 Bellaterra, Barcelona, Spain\\
$^{3}$Center for Cosmology and Astro-Particle Physics, The Ohio State University, Columbus, OH 43210, USA\\
$^{4}$Centro de Investigaciones Energ\'eticas, Medioambientales y Tecnol\'ogicas (CIEMAT), Madrid, Spain\\
$^{5}$Department of Astronomy, University of Illinois, 1002 W. Green Street, Urbana, IL 61801, USA\\
$^{6}$Kavli Institute for Cosmology Cambridge and Institute of Astronomy, University of Cambridge, Madingley Road, Cambridge CB3 0HA, UK\\
$^{7}$Fermi National Accelerator Laboratory, P. O. Box 500, Batavia, IL 60510, USA\\
$^{8}$Laborat\'orio Interinstitucional de e-Astronomia - LIneA, Rua Gal. Jos\'e Cristino 77, Rio de Janeiro, RJ - 20921-400, Brazil\\
$^{9}$Department of Physics \& Astronomy, University College London, Gower Street, London, WC1E 6BT, UK\\
$^{10}$National Center for Supercomputing Applications, 1205 West Clark St., Urbana, IL 61801, USA\\
$^{11}$Observat\'orio Nacional, Rua Gal. Jos\'e Cristino 77, Rio de Janeiro, RJ - 20921-400, Brazil\\
$^{12}$Kavli Institute for Cosmological Physics, University of Chicago, Chicago, IL 60637, USA\\
$^{13}$Department of Physics, ETH Zurich, Wolfgang-Pauli-Strasse 16, CH-8093 Zurich, Switzerland\\
$^{14}$Lawrence Berkeley National Laboratory, 1 Cyclotron Road, Berkeley, CA 94720, USA\\
$^{15}$Institute of Cosmology \& Gravitation, University of Portsmouth, Portsmouth, PO1 3FX, UK\\
$^{16}$Instituto de F{\'i}sica Te{\'o}rica and ICTP SAIFR, Universidade Estadual Paulista, Rua Dr. Bento T. Ferraz 271, Sao Paulo, SP 01140-070, Brazil \\
$^{17}$Kavli Institute for Particle Astrophysics \& Cosmology, P. O. Box 2450, Stanford University, Stanford, CA 94305, USA\\
$^{18}$SLAC National Accelerator Laboratory, Menlo Park, CA 94025, USA\\
$^{19}$Cerro Tololo Inter-American Observatory, National Optical Astronomy Observatory, Casilla 603, La Serena, Chile\\
$^{20}$Kavli Institute for Cosmology, University of Cambridge, Madingley Road, Cambridge CB3 0HA, UK\\
$^{21}$Department of Physics and Astronomy, University of Pennsylvania, Philadelphia, PA 19104, USA\\
$^{22}$CNRS, UMR 7095, Institut d'Astrophysique de Paris, F-75014, Paris, France\\
$^{23}$Sorbonne Universit\'es, UPMC Univ Paris 06, UMR 7095, Institut d'Astrophysique de Paris, F-75014, Paris, France\\
$^{24}$Excellence Cluster Universe, Boltzmannstr.\ 2, 85748 Garching, Germany\\
$^{25}$Faculty of Physics, Ludwig-Maximilians University, Scheinerstr. 1, 81679 Munich, Germany\\
$^{26}$Jet Propulsion Laboratory, California Institute of Technology, 4800 Oak Grove Dr., Pasadena, CA 91109, USA\\
$^{27}$Department of Astronomy, University of Michigan, Ann Arbor, MI 48109, USA\\
$^{28}$Department of Physics, University of Michigan, Ann Arbor, MI 48109, USA\\
$^{29}$Max Planck Institute for Extraterrestrial Physics, Giessenbachstrasse, 85748 Garching, Germany\\
$^{30}$Universit\"ats-Sternwarte, Fakult\"at f\"ur Physik, Ludwig-Maximilians Universit\"at M\"unchen, Scheinerstr. 1, 81679 M\"unchen, Germany\\
$^{31}$Department of Physics, The Ohio State University, Columbus, OH 43210, USA\\
$^{32}$Australian Astronomical Observatory, North Ryde, NSW 2113, Australia\\
$^{33}$George P. and Cynthia Woods Mitchell Institute for Fundamental Physics and Astronomy, and Department of Physics and Astronomy, Texas A\&M University, College Station, TX 77843,  USA\\
$^{34}$Departamento de F\'{\i}sica Matem\'atica,  Instituto de F\'{\i}sica, Universidade de S\~ao Paulo,  CP 66318, CEP 05314-970, S\~ao Paulo, SP,  Brazil\\
$^{35}$Department of Astronomy, The Ohio State University, Columbus, OH 43210, USA\\
$^{36}$Department of Physics and Astronomy, Pevensey Building, University of Sussex, Brighton, BN1 9QH, UK\\
$^{37}$Instituto de F\'\i sica, UFRGS, Caixa Postal 15051, Porto Alegre, RS - 91501-970, Brazil\\
$^{38}$Department of Physics, University of Illinois, 1110 W. Green St., Urbana, IL 61801, USA\\
$^{39}$Argonne National Laboratory, 9700 South Cass Avenue, Lemont, IL 60439, USA\\
$^{40}$Department of Physics, Stanford University, 382 Via Pueblo Mall, Stanford, CA 94305, USA\\
$^{41}$Jodrell Bank Center for Astrophysics, School of Physics and Astronomy, University of Manchester, Oxford Road, Manchester, M13 9PL, UK \\
$^{42}$Centre for Theoretical Cosmology, DAMTP, University of Cambridge, Wilberforce Road, Cambridge CB3 0WA, United Kingdom \\
$^{43}$Department of Physics and Electronics, Rhodes University, PO Box 94, Grahamstown, 6140, South Africa
}